\documentclass[11pt]{article}
\pdfoutput=1
\usepackage{graphicx}
\usepackage{hyperref}
\usepackage{epsfig}
\usepackage{amsmath}
\usepackage{amssymb}
\usepackage{multirow}
\usepackage{color}
\usepackage[nosort]{cite}
\usepackage{hyperref}
\usepackage[fontsize=11.45pt]{scrextend}
\usepackage{braket,mathrsfs,slashed}
\usepackage{float}
\usepackage{setspace}
\usepackage[caption=false]{subfig}
\newlength{\abstractwidth}
\newcommand{\be}{\begin{equation}}
\newcommand{\ee}{\end{equation}}

\renewcommand{\title}[1]{\vbox{\center\bf{\Large{#1}}}\vspace{5mm}}
\renewcommand{\author}[1]{\vbox{\center#1}\vspace{5mm}}
\newcommand{\address}[1]{\vbox{\center\em#1}}
\newcommand{\email}[1]{\vbox{\center\tt#1}\vspace{5mm}}

\usepackage{dsfont}

\usepackage[paper=a4paper,margin=1.5cm]{geometry}
\usepackage[utf8]{inputenc}
\usepackage{rotating}
\usepackage{soul}

\allowdisplaybreaks

\usepackage{mathrsfs} 
\usepackage{bbm}
\usepackage{etoolbox} 
\usepackage{multirow}
\renewcommand\[{\begin{equation}}
\renewcommand\]{\end{equation}}

\newcommand{\ba}{\begin{eqnarray}}
\newcommand{\ea}{\end{eqnarray}}

\hypersetup{pdfstartview=FitV,colorlinks=true,linkcolor=midblue,citecolor=midblue,filecolor=midblue,urlcolor=midblue}

\definecolor{midblue}{rgb}{0,0,0.5}

\begin{document}

\newgeometry{top=3cm,bottom=3.3cm,right=2.4cm,left=2.4cm}
		
	\begin{titlepage}
		\begin{center}
			\hfill \\
			\hfill \\
			\hfill \\
			\vskip 0.5cm

			\title{\Large On the contour prescriptions in \\[2mm]
				string-inspired nonlocal field theories}
			
			\author{\large Luca Buoninfante}
			
			\address{Nordita, KTH Royal Institute of Technology and Stockholm University\\
				Hannes Alfvéns väg 12, SE-106 91 Stockholm, Sweden}
			
			\email{\rm \href{mailto:luca.buoninfante@su.se}{luca.buoninfante@su.se}}

		\end{center}

\begin{abstract}
In quantum field theory, a consistent prescription to define and deform integration contours in the complex energy plane is needed to evaluate loop integrals and compute scattering amplitudes. In some nonlocal field theories, including string field theory, interaction vertices contain transcendental functions of momenta that can diverge along certain complex directions, thus making it impossible to use standard techniques, such as Wick rotation, to perform loop integrals. The aim of this paper is to investigate the viability of several contour prescriptions in the presence of nonlocal vertices. We consider three ``different'' prescriptions, and establish their (in)equivalence in local and nonlocal theories. In particular, we prove that all these prescriptions turn out to be equivalent in standard local theories, while this is not the case for nonlocal theories where amplitudes must be defined first in Euclidean space, and then analytically continued to Minkowski.
We work at one-loop level and focus on the bubble diagram. In addition to proving general results for a large class of nonlocal theories, we show explicit calculations in a string-inspired nonlocal scalar model. 
\end{abstract}

\end{titlepage}

{
	\hypersetup{linkcolor=black}
	\tableofcontents
}

\baselineskip=17.63pt



\newpage

\section{Introduction}

In perturbative quantum field theory several techniques are required to evaluate loop integrals and compute scattering amplitudes consistently with analyticity and unitarity.
Some of the analyticity properties of an amplitude can be linked to physical observables through the condition of unitarity (i.e. optical theorem), for instance discontinuities (branch cuts) are physically related to decay rates and cross sections.  Therefore, when calculating loop integrals it is crucial to prescribe the correct rules for the deformation of an integration contour in the complex energy plane in order to circumvent poles and pinchings.

In standard theories, interaction vertices are local (i.e. polynomials in momenta), thus loop integrands that are made of products of propagators and vertices usually converge to zero in the limit of large loop momenta, e.g. $|k^0|\rightarrow \infty$ where $|k^0|$ is the modulus of the loop energy. It is this property that allows the use of several techniques for the deformation of a contour in the complex energy plane and the evaluation of loop integrals; for instance, the Cauchy theorem applied to infinite-radius semicircles and the Wick rotation.  
However, if the starting bare Lagrangian contains nonlocal (non-polynomial) differential operators, standard properties may fail to be valid and the usual methods to compute loop integrals cannot be applied.

An example characterized by  nonlocal vertices is \textit{string field theory}~\cite{Witten:1985cc,Eliezer:1989cr,Arefeva:2001ps,Zwiebach:2011rg,Sen:2015uaa,Pius:2016jsl,DeLacroix:2018arq}: a nonlocal quantum field theory  whose Feynman rules can be shown to reproduce the same expressions of the perturbative amplitudes computed in the world-sheet approach. In string field theory the interaction among strings is described via vertices containing transcendental functions of the momenta of the following type:
\begin{equation}
V(k_1,\dots,k_n)\sim e^{\sum_{i,j=1}^n\alpha_{ij}k_i \cdot k_j}\,,
\end{equation}	
where $\alpha_{ij}$ are constant coefficients. The simplest case is the Witten three-vertex for the open tachyon~\cite{Witten:1985cc,Eliezer:1989cr,Arefeva:2001ps,Zwiebach:2011rg} whose interaction potential is given by\footnote{Throughout this work we work with the mostly positive convention for the metric signature, $\eta={\rm diag}(-1,+1,+1,+1),$ and adopt natural units, $\hbar=1=c.$ With these conventions the d'Alembertian operator is defined as $\Box=\partial_\mu\partial^\mu=-\partial_t^2+\nabla^2,$ $\nabla^2=\partial_i\partial^i$ being the Laplacian operator.} $V(\phi)\sim(e^{\alpha\Box}\phi)^3,$ or in momentum space $V(k_1,k_2,k_3)\sim e^{-\alpha(k_1^2+k^2_2+k^2_3)},$ $\alpha$ being a positive constant. Another example is given by \textit{$p$-adic string}~\cite{Freund:1987kt,Brekke:1988dg,Freund:1987ck,Frampton:1988kr,Dragovich:2020uwa} that provides an effective tree-level Lagrangian description of the Veneziano amplitude; in this case the operator $e^{-\alpha\Box}$ appears in the kinetic term.

More general Lagrangians containing similar nonlocal operators were analysed even before string theory in Ref.~\cite{Born1,Born3,Pais:1950za,Efimov:1967pjn,Alebastrov:1973np,Alebastrov:1973vw} (see also references therein) to achieve finiteness of loop integrals, and more recently intensive investigations have been made for both matter and gravity sectors~\cite{Krasnikov:1987yj,Kuzmin:1989sp,Moffat:1990jj,Tomboulis:1997gg,Tomboulis:2015gfa,Talaganis:2014ida,Buoninfante:2018mre,Boos:2019fbu,Kolar:2021oba}, especially in relation to the problem of ghosts and renormalizability in higher-derivative theories of gravity~\cite{Biswas:2005qr,Modesto:2011kw,Biswas:2011ar,Modesto:2014lga,Modesto:2015lna,Biswas:2016egy,Frolov:2015bta,Buoninfante:2018xiw,Buoninfante:2018rlq,Boos:2018bxf,Koshelev:2017tvv,Koshelev:2020foq}. 

A generic nonlocal scalar Lagrangian reads~\cite{Tomboulis:2015gfa,Buoninfante:2018mre}
\begin{equation}
\mathcal{L}=\frac{1}{2}\phi(\Box-m^2)\phi-V(\tilde{\phi})\,,\qquad \tilde{\phi}=e^{-\frac{1}{2}\gamma(-\Box)}\phi\,,\label{nonlocal-lag-1}
\end{equation}
where $m$ is a mass parameter, and $\gamma(z)$ is required to be an \textit{entire function} of $z=-\Box$ in order to avoid additional degrees of freedom other than the one corresponding to $\Box=m^2.$ Since the operator $e^{\gamma(-\Box)}$ is invertible one can make the field redefinition $\phi=e^{\frac{1}{2}\gamma(-\Box)}\tilde{\phi}$ and, after integrating by parts, the above Lagrangian can be written in the following equivalent form:
\begin{equation}
\mathcal{L}=\frac{1}{2}\tilde{\phi}e^{\gamma(-\Box)}(\Box-m^2)\tilde{\phi}-V(\phi)\,.\label{nonlocal-lag-2}
\end{equation}
It is clear that as long as $\gamma(-\Box)$ is an entire function the propagator does not have any additional poles despite the presence of higher (infinite) order derivatives. 

The transcendental operators can make loop integrals convergent in the ultraviolet regime but, at the same time, the integrands can diverge along some complex direction because $|e^{-\gamma(k^2)}|\rightarrow \infty$ in the limit $|k^0|\rightarrow \infty$ and for some angle $\vartheta$ of the complex loop energy $k^0=|k^0|e^{i\vartheta}.$ Thus, any integration contour that extends to infinity in the complex plane would give a divergent contribution to a loop integral. This means that standard techniques such as the Wick rotation, or usual choices of contours with semi-circles of infinite radius, are not viable for the type of nonlocal theories introduced above.

To overcome these difficulties in the context of string field theory, Pius and Sen~\cite{Pius:2016jsl} introduced a new prescription to define and deform the integration contour consistently with the conditions of analyticity and unitarity. Such a prescription is quite general and applies to a wider class of nonlocal quantum field theories. In fact, subsequent works were made by other authors who investigated analyticity and unitarity for more generic nonlocal Lagrangians~\cite{Carone:2016eyp,Briscese:2018oyx,Chin:2018puw,Koshelev:2021orf,Briscese:2021mob}. 
However, the explanation of some crucial details of the new prescription may sometime appear not entirely clear, for example several statements about  amplitudes defined in Minkowski or Euclidean signature can be misleading. Moreover, full analytic computations are often absent, and the equivalence of the new prescription to others in the standard local quantum field theory is usually given for granted and never proven. 

In this work we wish to investigate and clarify several aspects of viable contour prescriptions in a wide class of nonlocal field theories. 
We mainly work with the one-loop bubble diagram but also comment on other type of diagrams and higher loops. 
The paper is organized as follows.
\begin{description}
	
	\item[Sec.~\ref{sec-local-vertices}:] We first consider the case of local (polynomial) vertices. We introduce three ``different'' contour prescriptions that we call Minkowski, Euclidean and Schwinger, and show their equivalence. Discussing the local case first will turn out to be very instructive and useful for the subsequent analysis of the nonlocal case.
	
	\item[Sec.~\ref{sec-nonlocal-vertex}:] We discuss the same prescriptions in the context of field theories with nonlocal (non-polynomial) vertices. In this case, we show that not all the prescriptions are equivalent. Indeed, Euclidean and Schwinger are well-defined and equivalent, whereas the Minkowski one gives divergent results because of the presence of singularities at infinity along certain directions in the complex energy plane. Besides giving proofs for generic nonlocal theories, we also perform fully analytic and explicit computations in a string-inspired nonlocal model.
	
	\item[Sec.~\ref{sec-discussion}:] We summarize and discuss the relevance of our results, and how they can be extended to more complicated diagrams and higher loops. We discriminate between theory formulations in Minkowski and Euclidean space, and emphasize the importance of initially defining nonlocal quantum field theories in Euclidean signature. Finally, we make concluding remarks and comment on future works.
	
	\item[App.~\ref{sec-app}:] We briefly review the unitarity condition on the $S$-matrix and its formulation via the optical theorem. We prove one-loop unitarity with nonlocal vertices. This Appendix will also be useful to clarify convention and notations that we use for propagators, vertices, and amplitudes in the main text.
	
\end{description}


\section{Local vertices}\label{sec-local-vertices}

\begin{figure}[t]
	\centering
	\subfloat[Subfigure 1 list of figures text][]{
		\includegraphics[scale=0.5]{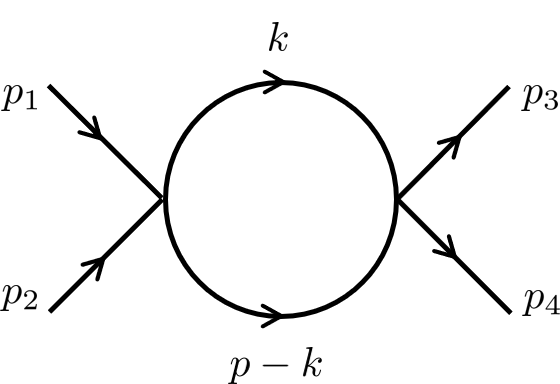}\label{fig1a}}\qquad\quad\,\,\,\,
	\subfloat[Subfigure 2 list of figures text][]{
		\includegraphics[scale=0.5]{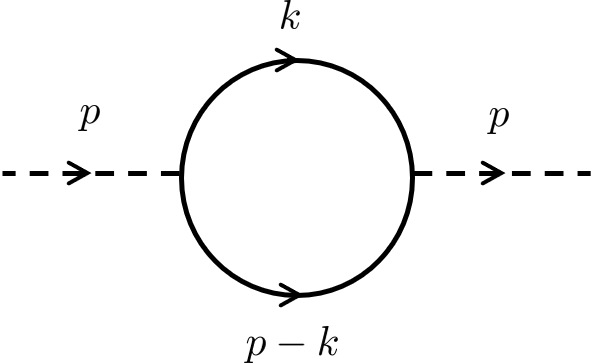}\label{fig1b}}
	\protect\caption{Type of diagrams corresponding to the one-loop integral in Eq.~\eqref{1-loop-two-gen-vert}. (a) is a one-loop diagram that can represent a scattering process, e.g. $\phi\phi\rightarrow \phi\phi$ with $\phi^4$ interaction;  $p=p_1+p_2=p_3+p_4$ is the total external momentum, and $k$ the loop momentum. The imaginary part of this diagram can correspond to an elastic cross section. (b) is a one-loop diagram that can represent a process in which the propagator of a scalar field $\phi$ (dashed line) is corrected by the one-loop contribution due to another scalar field $\chi$ (solid line). The imaginary part is related to a decay rate for the process $\phi\rightarrow \chi\chi.$}\label{fig1}
\end{figure}

In this Section we discuss standard and alternative methods to evaluate one-loop bubble diagrams in quantum field theories with local (polynomial) vertices. In particular, we are interested in the following type of scalar integral:
\begin{eqnarray}
\mathcal{M}(p_i,p_f)= -i\int_\mathcal{C} \frac{{\rm d}k^0}{2\pi}\int \frac{{\rm d}^3k}{(2\pi)^3} \frac{(-i)V(p_i,k,p-k)}{k^2+m^2-i\epsilon}\frac{(-i)V(k,p-k,p_f)}{(p-k)^2+m^2-i\epsilon}\,,
\label{1-loop-two-gen-vert}
\end{eqnarray}
where $p_i$ and $p_f$ are the ingoing and outgoing external four-momenta, $p$ is the total external momentum defined as the sum of all the ingoing (or, equivalently, outgoing) momenta,  $\mathcal{C}$ is the integration contour in the complex $k^0$ plane, and $-i\epsilon$ with $\epsilon>0$ is the usual Feynman shift of the poles. Since we are considering a theory with local vertices, $V(p_i,k,p-k)$ and $V(k,p-k,p_f)$ are polynomial functions of the external and internal momenta. For simplicity we are taking the two masses of the two internal propagators to be equal. In the case of a cubic or quartic vertex the one-loop integral~\eqref{1-loop-two-gen-vert} corresponds to Feynman diagrams of the type shown in Fig.~\ref{fig1}.

For local theories it is enough to focus on constant vertices because all the information about poles and pinching singularities is contained in the denominators of the internal propagators.
Therefore, in this Section we work with $V=-i\lambda,$ where $\lambda$ is some coupling constant, and analyse the following integral:
\begin{eqnarray}
\mathcal{M}(p^2)= (-i)\lambda^2\int_\mathcal{C} \frac{{\rm d}k^0}{2\pi}\int \frac{{\rm d}^3k}{(2\pi)^3} \frac{1}{k^2+m^2-i\epsilon}\frac{1}{(p-k)^2+m^2-i\epsilon}\,,
\label{1-loop-local}
\end{eqnarray}
where the dependence of the amplitude on $p^2$ follows from Lorentz invariance as we will explicitly show below.


The integrand contains four real poles in $k^0$, two for each propagator:
\begin{eqnarray}
Q_1=-\omega_{\vec{k}}+i\epsilon\,,\quad Q_2=p^0-\omega_{\vec{p}-\vec{k}}+i\epsilon\,,\quad Q_3=\omega_{\vec{k}}-i\epsilon\,,\quad Q_4=p^0+\omega_{\vec{p}-\vec{k}}-i\epsilon\,,
\label{real poles}
\end{eqnarray}
where $\omega_{\vec{k}}=\sqrt{\vec{k}^2+m^2}$ and $\omega_{\vec{p}-\vec{k}}=\sqrt{(\vec{p}-\vec{k})^2+m^2};$ see Fig.~\ref{fig2a} for a picture of their possible location in the complex $k^0$ plane.

When performing the integral some of the poles move around in the complex plane and can pinch the integration contour $\mathcal{C}$ from two opposite sides (after taking the $\epsilon\rightarrow 0$ limit). The only possible pinchings (for physical values of the external momenta) happen when either $Q_1=Q_4$ or $Q_2=Q_3,$ namely when
\begin{eqnarray}
p^0=\pm \big(\omega_{\vec{k}}+\omega_{\vec{p}-\vec{k}}\big)\,.\label{pinching-real}
\end{eqnarray}
Since we are interested in positive external energies ${\rm Re}[p^0]>0,$ then it is enough to discuss only the pinching with the ``$+$'' sign, i.e. $Q_2=Q_3.$ The same discussion will also apply to the other pinching.

The most delicate part in the evaluation of the one-loop integral~\eqref{1-loop-local} is the choice of contour $\mathcal{C},$ and of the prescription to be used for its deformation, in such a way that poles and pinchings can be circumvented in a consistent way. In what follows we investigate in detail three ``different'' prescriptions for the definition and the deformation of the integration contour in the complex energy plane, and also prove their equivalence in the presence of polynomial vertices. We name these prescriptions: \textit{Minkowski}, \textit{Euclidean}, and \textit{Schwinger}.

\subsection{Minkowski prescription}

The distinguishing feature of the Minkowski prescription is that all the external momenta (in particular the energy $p^0$) are kept real, i.e. do not have to be analytically continued to complex values, and the $k^0$-integral is performed along the real axis $\mathbb{R}$. The Feynman shift $p^2\rightarrow p^2-i\epsilon$  (or, equivalently, $m^2\rightarrow m^2-i\epsilon$ in the massive case) takes care of how to circumvent both poles and pinching singularities.

The Minkowski prescription for the evaluation of integrals such as~\eqref{1-loop-local} consists of the following set of rules:
\begin{enumerate}
	
	\item Keep all components of the external momenta real, i.e. $p^0\in \mathbb{R}$ and $\vec{p}\in \mathbb{R}^3,$ but analytically continue the internal energies to complex value, i.e. $k^0\in \mathbb{C},$ while keeping $\vec{k}\in \mathbb{R}^3.$ 
	
	\item Given that the integrand in Eq.~\eqref{1-loop-local} converges to zero in the limit $|k^0|\rightarrow \infty,$ recast the $k^0$-integral along $\mathbb{R}$ in a suitable form by making use of Cauchy theorem and/or Wick rotation. 
	
	\item Evaluate the resulting integral with a finite $\epsilon,$ and send it to zero ($\epsilon\rightarrow 0$) at the end of the computation. The Feynman shift $-i\epsilon$ will take care of how to circumvent the pinching singularities.
	
\end{enumerate}

We now show explicitly \textit{two} equivalent ways to implement the Minkowski prescription.

\begin{figure}[t!]
	\centering
	\subfloat[Subfigure 1 list of figures text][]{
		\includegraphics[scale=0.33]{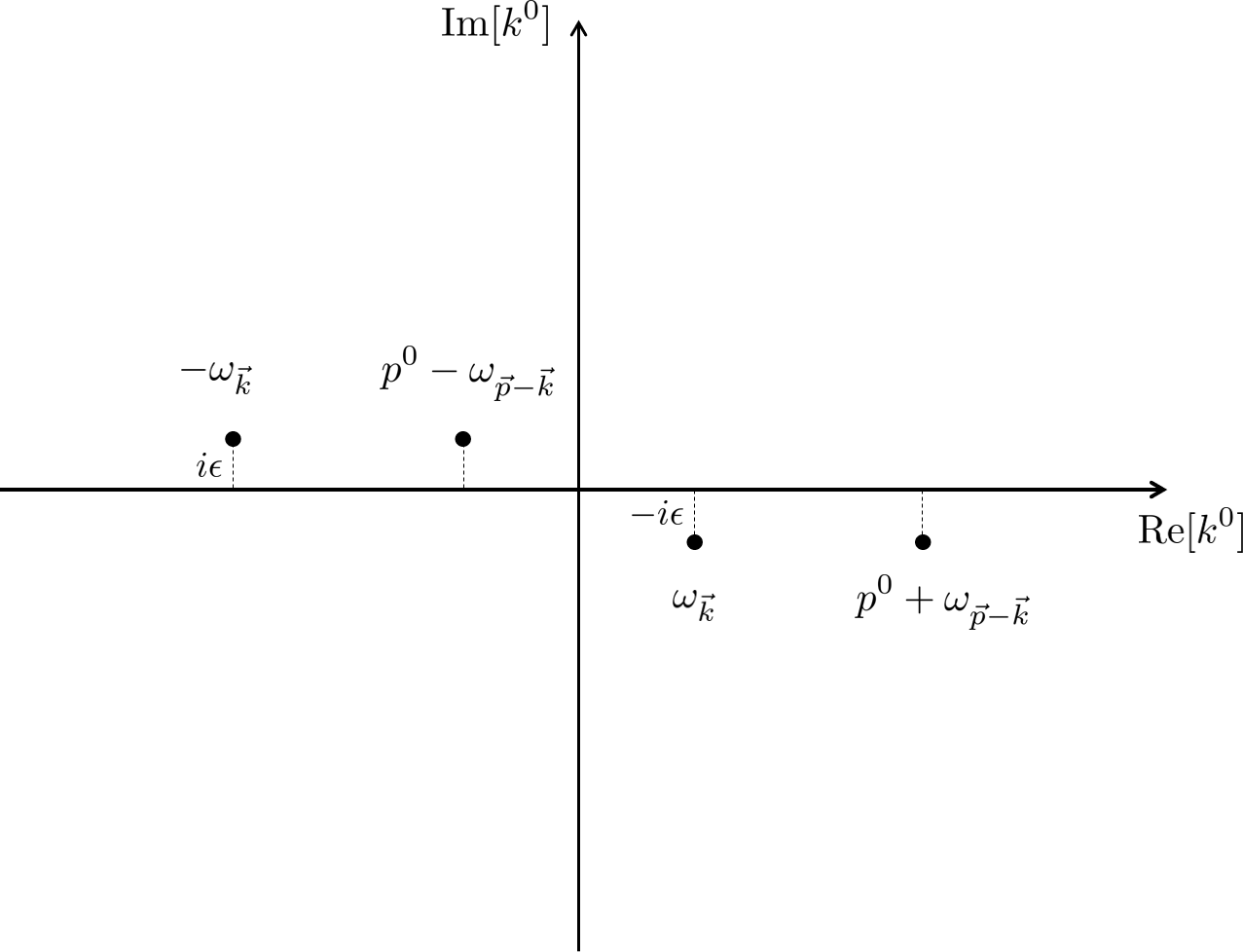}\label{fig2a}}\quad\,\,
	\subfloat[Subfigure 2 list of figures text][]{
		\includegraphics[scale=0.33]{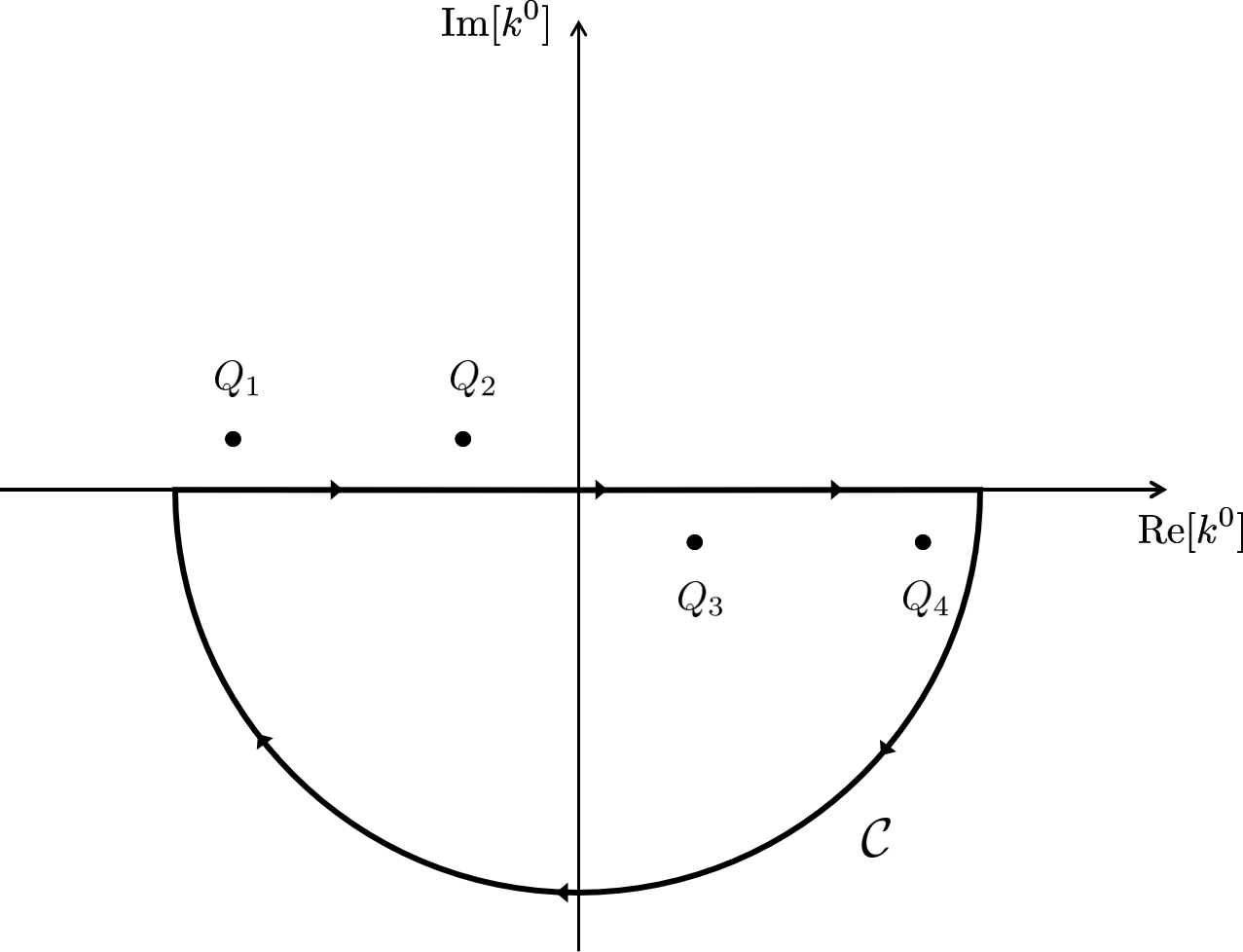}\label{fig2b}
	}
	\protect\caption{(a) Location of the four poles of the integrand in Eq.~\eqref{1-loop-local} in the complex $k^0$ plane. (b) Clockwise contour integration $\mathcal{C}$ for the evaluation of the $k^0$-integral in~\eqref{1-loop-local} according to the Minkowski prescription. The limit of infinite radius of the semicircle is understood.}\label{fig2}
\end{figure}

\paragraph{Minkowski (1).} We can evaluate the integral~\eqref{1-loop-local} considering the integration contour $\mathcal{C}$ in Fig.~\ref{fig2b}, and invoking the Cauchy theorem we can write
\begin{eqnarray}
\int_{\mathbb{R}}{\rm d}k^0 g(k^0)= \int_{\mathcal{C}}{\rm d}k^0 g(k^0)=-2\pi i\left[{\rm Res}\left\lbrace g(k^0)\right\rbrace_{k^0=Q_3}+{\rm Res}\left\lbrace g(k^0)\right\rbrace_{k^0=Q_4}  \right]\,,
\label{residue-Mink}
\end{eqnarray}
where $g(k^0)$ stands for the integrand of the $k^0$-integral in~\eqref{1-loop-local}, and we used the fact that the contribution of the semicircle vanishes at infinity due to the convergent behavior of the propagators. The overall minus sign comes from the clockwise orientation of the contour $\mathcal{C}$ in Fig.~\ref{fig2b}.
By computing the two residues we obtain
\begin{eqnarray}
\mathcal{M}(p^2)=\lambda^2 \int_{\mathbb{R}^3} \frac{{\rm d}^3k}{(2\pi)^3}\frac{1}{2\omega_{\vec{k}}2\omega_{\vec{p}-\vec{k}}} \left(  \frac{1}{p^0+\omega_{\vec{k}}+\omega_{\vec{p}-\vec{k}}}-\frac{1}{p^0-\omega_{\vec{k}}-\omega_{\vec{p}-\vec{k}}+i\epsilon}\right)\,,
\label{1-loop-two-resid}
\end{eqnarray}
from which the presence of pinching singularities becomes manifest. In the first denominator we took the limit $\epsilon\rightarrow 0$ as we are only interested in positive external energies $p^0>0,$ and in such a case only the second term can be singular and need to be treated with the Feynman shift.


By using the formula
\begin{equation}
\frac{1}{x\pm i\epsilon}={\rm P.V.}\left(\frac{1}{x}\right)\mp i\pi \delta(x)\,,\label{identity-delta}
\end{equation}
where ${\rm P.V.}$ stands for the Cauchy Principal value, we can write
\begin{eqnarray}
\mathcal{M}(p^2)&=&\lambda^2\int_{\mathbb{R}^3} \frac{{\rm d}^3k}{(2\pi)^3}\frac{1}{2\omega_{\vec{k}}2\omega_{\vec{p}-\vec{k}}} \left[ \frac{1}{p^0+\omega_{\vec{k}}+\omega_{\vec{p}-\vec{k}}}-{\rm P.V.}\left(\frac{1}{p^0-\omega_{\vec{k}}-\omega_{\vec{p}-\vec{k}}}\right)\right]\nonumber\\[2mm]
&&+i\pi\lambda^2 \int_{\mathbb{R}^3} \frac{{\rm d}^3k}{(2\pi)^3}\frac{1}{2\omega_{\vec{k}}2\omega_{\vec{p}-\vec{k}}}\delta(p^0-\omega_{\vec{k}}-\omega_{\vec{p}-\vec{k}})\,.
\label{real+imag+local-Mink}
\end{eqnarray}
The above integral has both real (first line) and imaginary (second line) contributions. The real part needs some further regularization because of the ultraviolet divergences for large loop momenta, while the imaginary component is finite and its value is constrained by the unitarity condition on the $S$-matrix (i.e. the optical theorem); see also App.~\ref{sec-app}. It is instructive to explicitly compute the imaginary part of the amplitude for all the prescriptions we investigate in this paper as a consistency check. For the sake of completeness, below we will compute the full amplitude~\eqref{1-loop-local} (both real and imaginary parts) with a more convenient method, see Eq.~\eqref{full-amplitude}.

We can easily evaluate the imaginary part of Eq.~\eqref{real+imag+local-Mink} by going to the centre-of-mass frame in which $\vec{p}=0,$ and making the change of variable $\omega_{\vec{k}}=\sqrt{\vec{k}^2-m^2}\equiv \omega$  we obtain:
\begin{eqnarray}
{\rm Im}[\mathcal{M}(p^2)]&=& \pi\lambda^2 \int_{\mathbb{R}^3} \frac{{\rm d}^3k}{(2\pi)^3}\frac{1}{4\omega^2}\delta(p^0-2\omega)\nonumber\\ [2mm]
&=& \frac{\lambda^2}{16\pi} \int_{-\infty}^\infty {\rm d}\omega \frac{\sqrt{\omega^2-m^2}}{\omega}\theta(\omega-m)\delta(p^0/2-\omega)\nonumber\\ [2mm]
&=& \frac{\lambda^2}{16\pi}\frac{\sqrt{(p^0)^2-4m^2}}{p^0}\theta(p^0-2m) \,,
\label{evaluated integral}
\end{eqnarray}
which coincides with the amplitude's discontinuity $(2i)^{-1}[\mathcal{M}(p^0+i\epsilon)-\mathcal{M}(p^0-i\epsilon)]$ in this case of the bubble diagram; the theta function $\theta(x)$ is equal to $1$ for $x\geq 0,$ and to $0$ for $x<0.$ From Eq.~\eqref{evaluated integral} it is clear that $\mathcal{M}(p^2)$ is analytic everywhere in the $p^0$-complex plane except on the real axis where there is a branch cut starting at the branch point $p^0=2m;$ in the case of negative external energies $p^0<0$ we would get a symmetric branch cut with branch point $p^0=-2m.$ The result~\eqref{evaluated integral} is consistent with the Cutkosky rules and unitarity (see App.~\ref{sec-app}).

The two conditions on the presence of branch cuts can be expressed through a single Lorentz invariant inequality $-p^2\geq 4m^2,$ and the imaginary part of the amplitude can be recast in the following Lorentz invariant form:
\begin{eqnarray}
{\rm Im}[\mathcal{M}(p^2)]= \frac{\lambda^2}{16\pi}\frac{\sqrt{-p^2-4m^2}}{\sqrt{-p^2}} \theta(-p^2-4m^2)\,.
\label{imag-lorentz-inv}
\end{eqnarray}

The discontinuity on the real axis is physical, and it can describe a decay rate or a cross section of processes that are kinematically allowed for external energies above the threshold $2m$.

\paragraph{Minkowski (2).} A second method to implement the Minkowski prescription is to express the $k^0$-integral along the real axis $\mathbb{R}$ as an integral over the imaginary axis $\mathcal{I}=[-i\infty,i\infty].$

By using the Feynman parametrization formula
\begin{eqnarray}
\frac{1}{AB}=\int_0^1 {\rm d}x \frac{1}{[A+(B-A)x]^2}\,,
\end{eqnarray}
with $A=(p-k)^2+m^2-i\epsilon$ and $B=k^2+m^2-i\epsilon,$ we can recast the integral~\eqref{1-loop-local} as
\begin{eqnarray}
\mathcal{M}(p^2)&=& (-i)\lambda^2\int_\mathcal{\mathbb{R}} \frac{{\rm d}k^0}{2\pi}\int \frac{{\rm d}^3k}{(2\pi)^3}\int_0^1{\rm d}x \frac{1}{[(p-k)^2+m^2+(k^2-(p-k)^2)x-i\epsilon]^2}\nonumber\\[2mm]
&=& (-i)\lambda^2\int_0^1{\rm d}x\int_\mathcal{\mathbb{R}} \frac{{\rm d}k^0}{2\pi}\int \frac{{\rm d}^3k}{(2\pi)^3} \frac{1}{[(k- p\,(1-x))^2+p^2x(1-x)+m^2-i\epsilon]^2}\nonumber\\[2mm]
&=&(-i)\lambda^2\int_0^1{\rm d}x\int_\mathcal{\mathbb{R}} \frac{{\rm d}k^0}{2\pi}\int \frac{{\rm d}^3k}{(2\pi)^3} \frac{1}{[k^2+\Delta-i\epsilon]^2}\,,
\end{eqnarray}
where in the last step we have made the change of variable $k\rightarrow k+p(1-x),$ and defined $\Delta\equiv p^2x(1-x)+m^2.$

\begin{figure}[t!]
	\centering
	\subfloat[Subfigure 1 list of figures text][]{
		\includegraphics[scale=0.33]{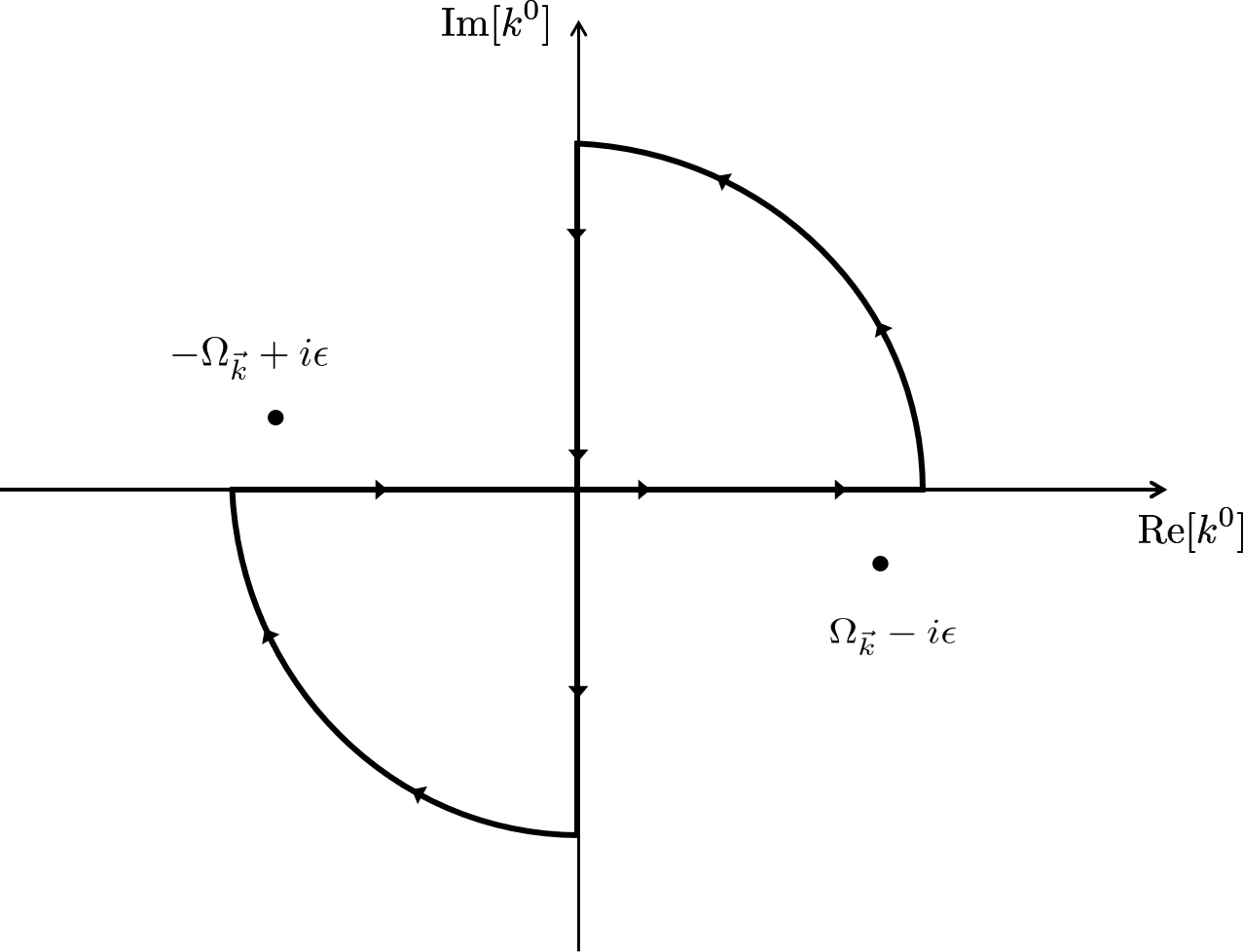}\label{fig2a-wick}}\quad\,\,
	\subfloat[Subfigure 2 list of figures text][]{
		\includegraphics[scale=0.33]{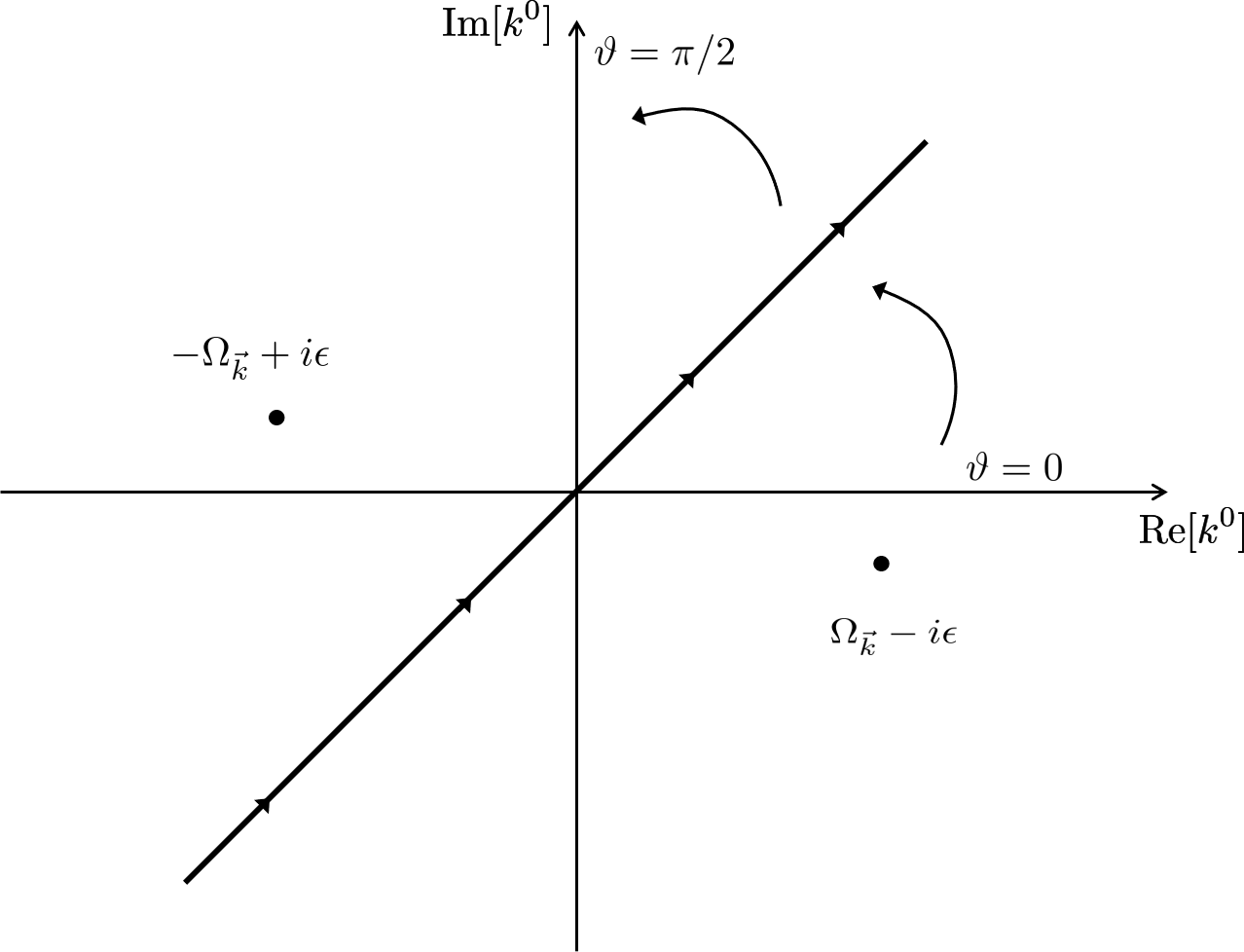}\label{fig2b-wick}
	}
	\protect\caption{(a) Location of the double poles $\pm \Omega_{\vec{k}}$ in the complex $k^0$ plane, and integration contour used in Eq.~\eqref{real-imag-int} to go from the real axis to the imaginary axis. The limit of infinite radius for the closed contours is understood. (b) Wick rotation from $\vartheta=0$ (real axis) to $\vartheta=\pi/2$ (imaginary axis) as an equivalent way to obtain Eq.~\eqref{real-imag-int}.}\label{fig2-wick}
\end{figure}

The integrand now has two double poles $\pm \Omega_{\vec{k}}\equiv\pm \sqrt{\vec{k}^2+\Delta}\mp i\epsilon$ whose location is shown in Fig.~\ref{fig2a-wick} for values of the external momenta below the threshold (i.e. $-p^2<4m^2$ which ensures $\vec{k}^2+\Delta>0$). Given such poles location, we consider the integration contour in Fig.~\ref{fig2a-wick}, and apply the Cauchy theorem to the two closed contours in the first and third quadrants which do not contain any pole. By doing so, and taking the infinite-radius limit, we get
\begin{eqnarray}
\int_{-\infty}^{\infty}{\rm d}k^0+\int_{+i\infty}^{-i\infty}{\rm d}k^0=0\quad \Leftrightarrow \quad \int_{-\infty}^{\infty}{\rm d}k^0=\int_{-i\infty}^{i\infty}{\rm d}k^0=i\int_{-\infty}^{\infty}{\rm d}k^4\,,
\label{real-imag-int}
\end{eqnarray}
where in the last step we have made the change of variable $k^0=ik^4$ such that $k^2=(k^4)^2+\vec{k}^2\geq0.$ The four-dimensional Euclidean integral over $k$ is ultraviolet divergent, and can be computed by implementing a regularization prescription. For instance, by using Pauli-Villars we obtain
\begin{eqnarray}
\mathcal{M}(p^2)=\int_0^1{\rm d}x\int\frac{{\rm d}k^4}{(2\pi)^4}\frac{1}{(k^2+\Delta)^2}
=-\frac{\lambda^2}{16\pi^2}\int_0^1{\rm d}x\log\left(\frac{\Delta}{\Lambda^2}\right)\,,
\end{eqnarray}
where $\Lambda$ is the cutoff energy scale (or renormalization scale). By working below the threshold, i.e. with $\Delta>0,$ we get~\cite{Itzykson:1980rh}
\begin{eqnarray}
\mathcal{M}(p^2)=\frac{\lambda^2}{16\pi^2}\left[ 2-\log\left(\frac{m^2}{\Lambda^2}\right) -\frac{\sqrt{p^2(p^2+4m^2)}}{p^2}\log\left(\frac{p^2+2m^2+\sqrt{p^2(p^2+4m^2)}}{2m^2}\right) \right]\,,
\label{full-amplitude}
\end{eqnarray}
which is the complete expression of the amplitude~\eqref{1-loop-local} containing both real and imaginary parts. The first term in the square brackets is unphysical as it can be absorbed in a redefinition of $\Lambda,$ the second is the divergent piece that can be eliminated through renormalization, while the third term is the physical finite contribution. 

We can analytically continue the logarithm to external momenta $-p^2>4m^2$ (which also implies $p^2+2m^2+\sqrt{p^2(p^2+4m^2)}<0$) by using the Feynman shift $p^2\rightarrow p^2-i\epsilon,$ i.e. $\log(x+i\epsilon)=\log(-x)+i\pi$ for $x<0.$ Then, we obtain the following expression for the real part valid above the threshold:
\begin{eqnarray}
{\rm Re}[\mathcal{M}(p^2)]=\frac{\lambda^2}{16\pi^2}\left[ 2-\log\left(\frac{m^2}{\Lambda^2}\right) +\frac{\sqrt{-p^2-4m^2}}{\sqrt{-p^2}}\log\left(\frac{\sqrt{-p^2-4m^2}+\sqrt{-p^2}}{\sqrt{-p^2-4m^2}-\sqrt{-p^2}}\right)\right]\,,
\end{eqnarray}
while the imaginary part coincides with~\eqref{imag-lorentz-inv} as expected.


In the massless case, the full expression for the amplitude~\eqref{full-amplitude} reduces to
\begin{eqnarray}
\mathcal{M}(p^2)=\frac{\lambda^2}{16\pi^2}\left[ 2-\log\left(\frac{p^2}{\Lambda^2}\right) \right]\,.
\label{full-amplitude-massless}
\end{eqnarray}
In this case, the branch point is $p^2=0,$ and the imaginary part of the amplitude reads
\begin{eqnarray}
{\rm Im}[\mathcal{M}(p^2)]= \frac{\lambda^2}{16\pi} \theta(-p^2)\,.
\end{eqnarray}
Hence, we have shown two ways of implementing the Minkowski prescription for the computation of the one-loop integral~\eqref{1-loop-local}.

Before concluding this Section, let us note that Eq.~\eqref{real-imag-int} for the $k^0$-integral can be equivalently obtained by performing the Wick rotation $k^0=e^{i\vartheta}k^4$ from $\vartheta=0$ to $\vartheta=\pi/2,$ see Fig.~\ref{fig2b-wick}. In this case, the convergence property of the integrand in the limit $|k^0|\rightarrow \infty$ guarantees that no contribution from infinity arises when performing the Wick rotation in the complex $k^0$ plane. As we will see in Sec.~\ref{sec-nonlocal-vertex}, the same technique is not well-defined in 
the presence of nonlocal vertices.


\subsection{Euclidean prescription}\label{sec-euclid}

We now introduce a second prescription to define and deform the integration contour $\mathcal{C}$ in~\eqref{1-loop-local}. The name ``Euclidean'' that we give to this prescription is motivated by the fact that, in this case, the external energies can be complex valued. In particular, the amplitude is initially defined as a function of purely imaginary external energies, and the integration contour is initially assumed to coincide with the imaginary axis $\mathcal{I}=[-i\infty,i\infty]$. 

The Euclidean prescription was elaborated in detail in the context of string field theory~\cite{Pius:2016jsl} but it is quite general and, obviously, also works in standard local quantum field theories. 
It consists of the following rules:
\begin{enumerate}
	
	\item Complexify both external and internal energies, i.e. $k^0\in \mathbb{C}$ and $p^0\in \mathbb{C},$ respectively, while keeping $\vec{k}\in \mathbb{R}^3$ and $\vec{p}\in \mathbb{R}^3.$ In particular, define the initial amplitude to be a function of purely imaginary external energies: $p^0=e^{i\vartheta}p^4$ with $p^4\in \mathbb{R},$ and $\vartheta=\pi/2$ initially.  \label{step1}
	
	\item Define the $k^0$-integration contour $\mathcal{C}$ to initially coincide with the imaginary axis $\mathcal{I}=[-i\infty,i\infty]$, and such that its ends are kept fixed at $\pm i \infty$.\footnote{To be more precise, it is sufficient that the imaginary parts of the ends are kept fixed at $\pm i\infty,$ while the real parts can also be non-zero but still finite, i.e. ${\rm ends}=A_\pm\pm i\infty$ with $0<A_{+}<C$ and $-C<A_{-}<0,$ where $C$ is a positive and finite real number. In this Subsection we only work with $A_\pm=0,$ but non-zero values of $A_\pm$ will be needed to prove the equivalence between the Euclidean and the Schwinger prescriptions in Sec.~\ref{sec-EP-SP-equiv}.\label{footnote-EP}} Any deformation of the contour must happen in finite-distance regions of the complex $k^0$ plane. \label{step2}
	
	\item Perform suitable deformations of the contour to circumvent poles and pinchings, while analytically continuing $p^0=e^{i\vartheta}p^4$ to physical real values, i.e. $\vartheta=\pi/2\rightarrow \vartheta=0,$ and take $\epsilon\rightarrow 0$ at the end of the computation. \label{step3}
	
\end{enumerate}
This prescription relies on the crucial assumption that the integrand of the amplitude is convergent along the imaginary axis in the limits $k^0\rightarrow \pm i\infty.$ This is indeed the case for Eq.~\eqref{1-loop-local}, and for all the types of nonlocal theories we will analyse in Sec.~\ref{sec-nonlocal-vertex}. According to the Euclidean prescription the Feynman shift $-i\epsilon$ is not enough to take care of poles and pinching singularities, the complexification of $p^0$ is necessary. Unlike the Minkowski prescription, in general the $k^0$-integral over $\mathcal{C}$ does not need to be equal to an integral over $\mathbb{R}$ (this will become more clear in the case of nonlocal vertices in Sec.~\ref{sec-nonlocal-vertex}). 

Let us remark that through the Euclidean prescription we are making an off-shell continuation from an (amputated) Green’s function to a scattering amplitude. This type of analytic continuation is expected to not work for theories in which the masslessness of external legs is protected by gauge invariance. These issues, together with more general analyticity properties of the amplitudes and the uniqueness of analytic continuation, were reviewed and generalized in the context of string field theory in Ref.~\cite{DeLacroix:2018arq}. 

Let us now evaluate the integral in Eq.~\eqref{1-loop-local} using the Euclidean prescription. We are going to consider two equivalent computations distinguished by a different choice of internal momenta: (i) $k$ and $p-k;$  (ii) $k+p/2$ and $k-p/2.$ Different choices of the internal momenta running through the loops correspond to different locations of the poles in the complex $k^0$ plane; this can alter the implementation of the prescription but without changing its main essence. It may happen that certain choices of internal momenta make some computation manifestly simpler (this will indeed be the case in Sec.~\ref{sec-explicit} when dealing with a specific nonlocal model). Therefore, it is worthwhile to show how the prescription works for at least two different choices of internal momenta.

\subsubsection{Internal momenta $k$ and $p-k$}\label{sec.k-p-k}

We start considering the choice of internal momenta $k$ and $p-k,$ whose corresponding integral exactly corresponds to the expression in~\eqref{1-loop-local}. According to the rules above, we define $\mathcal{M}(p^2)$ such that the integration contour $\mathcal{C}$ initially coincides with the imaginary axis $\mathcal{I}=[-i\infty,i\infty]$, and we take $p^0\in \mathbb{C}$ to be purely imaginary in such a way that the initial position of the poles satisfies the set of inequalities ${\rm Re}[Q_1]<{\rm Re}[Q_2]<0<{\rm Re}[Q_3]<{\rm Re}[Q_4],$ 
namely $Q_1,\,Q_2$ lie to the left of the imaginary axis, and  $Q_3,\,Q_4$ to the right.

Since we are interested in analytically continuing to positive real external energies, ${\rm Re}[p^0]>0,$ the only pinching we have to worry about is $p^0=\omega_{\vec{k}}+\omega_{\vec{p}-\vec{k}}$ (i.e. $Q_2=Q_3$). Let us divide the analysis into three regions; see also Refs.~\cite{Pius:2016jsl,Briscese:2018oyx}.

\begin{figure}[t!]
	\centering
	\subfloat[Subfigure 2 list of figures text][]{
		\includegraphics[scale=0.35]{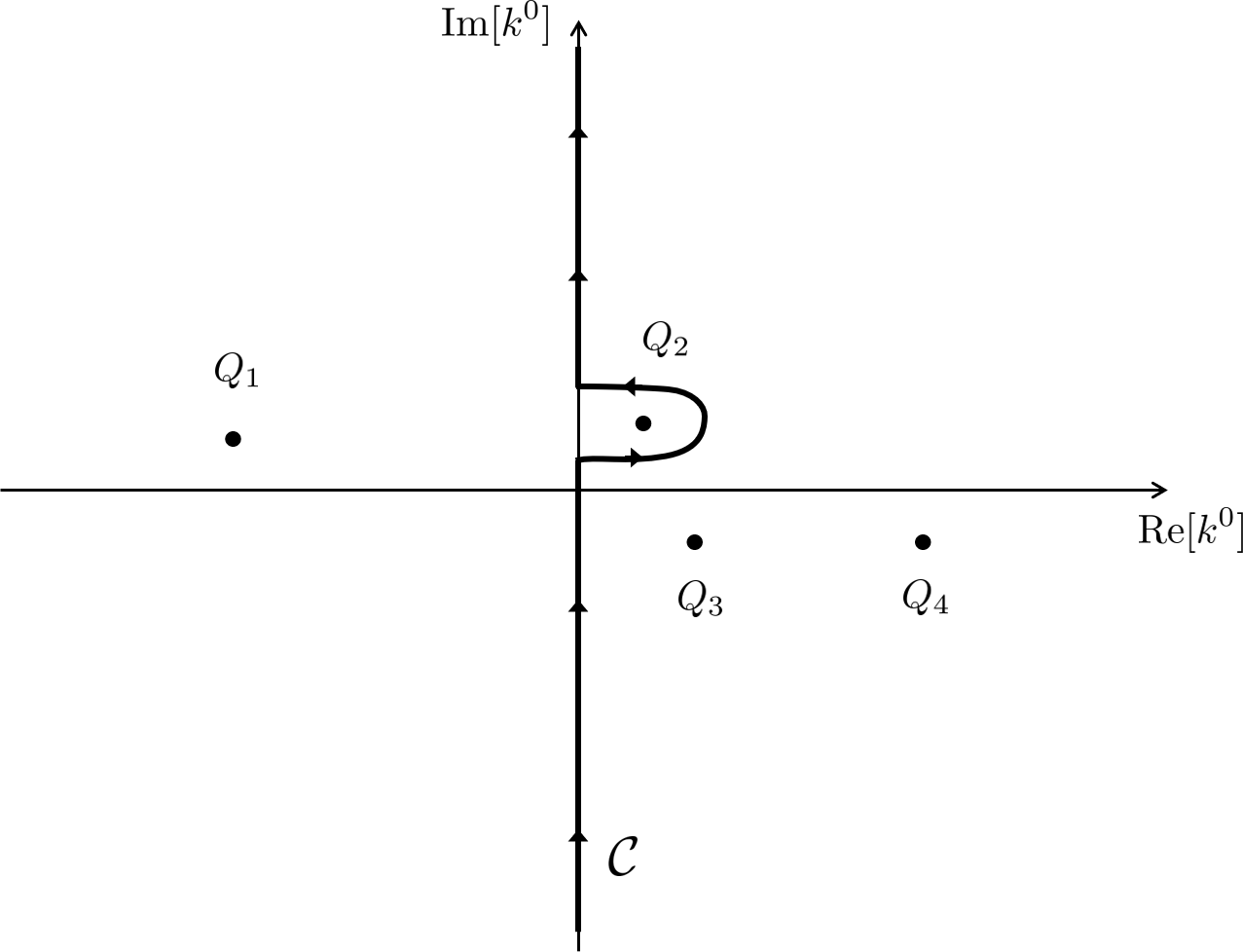}\label{fig4b}
	}\quad
	\subfloat[Subfigure 2 list of figures text][]{
		\includegraphics[scale=0.35]{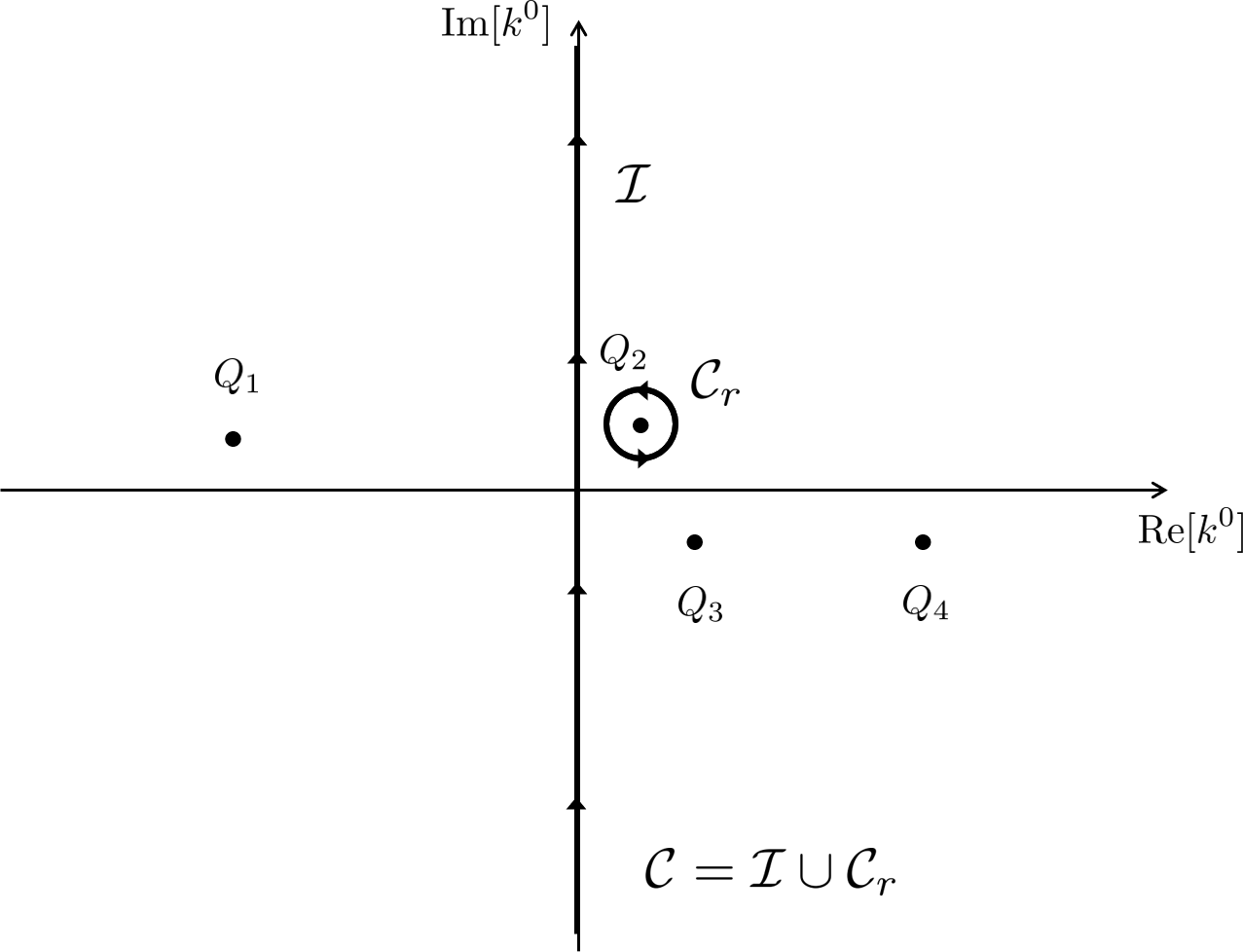}\label{fig4f}
	}
	\protect\caption{Illustration of the integration contour $\mathcal{C}$ in the complex $k^0$ plane according to the Euclidean prescription; the ends of $\mathcal{C}$ are kept fixed at $\pm i\infty.$ 
	The detailed description of the contour deformation is provided in the main text.}\label{fig4}
\end{figure}

\begin{itemize}
	
	\item In the region ${\rm Re}[p^0]<\omega_{\vec{p}-\vec{k}},$ the pole $Q_2$ is still to the left of the imaginary axis and no pinching singularity can appear, thus the integral can be computed with $\mathcal{C}=\mathcal{I}$ and turns out to be purely real.

	\item In the region ${\rm Re}[p^0]>\omega_{\vec{p}-\vec{k}},$ the pole $Q_2$ is to the right of the imaginary axis. In such a case, the integration contour $\mathcal{C}$ must be deformed according to the rules of the Euclidean prescription in order to keep $Q_1,$ $Q_2$ to the left and $Q_3,$ $Q_4$ to the right, and to maintain the ends of the contour fixed at $\pm i\infty;$ see Fig.~\ref{fig4b}. 
	
	Subsequently, we can topologically deform the contour as shown in Fig.~\ref{fig4f}, and  get two disconnected integration contours where $\mathcal{I}=[-i\infty,+i\infty]$ and $\mathcal{C}_r$ is an anticlockwise oriented circle around $Q_2.$ The total integration contour is now given by the union $\mathcal{C}=\mathcal{I}\cup\mathcal{C}_r.$ 
	
	The same type of deformation also works when $Q_2$ moves to the right of $Q_3,$ i.e. when ${\rm Re}[p^0]>\omega_{\vec{k}}+\omega_{\vec{p}-\vec{k}},$ and it remains valid after $p^0$ is analytically continued to real values and $\epsilon\rightarrow 0.$ 
	
	As already expected from the pinching condition~\eqref{pinching-real}, the region of the complex $k^0$ plane responsible for a non-vanishing imaginary part of the amplitude~\eqref{1-loop-local} is ${\rm Re}[Q_2]={\rm Re}[Q_3],$ namely $p^0=\omega_{\vec{k}}+\omega_{\vec{p}-\vec{k}}$ in the limit of real external energy and $\epsilon\rightarrow 0$.
	
	\item In the region  ${\rm Re}[p^0]=\omega_{\vec{p}-\vec{k}},$ the pole $Q_2$ lies on the imaginary axis, and the contour $\mathcal{C}=\mathcal{I}$ must be deformed in such a way that the pole $Q_2$ is circumvented with a semicircle of infinitesimal radius. Thus, in this case, the $k^0$-integral is given by the Cauchy principal value along $\mathcal{I}$ plus the contribution from the semicircle.

\end{itemize}
%


	Therefore, according to the Euclidean prescription the amplitude~\eqref{1-loop-local} can be split into two parts:
	\begin{eqnarray}
	\mathcal{M}(p^2)&=&(-i)\lambda^2\int_{\mathcal{I}\,\cup\, \mathcal{C}_r} \frac{{\rm d}k^0}{2\pi}\int_{\mathbb{R}^3} \frac{{\rm d}^3k}{(2\pi)^3} \frac{1}{k^2+m^2-i\epsilon}\frac{1}{(p-k)^2+m^2-i\epsilon}\nonumber\\[2mm]
	&=& \mathcal{M}_\mathcal{I}(p^2) +\mathcal{M}_{\mathcal{C}_r}(p^2)\,,\label{M-splitted}
	\end{eqnarray}
	where $\mathcal{M}_{\mathcal{I}}(p^2)$ is the contribution coming from the contour $\mathcal{I},$ while $\mathcal{M}_{\mathcal{C}_r}(p^2)$ is the one from $\mathcal{C}_r.$ 
	
	We can explicitly show that $\mathcal{M}_{\mathcal{I}}(p^2)$ is real. Since $\mathcal{I}$ is far from the poles, we can take real external energies $p^0\in\mathbb{R}$ and $\epsilon\rightarrow 0.$ Then, by changing the variable $k^0\rightarrow -ik^0$ we get
	\begin{eqnarray}
	\mathcal{M}_{\mathcal{I}}(p^2)
	=\lambda^2\int_{\mathbb{R}^4} \frac{{\rm d}^4k}{(2\pi)^4} \frac{1}{(k^0)^2+\vec{k}^2+m^2}\frac{1}{-(p^0+ik^0)^2+(\vec{p}-\vec{k})^2+m^2}\,.
	\label{real_M_I}
	\end{eqnarray}
	The complex conjugate is given by
	\begin{eqnarray}
	\mathcal{M}^*_{\mathcal{I}}(p^2)&=&\lambda^2\int_{\mathbb{R}^4} \frac{{\rm d}^4k}{(2\pi)^4} \frac{1}{(k^0)^2+\vec{k}^2+m^2}\frac{1}{-(p^0-ik^0)^2+(\vec{p}-\vec{k})^2+m^2}\nonumber\\[2mm]
	&=&\lambda^2\int_{\mathbb{R}^4} \frac{{\rm d}^4k}{(2\pi)^4} \frac{1}{(k^0)^2+\vec{k}^2+m^2}\frac{1}{-(p^0+ik^0)^2+(\vec{p}-\vec{k})^2+m^2}\nonumber\\[2mm]
	&=&\mathcal{M}_{\mathcal{I}}(p^2)\,,
	\label{real-I}
	\end{eqnarray}
	where in the last step we have made the change of variable $k^0\rightarrow -k^0.$ 
	
	We now evaluate the contribution coming from $\mathcal{C}_r$ using the residue theorem applied to the pole $Q_2=p^0-\omega_{\vec{p}-\vec{k}},$ and obtain:
	\begin{eqnarray}
	\mathcal{M}_{\mathcal{C}_r}(p^2)=- \lambda^2\int_{\mathbb{R}^3} \frac{{\rm d}^3k}{(2\pi)^3}\frac{\Theta\big(p^0-\omega_{\vec{p}-\vec{k}}\big)}{2\omega_{\vec{p}-\vec{k}}}\frac{1}{p^0+\omega_{\vec{k}}-\omega_{\vec{p}-\vec{k}}}\frac{1}{p^0-\omega_{\vec{k}}-\omega_{\vec{p}-\vec{k}}+i\epsilon}\,,
	\end{eqnarray}
	where the Heaviside theta function $\Theta\big(p^0-\omega_{\vec{p}-\vec{k}}\big)$ takes into account the fact that the integral over $\mathcal{C}_r$ is non-zero only when ${\rm Re}[Q_2]\geq 0,$ and it is defined as
	\begin{equation}
	\Theta(x)=\left\lbrace \begin{array}{cc}
	1\,,& x>0\,;\\[2mm]
	1/2\,,& x=0\,;\\[2mm]
	0\,,& x<0\,.
	\end{array}	\right.
	\label{theta-H}
	\end{equation}
	Note that, when the simple pole $Q_2$ is on the imaginary axis it only contributes half as compared to the case in which it lies entirely in the first quadrant. 
    
    By using the identity~\eqref{identity-delta} we can write
	\begin{eqnarray}
	\mathcal{M}_{\mathcal{C}_r}(p^2)&=&-\lambda^2\int_{\mathbb{R}^3} \frac{{\rm d}^3k}{(2\pi)^3}\frac{\Theta(p^0-\omega_{\vec{p}-\vec{k}})}{2\omega_{\vec{p}-\vec{k}}}  \frac{1}{p^0+\omega_{\vec{k}}+\omega_{\vec{p}-\vec{k}}}{\rm P.V.}\left(\frac{1}{p^0-\omega_{\vec{k}}-\omega_{\vec{p}-\vec{k}}}\right)\nonumber\\[2mm]
	&&+i\pi\lambda^2 \int_{\mathbb{R}^3} \frac{{\rm d}^3k}{(2\pi)^3}\frac{1}{2\omega_{\vec{k}}2\omega_{\vec{p}-\vec{k}}}\delta(p^0-\omega_{\vec{k}}-\omega_{\vec{p}-\vec{k}})\,.
	\label{real+imag+local-Eucl-C_r}
	\end{eqnarray}
	For $p^0>\omega_{\vec{p}-\vec{k}}$ the contribution in the first line is real, while the one in the second line is imaginary. We can immediately notice that the expression of the imaginary part coincides with the one obtained via the Minkowski prescription in Eq.~\eqref{real+imag+local-Mink}. It is less trivial to explicitly check the same for the real part, but below we will prove the equivalence between the two prescriptions which will definitely confirm that the real contributions should indeed coincide.
	
	\paragraph{Remark 1.} The expression~\eqref{real+imag+local-Eucl-C_r} is valid only when $p^0>\omega_{\vec{p}-\vec{k}},$ but not if $p^0=\omega_{\vec{p}-\vec{k}}.$ When the pole $Q_2$ lies on $\mathcal{I}$ we have $p^0=\omega_{\vec{p}-\vec{k}}$ and $\Theta(x=0)=1/2;$ this corresponds to a zero-measure subset of $\mathbb{R}^3$ which does not contribute to~\eqref{real+imag+local-Eucl-C_r}. In particular, the condition $p^0=\omega_{\vec{p}-\vec{k}}$ makes the delta in the second term of Eq.~\eqref{real+imag+local-Eucl-C_r} vanish. Therefore, in this case, the contribution to the imaginary part of the amplitude is zero, and the integral over $\mathcal{I}$ in Eq.~\eqref{real_M_I} must be interpreted as a Cauchy principal value.

\subsubsection{Internal momenta $k+p/2$ and $k-p/2$}\label{sec-euclid-k+p/2}

\begin{figure}[t!]
	\centering
	\subfloat[Subfigure 2 list of figures text][]{
	\includegraphics[scale=0.35]{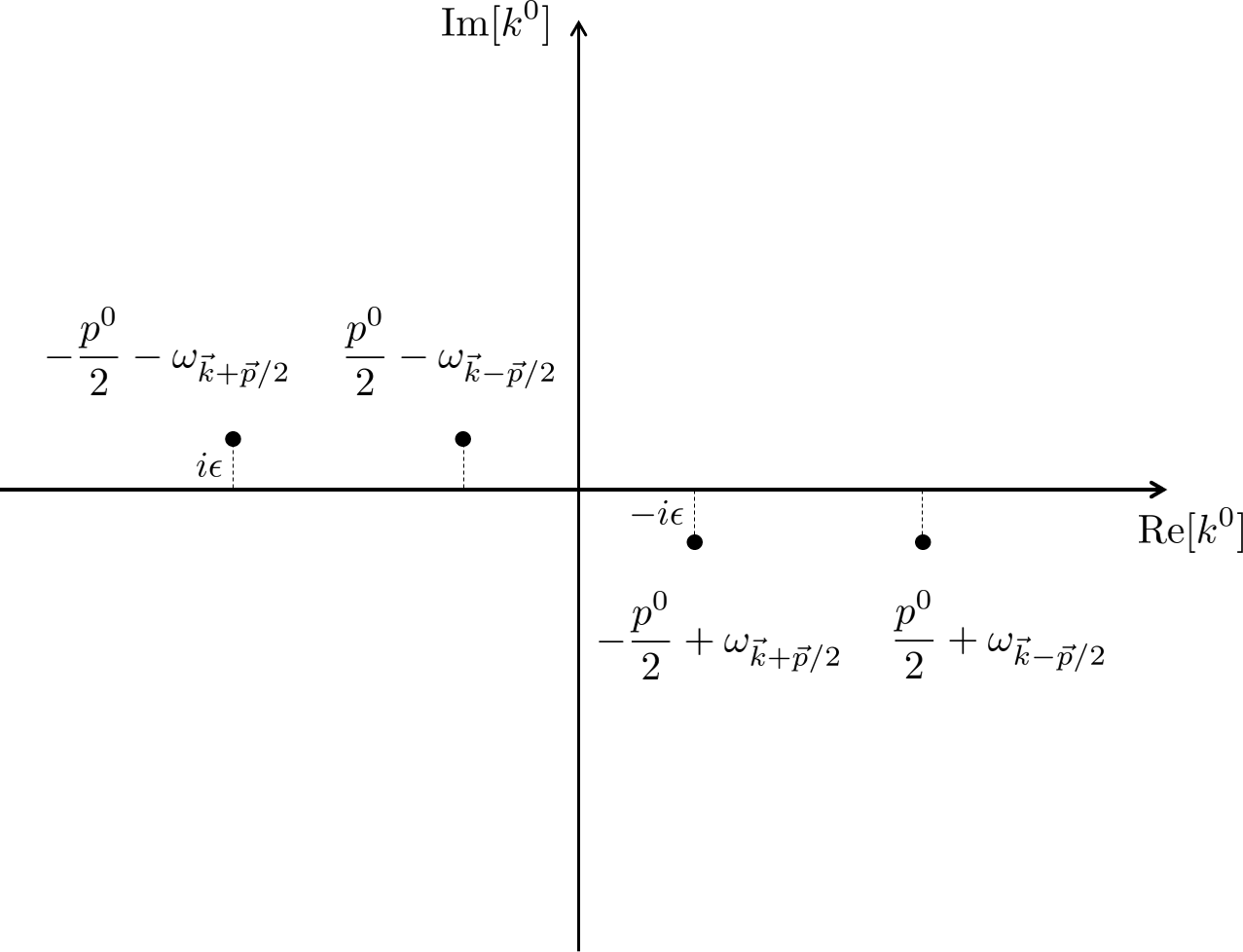}\label{fig2a-2}
	}\quad
	\subfloat[Subfigure 2 list of figures text][]{
		\includegraphics[scale=0.35]{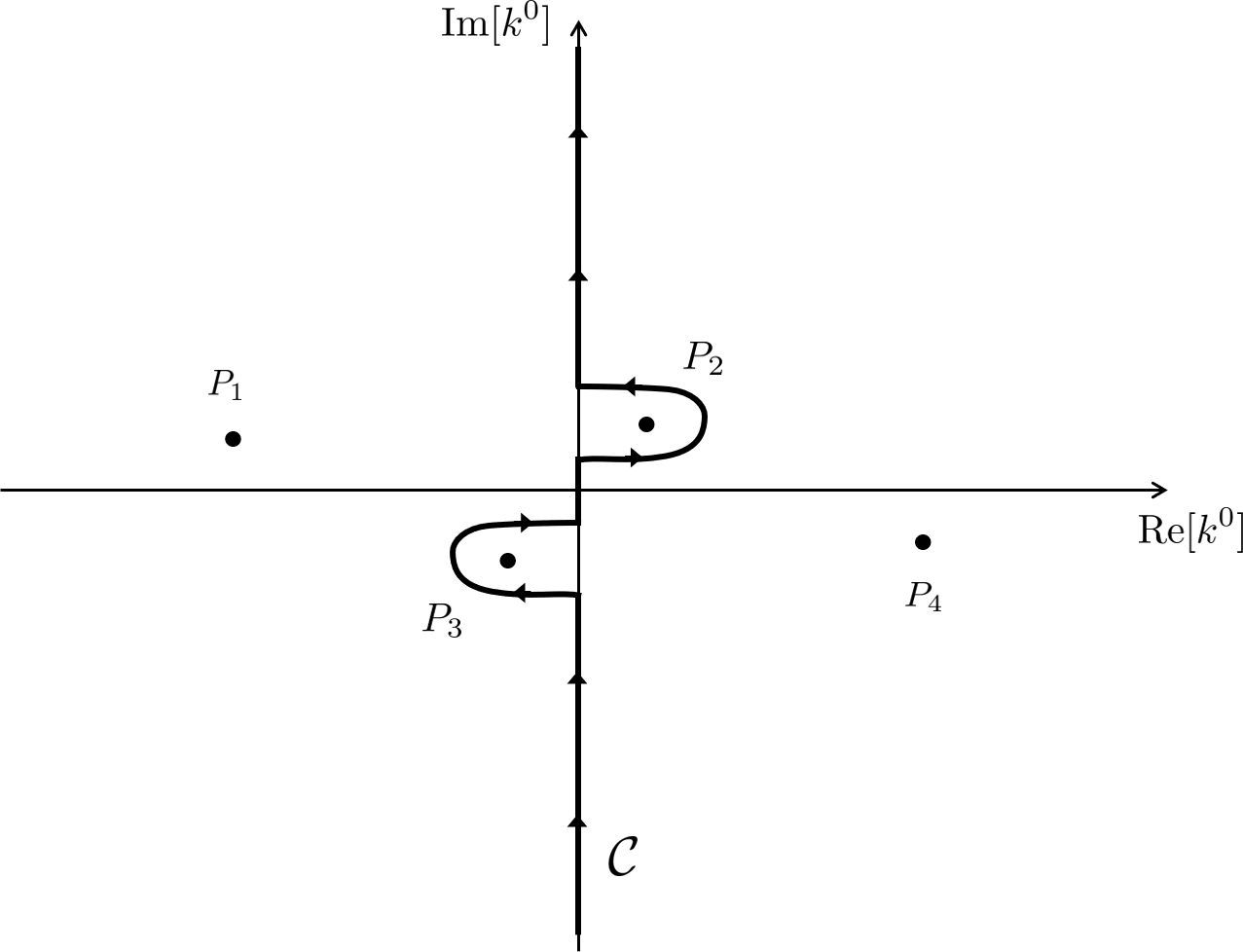}\label{fig4b-2}
	}\quad
	\subfloat[Subfigure 2 list of figures text][]{
		\includegraphics[scale=0.35]{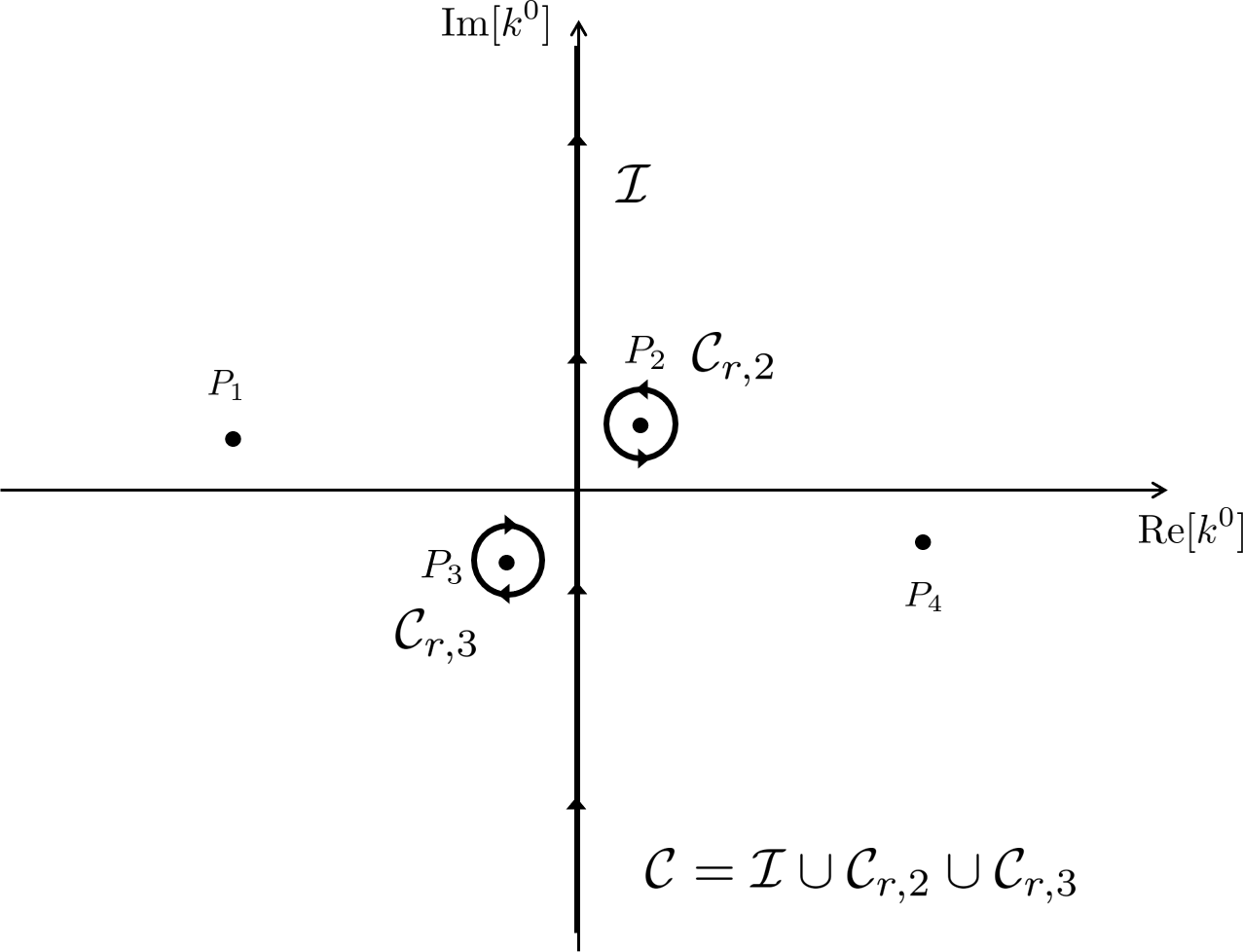}\label{fig4f-2}
	}\quad
\subfloat[Subfigure 2 list of figures text][]{
\includegraphics[scale=0.35]{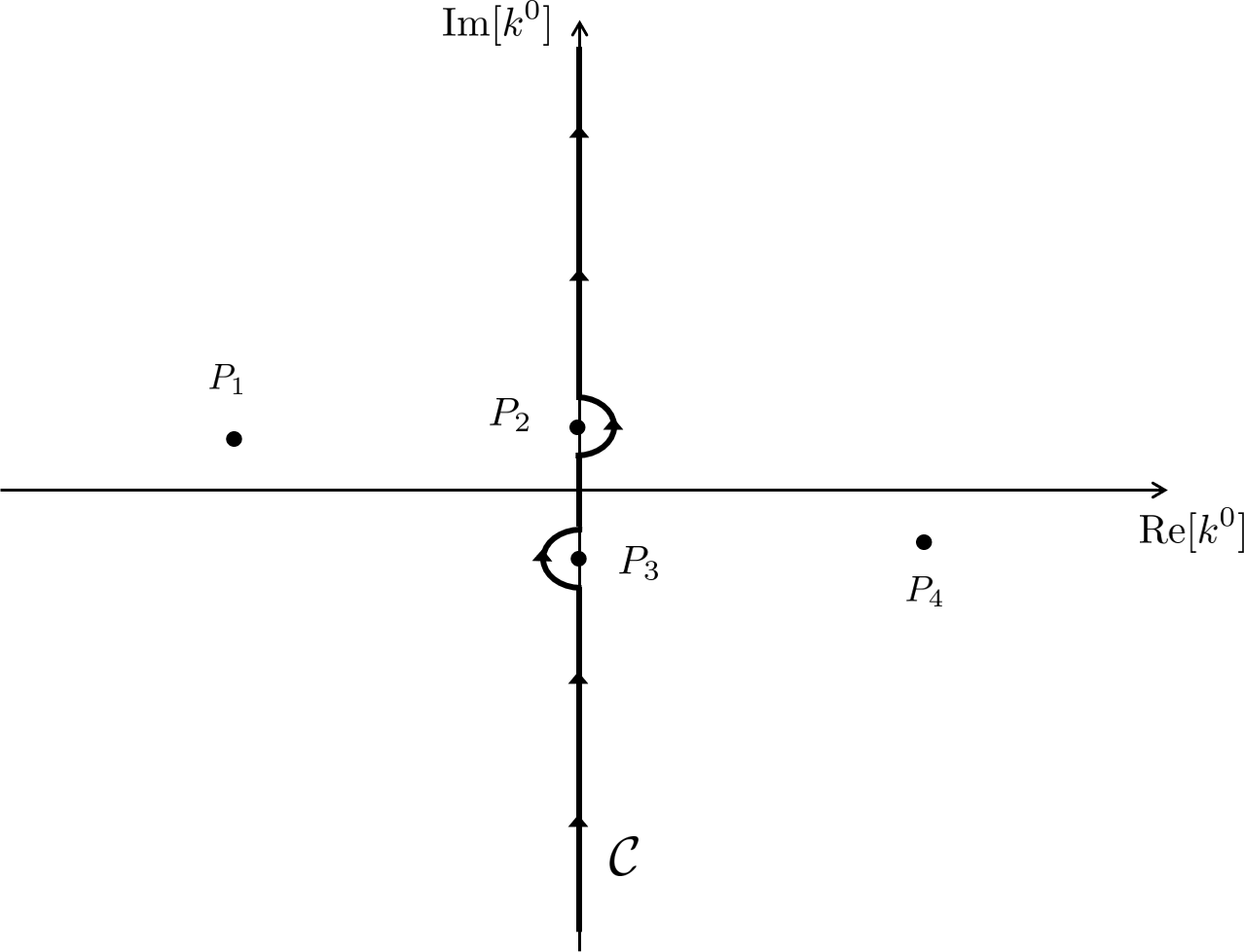}\label{fig-cauchy}
}	
	\protect\caption{(a) Location of the poles~\eqref{real poles-2} of the integrand in Eq.~\eqref{1-loop-local-2}. (b)-(c) Deformation of the contour $\mathcal{C}$ in the complex $k^0$ plane according to the Euclidean prescription applied to the amplitude~\eqref{1-loop-local-2} with internal momenta $k+p/2$ and $k-p/2$. (d) Integration contour when the poles lie on the imaginary axis; in this case the $k^0$-integral must be interpreted as the Cauchy principal value. This happens, for instance, when working in the centre-of-mass frame where the pinching condition reads ${\rm Re}[P_2]=0={\rm Re}[P_3].$}\label{fig4-2}
\end{figure}

Let us now make the same computation with a different choice of internal momenta. By changing the variable $k\rightarrow k+p/2,$ we can recast~\eqref{1-loop-local} in the following equivalent form:
\begin{eqnarray}
\mathcal{M}(p^2)= (-i)\lambda^2\int_\mathcal{C} \frac{{\rm d}k^0}{2\pi}\int_{\mathbb{R}^3} \frac{{\rm d}^3k}{(2\pi)^3} \frac{1}{(k+p/2)^2+m^2-i\epsilon}\frac{1}{(k-p/2)^2+m^2-i\epsilon}\,.
\label{1-loop-local-2}
\end{eqnarray}
The four poles of the integrand are now given by
\begin{eqnarray}
&&P_1=-\frac{p^0}{2}-\omega_{\vec{k}+\vec{p}/2}+i\epsilon\,,\qquad P_2=\frac{p^0}{2}-\omega_{\vec{k}-\vec{p}/2}+i\epsilon\,,\nonumber\\[2mm]
&&P_3=-\frac{p^0}{2}+\omega_{\vec{k}+\vec{p}/2}-i\epsilon\,,\qquad P_4=\frac{p^0}{2}+\omega_{\vec{k}-\vec{p}/2}-i\epsilon\,,
\label{real poles-2}
\end{eqnarray}
where $\omega_{\vec{k}\pm\vec{p}/2}=\sqrt{(\vec{k}\pm \vec{p}/2)^2+m^2};$  see Fig.~\ref{fig2a-2} for a picture of their possible location in the complex $k^0$ plane. Pinching singularities can happen when $P_1=P_4$ or $P_2=P_3,$ namely when
\begin{eqnarray}
p^0=\pm \big(\omega_{\vec{k}+\vec{p}/2}+\omega_{\vec{k}-\vec{p}/2}\big)\,.\label{pinching-real-2}
\end{eqnarray}
Since we are interested in positive external energies ${\rm Re}[p^0]>0,$ then it is enough to only discuss the pinching  $P_2=P_3;$  an analog discussion will apply to $P_1=P_4.$ 

According to the rules of the Euclidean prescription outlined above, and analogously to the other choice of internal momenta ($k$ and $p-k$) discussed in the previous Subsection, we define $\mathcal{M}(p^2)$ such that the integration contour $\mathcal{C}$ initially coincides with the imaginary axis $\mathcal{I}=[-i\infty,i\infty].$ Moreover, we take $p^0\in \mathbb{C}$ to be purely imaginary in such a way that initially $P_1,\,P_2$ lie to the left of the imaginary axis, and  $P_3,\,P_4$ to the right. Then, the contour must be deformed by circumventing poles and pinchings while keeping the ends fixed at $\pm i\infty$ as shown in Fig.~\ref{fig4b-2}. Subsequently, we can topologically deform $\mathcal{C}$ and transform it to the union of three disconnected contours $\mathcal{C}=\mathcal{I}\cup \mathcal{C}_{r,2}\cup \mathcal{C}_{r,3},$ where $\mathcal{C}_{r,2}$ is an anticlockwise-oriented circle around the pole $P_2,$ and $\mathcal{C}_{r,3}$ is a clockwise-oriented circle around the pole $P_3;$ see Fig.~\ref{fig4f-2}.


Therefore, according to the Euclidean prescription the amplitude~\eqref{1-loop-local-2} can be split into three contributions:
\begin{eqnarray}
\mathcal{M}(p^2)&=&(-i)\lambda^2\int_{\mathcal{I}\cup \mathcal{C}_{r,2}\cup \mathcal{C}_{r,3}} \frac{{\rm d}k^0}{2\pi}\int_{\mathbb{R}^3} \frac{{\rm d}^3k}{(2\pi)^3} \frac{1}{(k+p/2)^2+m^2-i\epsilon}\frac{1}{(k-p/2)^2+m^2-i\epsilon}\nonumber\\[2mm]
&=& \mathcal{M}_\mathcal{I}(p^2) +\mathcal{M}_{\mathcal{C}_{r,2}}(p^2)+\mathcal{M}_{\mathcal{C}_{r,3}}(p^2)\,,\label{M-splitted-2}
\end{eqnarray}
where $\mathcal{M}_{\mathcal{I}}(p^2)$ is the contribution coming from the contour $\mathcal{I},$ while $\mathcal{M}_{\mathcal{C}_{r,2}}(p^2)$ and $\mathcal{M}_{\mathcal{C}_{r,3}}(p^2)$ are the ones coming from $\mathcal{C}_{r,2}$ and $\mathcal{C}_{r,3},$ respectively. Their explicit expressions are
\begin{eqnarray}
\mathcal{M}_{\mathcal{I}}(p^2)&=&(-i)\lambda^2\int_{\mathcal{I}} \frac{{\rm d}k^0}{2\pi}\int \frac{{\rm d}^3k}{(2\pi)^3} \frac{1}{(k+p/2)^2+m^2}\frac{1}{(k-p/2)^2+m^2}\,,
\label{real_M_I-2}
\end{eqnarray}
\begin{eqnarray}
\mathcal{M}_{\mathcal{C}_{r,2}}(p^2)\!=\!- \lambda^2\!\!\int_{\mathbb{R}^3}\! \frac{{\rm d}^3k}{(2\pi)^3}\frac{\Theta\big(p^0/2-\omega_{\vec{k}-\vec{p}/2}\big)}{2\omega_{\vec{k}-\vec{p}/2}}\frac{1}{p^0+\omega_{\vec{k}+\vec{p}/2}-\omega_{\vec{k}-\vec{p}/2}}\frac{1}{p^0-\omega_{\vec{k}+\vec{p}/2}-\omega_{\vec{k}-\vec{p}/2}\!+i\epsilon}\,,\,\,
\end{eqnarray}
and
\begin{eqnarray}
\mathcal{M}_{\mathcal{C}_{r,3}}(p^2)\!=\! -\lambda^2\!\!\int_{\mathbb{R}^3} \!\frac{{\rm d}^3k}{(2\pi)^3}\frac{\Theta\big(p^0/2-\omega_{\vec{k}+\vec{p}/2}\big)}{2\omega_{\vec{k}+\vec{p}/2}}\frac{1}{p^0-\omega_{\vec{k}+\vec{p}/2}+\omega_{\vec{k}-\vec{p}/2}}\frac{1}{p^0-\omega_{\vec{k}+\vec{p}/2}-\omega_{\vec{k}-\vec{p}/2}\!+i\epsilon}\,.\,\,
\end{eqnarray}
By using the formula~\eqref{identity-delta}, we can extract the imaginary part of the amplitude:
\begin{eqnarray}
{\rm Im}[\mathcal{M}(p^2)]={\rm Im}[\mathcal{M}_{\mathcal{C}_{r,2}}(p^2)+\mathcal{M}_{\mathcal{C}_{r,3}}(p^2)]\!\!&\!\!=\!\!&\!\! \lambda^2 \int_{\mathbb{R}^3} \frac{{\rm d}^3k}{(2\pi)^3}\frac{1}{2\omega_{\vec{k}+\vec{p}/2}2\omega_{\vec{k}-\vec{p}/2}}\delta(p^0-\omega_{\vec{k}+\vec{p}/2}-\omega_{\vec{k}-\vec{p}/2})\nonumber\\[2mm]
&&\!\!\!\times \left[\Theta\big(\omega_{\vec{k}+\vec{p}/2}-\omega_{\vec{k}-\vec{p}/2}\big)+\Theta\big(\omega_{\vec{k}-\vec{p}/2}-\omega_{\vec{k}+\vec{p}/2}\big)\right]\,.
\label{two-thetas}
\end{eqnarray}
If $\vec{k}\cdot\vec{p}\neq 0,$ we can have the two possibilities $\omega_{\vec{k}+\vec{p}/2}>\omega_{\vec{k}-\vec{p}/2}$ or $\omega_{\vec{k}+\vec{p}/2}<\omega_{\vec{k}-\vec{p}/2}$, and in either cases the imaginary part of the amplitude reads
\begin{eqnarray}
{\rm Im}[\mathcal{M}(p^2)]= \lambda^2 \int_{\mathbb{R}^3} \frac{{\rm d}^3k}{(2\pi)^3}\frac{1}{2\omega_{\vec{k}+\vec{p}/2}2\omega_{\vec{k}-\vec{p}/2}}\delta(p^0-\omega_{\vec{k}+\vec{p}/2}-\omega_{\vec{k}-\vec{p}/2})\,,
\label{imag-2-ways}
\end{eqnarray}
and, by changing variable $\vec{k}\rightarrow \vec{k}-\vec{p}/2,$ it will coincide with the expressions in Eqs.~\eqref{real+imag+local-Mink} and~\eqref{real+imag+local-Eucl-C_r}. 

\paragraph{Remark 2.} The formula~\eqref{imag-2-ways} was obtained assuming that the poles $P_2$ and $P_3$ do not lie on the imaginary axis. In the opposite scenario, the integral~\eqref{1-loop-local-2} should be interpreted as the Cauchy principal value plus two infinitesimal semi-circles around the poles; see Fig.~\ref{fig-cauchy}. This happens when working in the centre-of-mass frame because  $\vec{p}=0$ implies $\vec{k}\cdot\vec{p}=0,$ and the pinching condition reduces to
\begin{eqnarray}
p^0=2\omega_{\vec{k}}\,,
\label{pinching-COM}
\end{eqnarray}
thus we have ${\rm Re}[P_2]=0={\rm Re}[P_3]$ at the pinching.
In this case, $\mathcal{M}_{\mathcal{C}_{r,2}}$ and $\mathcal{M}_{\mathcal{C}_{r,3}}$ take into account the contributions from the two semicircles in Fig.~\ref{fig-cauchy}, and the arguments of both Heaviside thetas in Eq.~\eqref{two-thetas} are zero, each of them contributes with half value, i.e. $\Theta(0)=1/2.$ Therefore, the total contribution to the imaginary part of the amplitude will still be the same (because $\Theta(0)+\Theta(0)=1$). 

It is also worthwhile to note that this scenario is not realized for the other choice of internal momenta, $k$ and $p-k,$ discussed before; see the remark at the end of Sec.~\ref{sec.k-p-k}. In that case, when the poles lie on the imaginary axis the contribution to the imaginary part of the amplitude is always zero. Whereas, for this second choice of internal momenta the imaginary part can be non-zero even when the poles lie on the imaginary axis because both the Heaviside theta function $\Theta(x)$ and the delta $\delta(y)$ can have the same argument $x=y$, thus when $x=0$ the delta $\delta(0)$ still contributes to the integral in Eq.~\eqref{two-thetas}. Indeed, this happens in the centre-of-mass frame as $\Theta(p^0/2-\omega_{\vec{k}})\delta(p^0-2\omega_{\vec{k}})\rightarrow \Theta(0)\delta(0)=1/2$ (under the integral sign).

The remark just made will prove to be very important in Sec.~\ref{sec-explicit}, where we will perform a full analytic computation in a specific nonlocal model by implementing the Euclidean prescription with the choice of internal momenta $k+p/2$ and $k-p/2.$

\subsection{Schwinger prescription}\label{sec-sch}

We now introduce a third prescription to evaluate the integral in Eq.~\eqref{1-loop-local}. We call it Schwinger prescription\footnote{It is worthwhile to mention that in the context of string theory some authors refer to this prescription with the name of $i\epsilon$-prescription~\cite{Witten:2013pra,Sen:2016gqt,Sen:2016ubf} because the Feynman shift $p^2\rightarrow p^2-i\epsilon$ (or $m^2\rightarrow m^2-i\epsilon$) is needed to make the Schwinger parametrization well-defined and the integral convergent.} because it involves the use of the Schwinger parametrization, and it consists of the following rules.

\begin{enumerate}
	
	\item  Rewrite the propagators in integral form via the Schwinger parametrization, taking $k^0\in \mathbb{C}$ and $\vec{k}\in \mathbb{R}^3$, while the external energy does not necessarily need to be complexified.
	
	\item Recast the amplitude as an integral over the Schwinger parameters. To circumvent the pinching singularity and make the integral convergent above the threshold, analytically continue the Schwinger parameters to complex values and deform the integration contour in a suitable way. 

	\item Evaluate the resulting integral and analytically continue the external energy to real physical values by implementing the Feynman shift $p^2\rightarrow p^2-i\epsilon$ or by complexifying $p^0$.
	
\end{enumerate}
Unlike the Euclidean prescription, the Schwinger does not necessarily require that the amplitude initially depends on purely imaginary external energies.

Let us now apply this prescription to the integral~\eqref{1-loop-local}. We rewrite the propagators by using the Schwinger parametrization:
\begin{eqnarray}
\frac{1}{k^2+m^2}&=&\int_0^\infty {\rm d}t_1\,e^{-t_1(k^2+m^2)}\,, \nonumber\\[2mm]
\frac{1}{(p-k)^2+m^2}&=&\int_0^\infty {\rm d}t_2\,e^{-t_2((p-k)^2+m^2)}\,,
\label{schwinger-param}
\end{eqnarray}
where $t_1$ and $t_2$ are sometime called Schwinger parameters. The two integrals are convergent as long as ${\rm Re}[k^2+m^2]>0$ and ${\rm Re}[(p-k)^2+m^2]>0.$ By using the above formula for the propagators, assuming that $\mathcal{C}=[-i\infty,i\infty],$\footnote{Note that, in the presence of local vertices it is not necessarily needed that the integration contour $\mathcal{C}$ initially coincides with the imaginary axis. For instance, one can start from $\mathcal{C}=\mathbb{R},$ and then Wick rotate to $\mathcal{I}.$ However, this is not possible in the presence of nonlocal vertices as we will discuss in Sec.~\ref{sec-nonlocal-vertex}.\label{footnote-2}} making the change of variable $k^0\rightarrow ik^0,$ and integrating on the full loop momentum $k$, we can recast the amplitude~\eqref{1-loop-local} in the following form
\begin{eqnarray} 
\mathcal{M}(p^2)&=&  \nonumber\lambda^2 \displaystyle \int_{0}^{\infty}{\rm d}t_1 \int_{0}^{\infty}{\rm d}t_2 \int \frac{{\rm d}^4k}{(2\pi)^4}e^{-t_1(k^2+m^2)}e^{-t_2((p-k)^2+m^2)}\\[2mm]
&=& \lambda^2 \int_{0}^{\infty}{\rm d}t_1 \int_{0}^{\infty}{\rm d}t_2\, e^{-p^2 t_2}e^{-m^2(t_1+t_2)}\int \frac{{\rm d}^4k}{(2\pi)^4}e^{-k^2(t_1+t_2)}e^{2k\cdot p\,t_2}\nonumber\\[2mm]
&=& \frac{\lambda^2}{16\pi^2}\int_0^\infty{\rm d}t_1\int_0^\infty{\rm d}t_2\,\frac{e^{-p^2\frac{t_1t_2}{t_1+t_2}}e^{-m^2(t_1+t_2)}}{(t_1+t_2)^2}\,,
\label{t1t2}
\end{eqnarray}
where we have used the four-dimensional spherical coordinates to go from the second to the third line, i.e. ${\rm d}^4k=k^3\sin\alpha^2\cos\theta\,{\rm d}k{\rm d}\alpha{\rm d}\theta{\rm d}\varphi,$ with $0\leq k\leq \infty,$ $0\leq \alpha,\theta\leq \pi,$ and $0\leq \varphi \leq 2\pi$.

Besides the ultraviolet divergences that are now translated to small values of the $t_i$'s, it is easy to check that the integral~\eqref{t1t2} diverges also in the double limit $t_1,t_2\rightarrow \infty$ when the inequality $-p^2>4m^2$ holds true, and this divergence is related to the pinching singularity and the imaginary part of the amplitude. The inequality $-p^2>4m^2$ also implies that the real part of $(p-k)^2+m^2$ cannot be kept always positive, thus going against the validity of the Schwinger parametrization~\eqref{schwinger-param}.  

\begin{figure}[t!]
	\centering
		\includegraphics[scale=0.35]{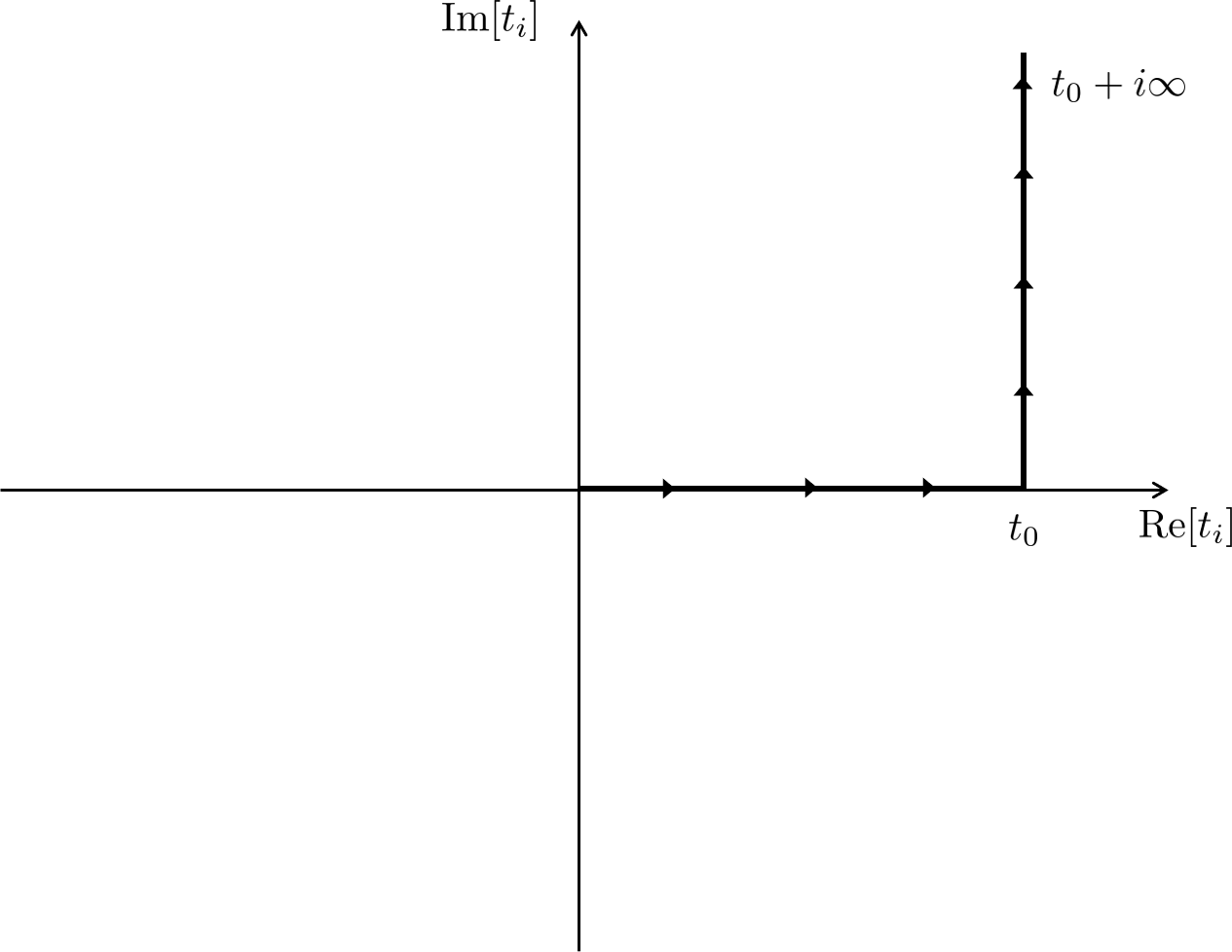}
	\protect\caption{Illustration of the integration contour for the integral~\eqref{t1t2} over the Schwinger parameters $t_1,\,t_2\in \mathbb{C}.$}\label{fig-schwinger}
\end{figure}

According to the Schwinger prescription outlined above, the divergence for external momenta above the threshold ($-p^2>4m^2$) can be avoided by analytically continuing the Schwinger parameters to complex values,  $t_i\in\mathbb{C},$ and deforming the contour as shown in  Fig.~\ref{fig-schwinger}. The $t_i$'s run from $0$ to $t_0,$ and from $t_0$ to $t_0+i\infty,$ where $t_0$ is some large positive real number such that the final result should not depend on its explicit value. See Refs.~\cite{Witten:2013pra,Sen:2016gqt,Sen:2016ubf} for the implementation of this prescription in the context of string theory.


To understand why this choice of contour can make the integral convergent for any value of the internal and external momenta, we can first apply it to one of the propagators in Eq.~\eqref{schwinger-param}. We have
\begin{eqnarray} 
\frac{1}{(p-k)^2+m^2}&=&\int_0^{t_0} {\rm d}t_2\,e^{-t_2((p-k)^2+m^2)}+\int_{t_0}^{t_0+i\infty} {\rm d}t_2\,e^{-t_2((p-k)^2+m^2)}\nonumber\\[2mm]
&=&\int_0^{t_0} {\rm d}t_2\,e^{-t_2((p-k)^2+m^2)}+i\,e^{-t_0((p-k)^2+m^2)}\int_{0}^{\infty} {\rm d}\tau\,e^{-i\tau((p-k)^2+m^2)}\,,
\end{eqnarray}
where in the second integral we have made the change of variable $t_2=t_0+i\tau$. Notice that even when ${\rm Re}[(p-k)^2+m^2]<0$ the second integral can still be convergent if ${\rm Im}[(p-k)^2+m^2]<0,$ and this can happen either by complexifying $p^0$ such that ${\rm Im}[(p^0-k^0)^2]>0,$ or using the Feynman shift $m^2\rightarrow m^2-i\epsilon$ (with $\epsilon>0$). The integrals can be computed, and the final result will give back the rational form of the propagator without any dependence on $t_0,$ as expected.

The same logic for the choice of contour applies to the one-loop amplitude in Eq.~\eqref{t1t2}. We show the explicit computation only for the imaginary part that is ultraviolet finite. Since an imaginary contribution appears only above the threshold, we directly work in the regime $-p^2>4m^2$, and choose new integration variables that are more suitable for those values of the external momentum.

By making the additional change of variable
\begin{eqnarray} 
u=t_1+t_2\,,\qquad v=t_1-t_2\,,
\label{uv-coord}
\end{eqnarray}
whose Jacobian determinant is equal to $1/2,$  the integral~\eqref{t1t2} can be recast as
\begin{eqnarray} 
\mathcal{M}(p^2)= \frac{\lambda^2}{32\pi^2}\int_0^\infty{\rm d}u\int_{-\infty}^\infty{\rm d}v\,\frac{e^{-u(p^2/4+m^2)}}{u^2}e^{\frac{p^2v^2}{4u}}\,,
\label{uv-int}
\end{eqnarray}
The integral in $v$ is a Gaussian $(-p^2>4m^2>0),$ and gives
\begin{eqnarray} 
\int_{-\infty}^\infty{\rm d}v\,e^{\frac{p^2v^2}{4u}}=\frac{2\sqrt{\pi}}{\sqrt{-p^2}}\sqrt{u}\,,
\end{eqnarray}
thus~\eqref{uv-int} becomes
\begin{eqnarray} 
\mathcal{M}(p^2)= \frac{\lambda^2}{16\pi^2}\frac{\sqrt{\pi}}{\sqrt{-p^2}}\int_0^\infty{\rm d}u\,\frac{e^{-u(p^2/4+m^2)}}{u^{3/2}}\theta(-p^2-4m^2)\,.
\label{uv-int-2}
\end{eqnarray}
We now deform the integration contour $u\in [0,\infty]$ as done for the $t_i$s in Fig.~\ref{fig-schwinger} by integrating from $0$ to $u_0,$ and from $u_0$ to $u_0+i\infty,$ where $u_0$ is some positive large real number, so that we obtain
\begin{eqnarray} 
\int_0^{u_0}{\rm d}u\,\frac{e^{-u(p^2/4+m^2)}}{u^{3/2}}+\int_{u_0}^{u_0+i\infty}{\rm d}u\,\frac{e^{-u(p^2/4+m^2)}}{u^{3/2}}\,.
\label{uv-int-3}
\end{eqnarray}
We are only interested in the imaginary part which is finite, thus we can neglect all the real terms. Moreover, only some of the contributions from the second integral in Eq.~\eqref{uv-int-3} are imaginary.
By making a further change of variable $u=u_0+iw,$ and shifting $p^2\rightarrow p^2-i\epsilon $ (or $m^2\rightarrow m^2-i\epsilon$), we can perform the second integral in~\eqref{uv-int-3} and pick its imaginary part:
\begin{eqnarray} 
i\,e^{-u_0(p^2/4+m^2)}\int_{0}^{\infty}{\rm d}w\,\frac{e^{-w(p^2/4+m^2)}e^{-w\epsilon}}{(u_0+iw)^{3/2}}\quad \rightarrow \quad i\sqrt{\pi}\sqrt{-p^2-4m^2}\,,
\label{uv-int-4}
\end{eqnarray}
where the arrow means that we are neglecting the real contributions, and the limit $\epsilon\rightarrow 0^+$ is understood in the calculation. 

Thus, inserting the last expression in Eq.~\eqref{uv-int-2}, the imaginary part of the one-loop amplitude~\eqref{t1t2} matches the one in Eq.~\eqref{imag-lorentz-inv}, consistently with Minkowski and Euclidean prescriptions.

\paragraph{Remark 3.} The choice of contour just discussed may appear quite involved. However, an equivalent and more practical way to apply the Schwinger prescription is the following: work with imaginary external energies, i.e. $p^0=ip^4$ ($p^4\in \mathbb{R}$) such that $p^2>0;$ perform the integral in Eq.~\eqref{t1t2} or~\eqref{uv-int-2} that is manifestly convergent for space-like external momenta; analytically continue $p^0$ to real physical values at the end of the calculation. One can check that the application of this alternative procedure to~\eqref{uv-int-2} would give exactly the same result for the imaginary part.
In Sec.~\ref{sec-explicit} we will implement the Schwinger prescription in this alternative way to explicitly compute both real and imaginary parts of the  bubble amplitude in a nonlocal model, and we will also verify the consistency of the result in the local limit.

\subsection{Equivalence between the prescriptions}\label{sec-equiv-local}

We discussed three prescriptions to define and deform the integration contour in the complex $k^0$ plane for the amplitude in Eq.~\eqref{1-loop-local}. We explicitly showed that all of them give the same result for the imaginary part, but we have not 
yet checked the consistency for the real part. More generally, we are now going to prove that the three prescriptions are all equivalent in the presence of local vertices. In this Subsection we will refer to the Minkowski, Euclidean, and Schwinger prescriptions with the abbreviations MP, EP and SP, respectively.

\subsubsection{Equivalence between MP and EP}

Let us first show that MP and EP are equivalent.  MP gives the result in Eq.~\eqref{residue-Mink} for the amplitude~\eqref{1-loop-local}, namely
\begin{eqnarray}
\mathcal{M}(p^2)=-2\pi i\left[{\rm Res}\left\lbrace g(k^0)\right\rbrace_{k^0=Q_3}+{\rm Res}\left\lbrace g(k^0)\right\rbrace_{k^0=Q_4}  \right]\,,
\label{Mink-result}
\end{eqnarray}
where $g(k^0)$ stands for the integrand of the $k^0$-integral in Eq.~\eqref{1-loop-local}; the integral in ${\rm d}^3k$ is included in the definition of $g(k^0).$

\begin{figure}[t!]
	\centering
	\subfloat[Subfigure 2 list of figures text][]{
		\includegraphics[scale=0.35]{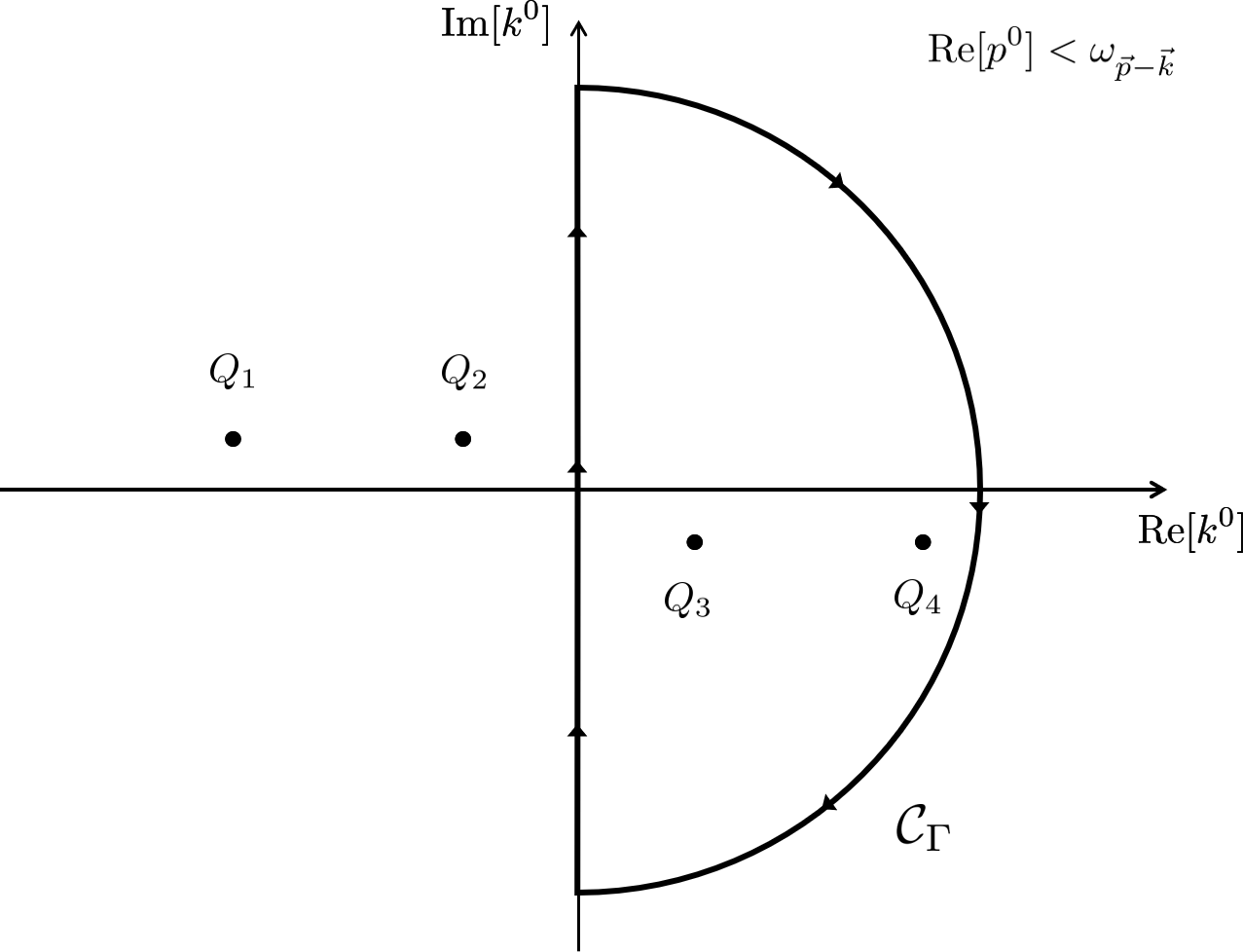}\label{fig-equiv-1}
	}\quad
	\subfloat[Subfigure 2 list of figures text][]{
		\includegraphics[scale=0.35]{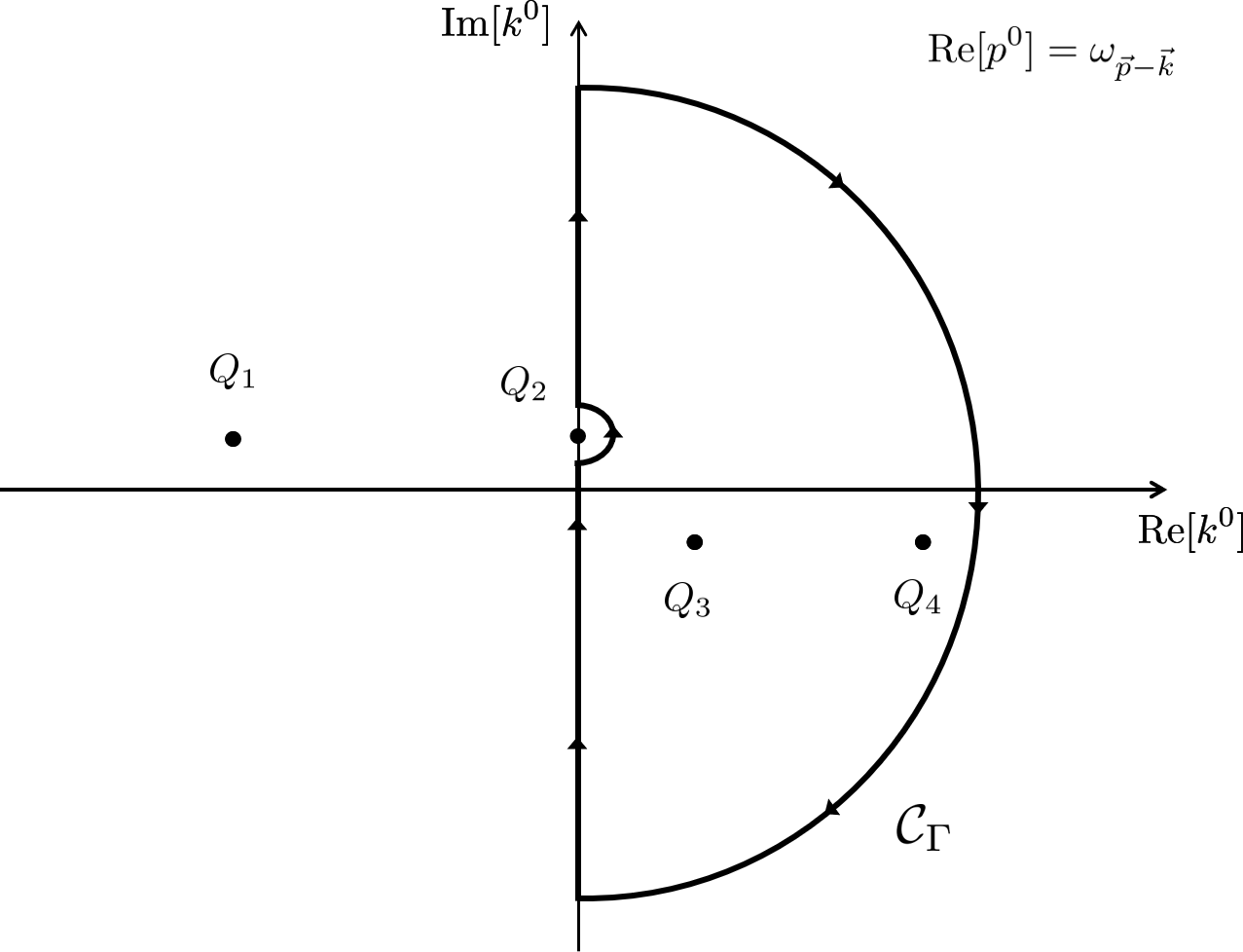}\label{fig-equiv-2}
	}\quad
	\subfloat[Subfigure 2 list of figures text][]{
		\includegraphics[scale=0.35]{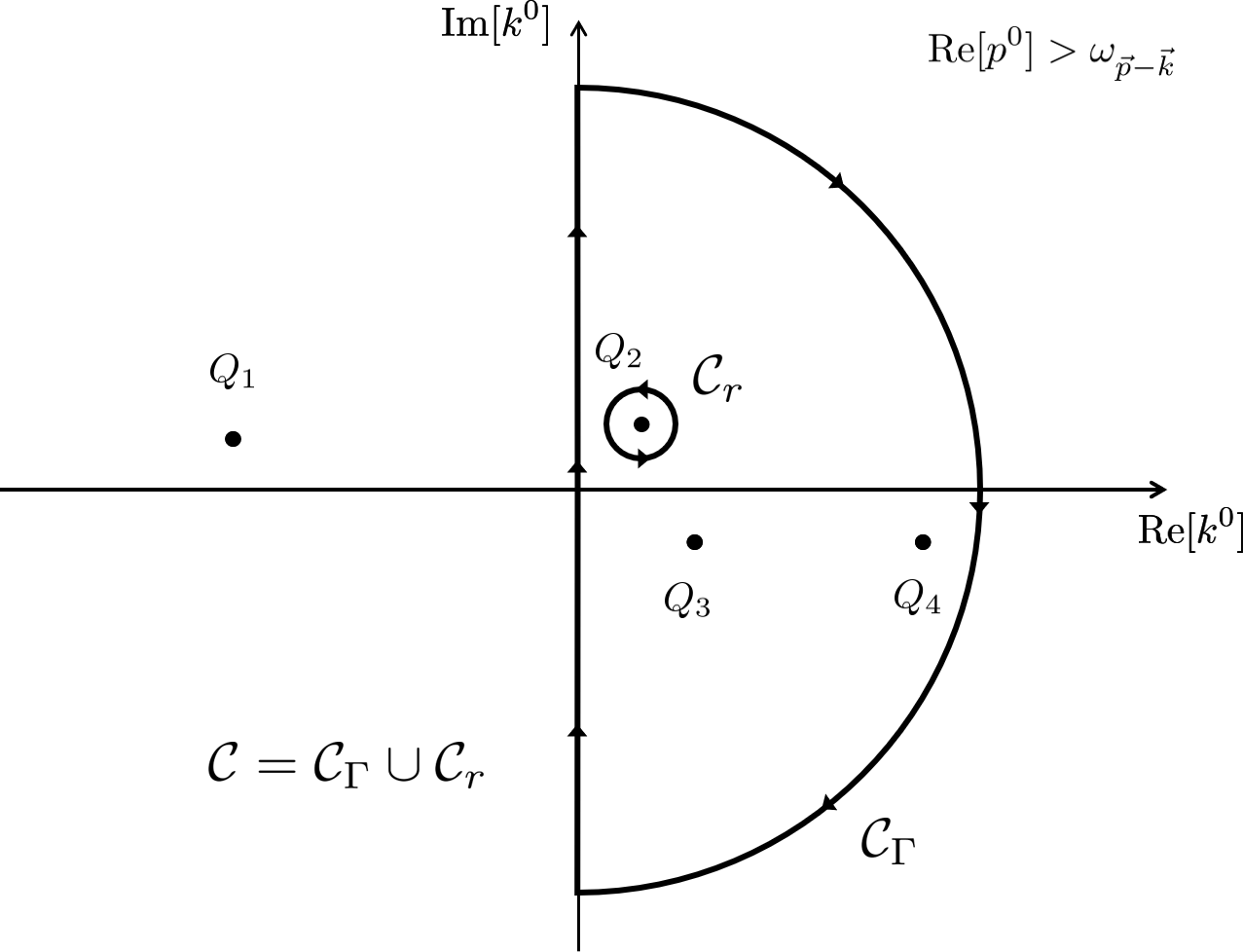}\label{fig-equiv-3}
	}\quad
	\subfloat[Subfigure 2 list of figures text][]{
		\includegraphics[scale=0.35]{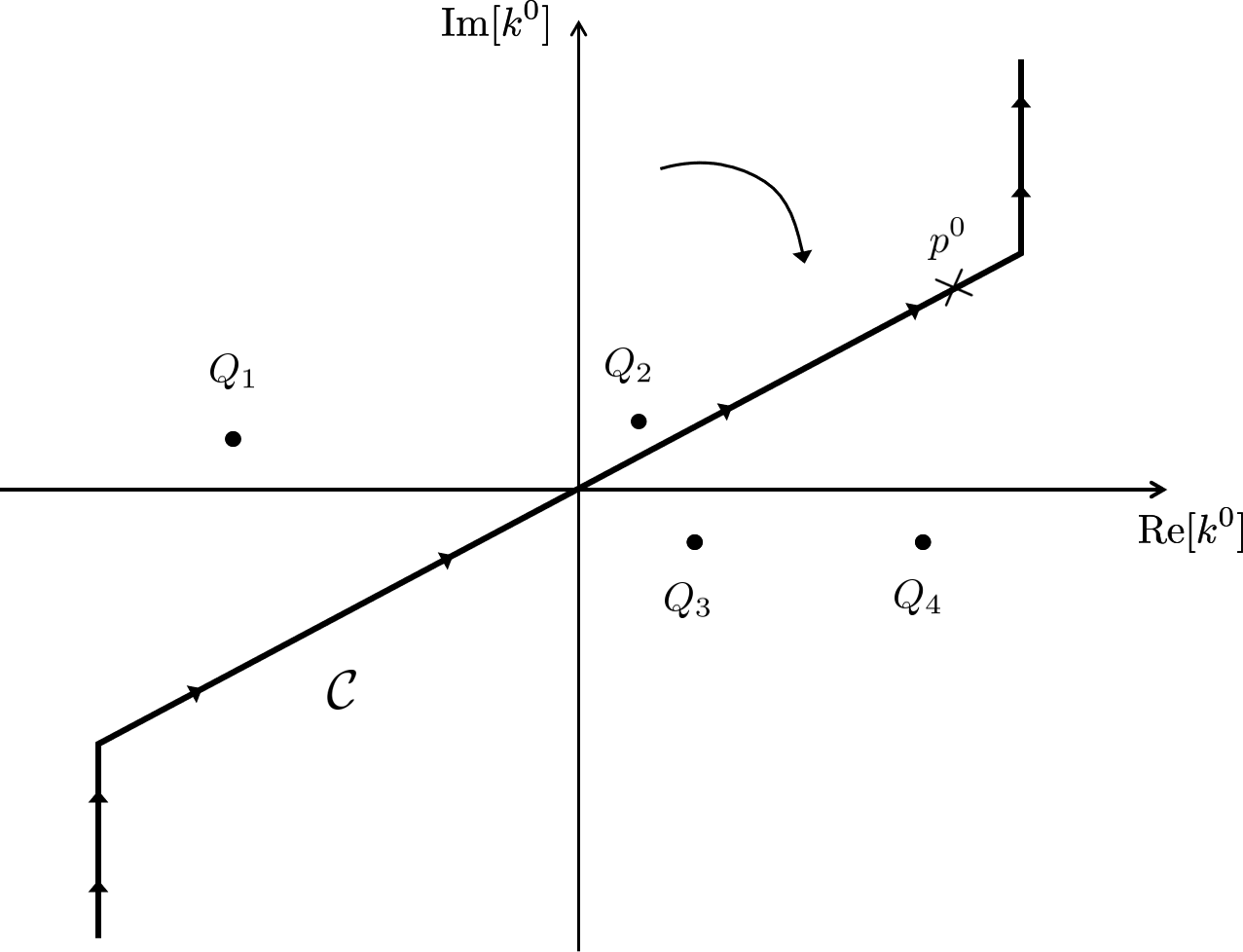}\label{fig-equiv-4}
	}	
	\protect\caption{(a)-(b)-(c) illustration of the integration contours in complex $k^0$ plane used to prove the equivalence between Minkowski (MP) and Euclidean (EP) prescriptions. Each of them corresponds to a different region of values for external and internal momenta: (a) ${\rm Re}[p^0]<\omega_{\vec{p}-\vec{k}};$ (b) ${\rm Re}[p^0]=\omega_{\vec{p}-\vec{k}};$ (c) ${\rm Re}[p^0]>\omega_{\vec{p}-\vec{k}}.$ (d) Illustration of the integration contour in complex $k^0$ plane used to prove the equivalence between Euclidean (EP) and Schwinger (SP) prescriptions. The cross represents the external energy  $p^0.$}\label{fig-equiv}
\end{figure}

According to EP the integration contour $\mathcal{C}$ must initially coincide with the imaginary axis $\mathcal{I}=[-i\infty,i\infty],$ and its ends must be kept fixed while deforming it and analytically continuing the external energy to real value. Since in the case of local vertices the integrand $g(k^0)$ converges to zero in the limit $|k^0|\rightarrow \infty,$ we can recast the integral over $\mathcal{I}$ as an integral over a closed semicircle $\mathcal{C}_\Gamma$ as shown in Figs.~\ref{fig-equiv-1},~\ref{fig-equiv-2} and~\ref{fig-equiv-3}. Let us divide the proof in three regions.
\begin{itemize}
	
	\item In the region ${\rm Re}[p^0]<\omega_{\vec{p}-\vec{k}}$ illustrated in Fig.~\ref{fig-equiv-1}, the pole $Q_2$ is still to the left of the imaginary axis, and we can write
	\begin{equation}
	\mathcal{M}(p^2)=\int_{\mathcal{I}}{\rm d}k^0g(k^0)=\int_{\mathcal{C}_\Gamma}{\rm d}k^0g(k^0)=-2\pi i\left[{\rm Res}\left\lbrace g(k^0)\right\rbrace_{k^0=Q_3}+{\rm Res}\left\lbrace g(k^0)\right\rbrace_{k^0=Q_4}  \right]\,,
	\label{E-M-1}
	\end{equation}
	which coincides with~\eqref{Mink-result}.
		
	\item In the region  ${\rm Re}[p^0]=\omega_{\vec{p}-\vec{k}}$ illustrated in Fig.~\ref{fig-equiv-2}, the pole $Q_2$ lies on the imaginary axis. In this case, the contour $\mathcal{I}$ must be deformed as shown in Fig.~\ref{fig-equiv-2}. By using the residue theorem we obtain again the same expression as in~\eqref{E-M-1} which, in turn, coincides with~\eqref{Mink-result}.
	
	\item In the region ${\rm Re}[p^0]>\omega_{\vec{p}-\vec{k}}$ illustrated in Fig.~\ref{fig-equiv-3}, the pole $Q_2$ lies to the right of the imaginary axis. We can still rewrite the integral over $\mathcal{I}$ as one over $\mathcal{C}_\Gamma$, but we need to take into account the additional anticlockwise-oriented contour $\mathcal{C}_r$ encircling $Q_2.$ Thus, the total integration contour is now given by $\mathcal{C}=\mathcal{C}_\Gamma\cup\mathcal{C}_r,$ and we get
	\begin{eqnarray}
	\mathcal{M}(p^2)&=&\int_{\mathcal{C}_\Gamma}{\rm d}k^0g(k^0)+\int_{\mathcal{C}_r}{\rm d}k^0g(k^0)\nonumber\\[2mm]
	&=&-2\pi i\left[{\rm Res}\left\lbrace g(k^0)\right\rbrace_{k^0=Q_3}+{\rm Res}\left\lbrace g(k^0)\right\rbrace_{k^0=Q_4} +{\rm Res}\left\lbrace g(k^0)\right\rbrace_{k^0=Q_2}  \right]\nonumber\\[2mm]
	&&+2\pi i\left[{\rm Res}\left\lbrace g(k^0)\right\rbrace_{k^0=Q_2}  \right] \nonumber\\[2mm]
	&=&-2\pi i\left[{\rm Res}\left\lbrace g(k^0)\right\rbrace_{k^0=Q_3}+{\rm Res}\left\lbrace g(k^0)\right\rbrace_{k^0=Q_4}  \right]\,,
	\label{E-M-3}
	\end{eqnarray}
	which again coincides with the MP result in Eq.~\eqref{Mink-result}.	
	
\end{itemize}

We showed that EP is equivalent to MP. To prove the equivalence we could also start from MP and use Wick rotation to transform the integral over $\mathbb{R}$ as an integral over $\mathcal{I},$ and then apply the rules of EP to compute~\eqref{Mink-result}.
Hence, both prescriptions give the same result for both real and imaginary parts of the amplitude~\eqref{1-loop-local}.


\subsubsection{Equivalence between EP and SP}\label{sec-EP-SP-equiv}

Let us now show that EP is equivalent to SP. According to SP the integration contour in the complex $t_i$ planes must be deformed as discussed in the previous Subsection and shown in Fig.~\ref{fig-schwinger}. When applying this deformation to the integral~\eqref{t1t2}, the terms involving the integration from $t_0$ to $t_0+i\infty$ are convergent as long as $k^2+m^2$ and $(p-k)^2+m^2$ have negative imaginary parts. This happens if $k^0$ and $p^0-k^0$ lie in either the first or third quadrant, namely when
\begin{equation}
{\rm Im}[k^0]\,{\rm Re}[k^0]>0\quad {\rm and}\quad {\rm Im}[p^0-k^0]\,{\rm Re}[p^0-k^0]>0\,.
\label{condition-E-P-equiv}
\end{equation}

If we can deform the contour prescribed by EP in such a way that the conditions in Eq.~\eqref{condition-E-P-equiv} are always satisfied, then we have proven that EP is equivalent to SP. Such a contour deformation indeed exists, and we have shown a possible one in Fig.~\ref{fig-equiv-4} (see also Ref.~\cite{Sen:2016gqt}). This deformation requires that the external energy $p^0$ takes complex values in order to keep the conditions~\eqref{condition-E-P-equiv} always true. The imaginary parts of the ends of the integration contour are kept fixed at $\pm i\infty,$ while the real parts can acquire finite real values consistently with the rules of EP, see also the remark in footnote~\ref{footnote-EP}.

We also proved that EP and SP are equivalent, and thus the two computations of integral~\eqref{1-loop-local}  made with the rules of EP and SP must give the same result for both real and imaginary parts.

\subsubsection{Equivalence between MP and SP}

Since we proved the equivalence between MP and EP, and between EP and SP, then it follows that also the prescriptions MP and SP must be equivalent. Therefore, there is no need to show an explicit proof for this third equivalence.

One important point to emphasize is that the proofs of equivalence between MP and the other two prescriptions strongly rely on the property that the integrand in~\eqref{1-loop-local} converges in the limit $|k^0|\rightarrow \infty$ for any complex direction. Indeed, thanks to this property we can use the contours in Figs.~\ref{fig-equiv-1},~\ref{fig-equiv-2},~\ref{fig-equiv-3}, and we can Wick rotate the real axis $\mathbb{R}$ in the complex plane. However, as we explain below, this is not possible in the presence of some nonlocal vertices.

\section{Nonlocal vertices}\label{sec-nonlocal-vertex}

So far we have analysed in great detail the computation of the bubble diagram in the presence of local vertices with three equivalent prescriptions for the choice and deformation of the integration contour in the complex $k^0$ plane. In this Section we are going to perform an analog study in the context of quantum field theories of the type shown in Eq.~\eqref{nonlocal-lag-1}, in which the interaction vertices are nonlocal (non-polynomial) functions of derivatives or momenta.

In particular, we focus on the following one-loop bubble diagram:
\begin{eqnarray}
\mathcal{M}(p^2)= (-i)\lambda^2\int_\mathcal{C} \frac{{\rm d}k^0}{2\pi}\int_{\mathbb{R}^3} \frac{{\rm d}^3k}{(2\pi)^3} \frac{e^{-\gamma(k^2)}}{k^2+m^2-i\epsilon}\frac{e^{-\gamma((p-k)^2)}}{(p-k)^2+m^2-i\epsilon}\,,
\label{1-loop-nonlocal}
\end{eqnarray}
where we have chosen the vertices in Eq.~\eqref{1-loop-two-gen-vert} to be proportional to the exponentials $e^{-\gamma(k^2)}$ and $e^{-\gamma((p-k)^2)}.$ 
The function $\gamma$ is assumed to be an \textit{entire function} to ensure that no additional pole appears in the integrand of~\eqref{1-loop-nonlocal}, and it is such that $e^{-\gamma(k^2)}$ converges to zero in the limit $k^0\rightarrow A\pm i\infty,$ where $A$ is some finite real constant. This means that the singularity structure of~\eqref{1-loop-nonlocal} is the same as the one in the local case~\eqref{1-loop-local}~\cite{Chin:2018puw}, i.e. poles and pinching singularities are still given by Eqs.~\eqref{real poles} and~\eqref{pinching-real}, respectively.
For simplicity, but without any loss of generality, we work with the normalization $\gamma(-m^2)=0.$

We are now going to discuss the application of Minkowski, Euclidean and Schwinger prescriptions to the computation of the integral~\eqref{1-loop-nonlocal}. We should immediately notice that the property of the function $\gamma$ of being entire is crucial for our next discussion. Indeed, from the well-known Liouville theorem in complex analysis it follows that a non-constant entire function must diverge along some directions in the complex $k^0$ plane because of essential singularities at infinity. This means that techniques involving Wick rotation or choices of contours that extend to infinity are \textit{not} well-defined in the presence of nonlocal vertices because they introduce unwanted divergences. This creates problems for the implementation of the Minkowski prescription.

\subsection{Failure of Minkowski prescription}\label{sec-failure}

\begin{figure}[t!]
	\centering
	\subfloat[Subfigure 1 list of figures text][]{
		\includegraphics[scale=0.43]{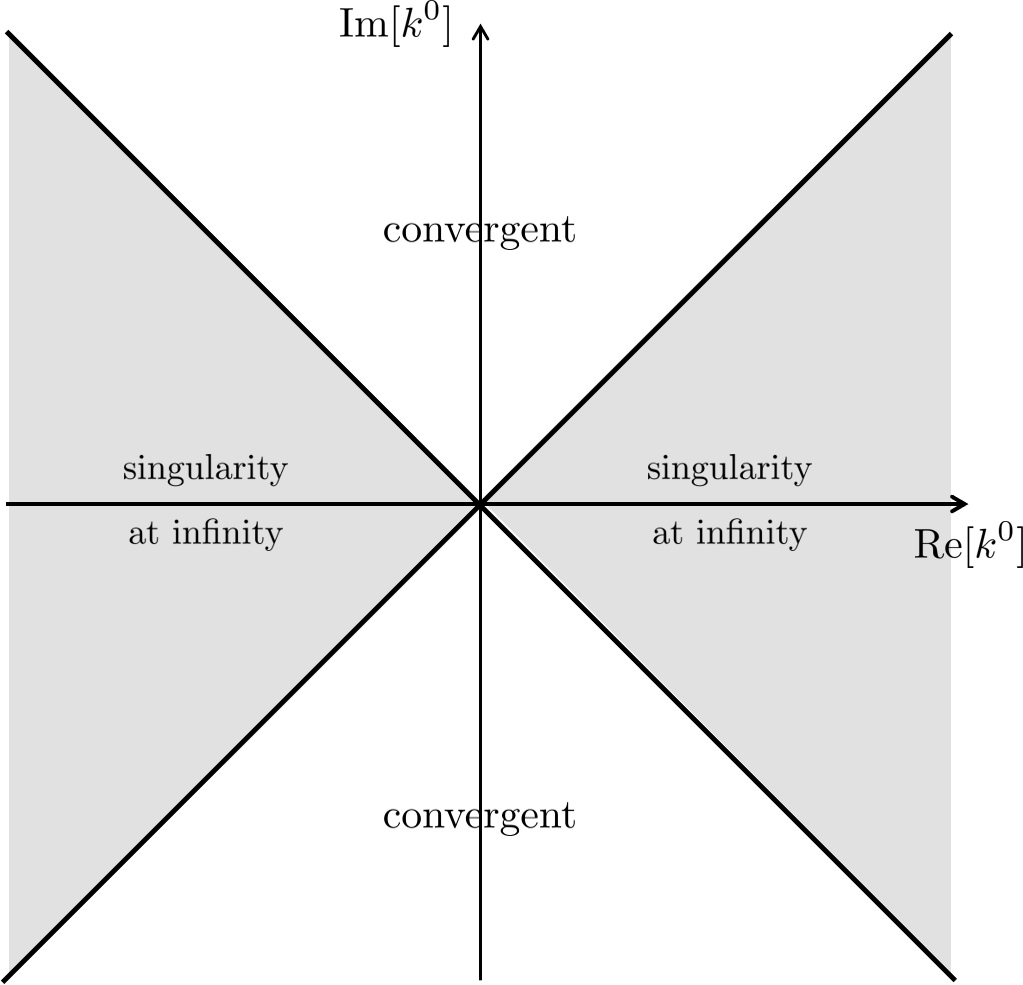}\label{fig3a}}\qquad
	\subfloat[Subfigure 2 list of figures text][]{
		\includegraphics[scale=0.43]{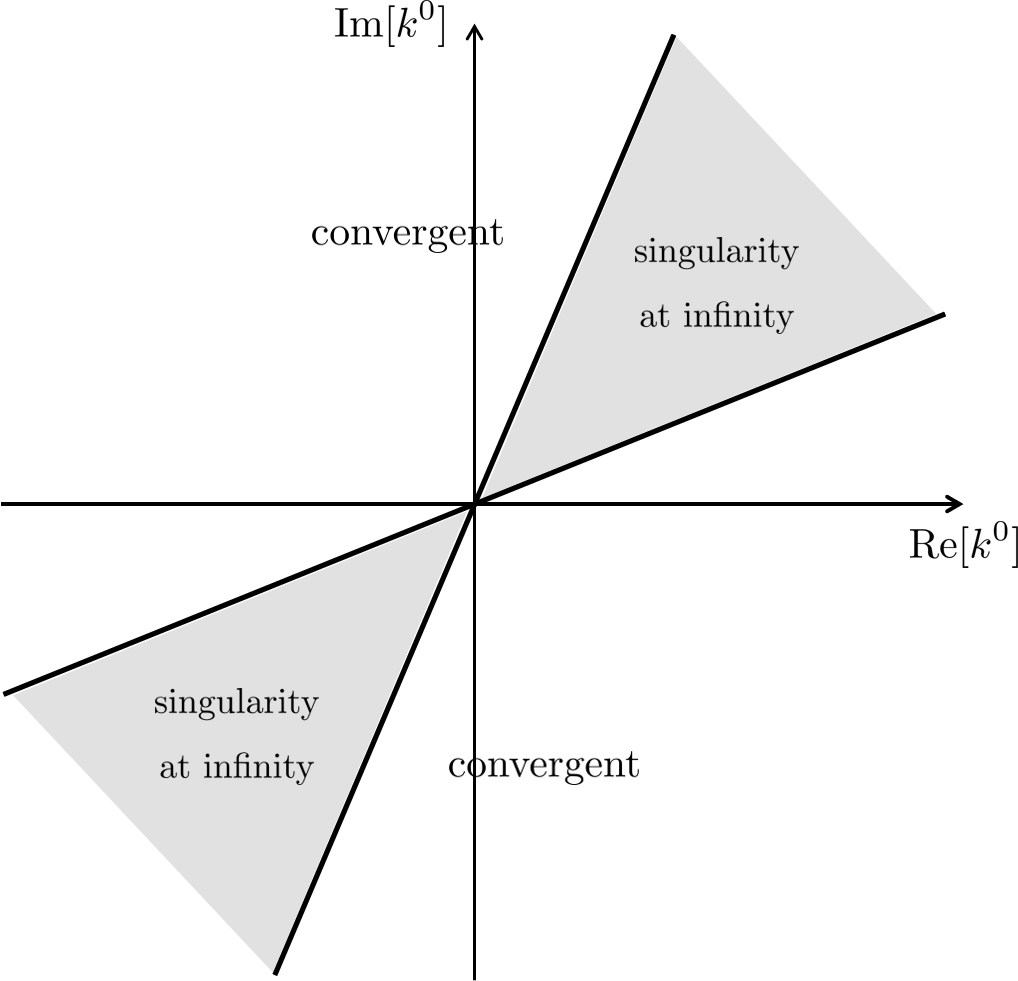}\label{fig3b}
	}
	\protect\caption{(a) Regions of convergence (white color) and divergence (gray color) for the entire function $e^{-k^2/M_s^2};$ it diverges at infinity along the complex directions $-\pi/4<\vartheta<\pi/4$ and $3\pi/4<\vartheta< 5\pi/4.$  (b)  Regions of convergence (white color) and divergence (gray color) for the entire function $e^{-k^4/M_s^4};$ it diverges at infinity along the complex directions $\pi/8<\vartheta<3\pi/8$ and $9\pi/8<\vartheta< 11\pi/8.$ }\label{fig3}
\end{figure}

To better understand the problem of singularities at infinity, let us consider two explicit examples with the entire functions $\gamma(k^2)=(k^2+m^2)/M_s^2$ and $\gamma(k^2)=(k^2+m^2)^2/M_s^4,$ where $M_s$ is an energy scale at which nonlocal effects are expected to become important, and it is mathematically needed to make the exponent dimensionless.
\begin{itemize}
	
	\item By considering complex energies $k^0= \kappa e^{i\vartheta}$ with $\kappa\geq0,$ for the first choice of entire function we have
	\begin{eqnarray}
	e^{-m^2/M_s^2}e^{-k^2/M_s^2}&=& e^{-m^2/M_s^2}e^{(k^0)^2/M_s^2}e^{-\vec{k}^2/M_s^2}\nonumber\\[2mm]
	&=&e^{-m^2/M_s^2}e^{(\kappa^2 \cos 2\vartheta) /M_s^2}e^{i(\kappa^2 \sin 2\vartheta)/M_s^2}e^{-\vec{k}^2/M_s^2}\,,
	\end{eqnarray}
	which diverges in the limit $\kappa \rightarrow \infty$ along the directions $-\pi/4<\vartheta<\pi/4$ and $3\pi/4<\vartheta< 5\pi/4;$ see Fig.~\ref{fig3a}.
	
	\item  For the second entire function, the dominant factor in the $\kappa\rightarrow \infty$ limit is given by
	\begin{eqnarray}
	e^{-(\kappa^4\cos 4\vartheta)/M_s^4}\,,
	\end{eqnarray}
	which diverges at infinity along the directions $\pi/8<\vartheta<3\pi/8$ and $9\pi/8<\vartheta< 11\pi/8;$ see Fig.~\ref{fig3b}.
	
\end{itemize}


This feature makes it impossible to use neither the contours in Figs.~\ref{fig2b},~\ref{fig2a-wick},~\ref{fig-equiv-1},~\ref{fig-equiv-2},~\ref{fig-equiv-3}, nor the Wick rotation in Fig.~\ref{fig2b-wick}. This shows the \textit{failure} of Minkowski prescription to evaluate loop integrals in the context of the nonlocal quantum field theories under investigation.

Although this seems to be a very serious problem, it so happens that well-defined prescriptions to define and compute amplitudes with nonlocal vertices can still be found. In fact, as mentioned in the Introduction, this type of non-polynomial functions appear in string field theory, and in that context the Euclidean prescription was introduced and shown to give unitary amplitudes~\cite{Pius:2016jsl}. Consequently, several works~\cite{Briscese:2018oyx,Chin:2018puw,Koshelev:2021orf,Briscese:2021mob} have implemented the same prescription to show unitarity for more general nonlocal models such as the ones in Eqs.~\eqref{nonlocal-lag-1} and~\eqref{nonlocal-lag-2}.

We will now show that both Euclidean and Schwinger prescriptions can be applied to compute the integral in Eq.~\eqref{1-loop-nonlocal}.

\subsection{Euclidean prescription}

The rules of the Euclidean prescription introduced in Sec.~\ref{sec-euclid} can be applied to~\eqref{1-loop-nonlocal}, indeed all the required properties are satisfied also in the presence of nonlocal vertices. According to this prescription the contour has fixed ends with imaginary parts $\pm i\infty$ (see also footnote~\ref{footnote-EP}), and any deformation happens in finite distance regions of the complex $k^0$ plane. This ensures that no unwanted divergence appears during the evaluation of the integral~\eqref{1-loop-nonlocal}.

According to the Euclidean rules in Sec.~\ref{sec-euclid} the amplitude~\eqref{1-loop-nonlocal} can be split into two parts:
\begin{eqnarray}
\mathcal{M}(p^2)&=&(-i)\lambda^2\int_{\mathcal{I}\,\cup\, \mathcal{C}_r} \frac{{\rm d}k^0}{2\pi}\int_{\mathbb{R}^3} \frac{{\rm d}^3k}{(2\pi)^3} \frac{e^{-\gamma(k^2)}}{k^2+m^2-i\epsilon}\frac{e^{-\gamma((p-k)^2)}}{(p-k)^2+m^2-i\epsilon}\nonumber\\[2mm]
&=& \mathcal{M}_\mathcal{I}(p^2) +\mathcal{M}_{\mathcal{C}_r}(p^2)\,,\label{M-splitted-NL}
\end{eqnarray}
where $\mathcal{M}_{\mathcal{I}}(p^2)$ is the contribution coming from the contour $\mathcal{I},$ while $\mathcal{M}_{\mathcal{C}_r}(p^2)$ is the one from $\mathcal{C}_r.$ 

Also in the nonlocal case we can explicitly show that $\mathcal{M}_{\mathcal{I}}(p^2)$ is real. Since $\mathcal{I}$ is far from the poles, we can take real external energies $p^0\in\mathbb{R}$ and $\epsilon\rightarrow 0.$ Then, by making the change of variable $k^0\rightarrow -ik^0$ we get
\begin{eqnarray}
\mathcal{M}_{\mathcal{I}}(p^2)=\lambda^2\int_{\mathbb{R}^4} \frac{{\rm d}^4k}{(2\pi)^4} \frac{e^{-\gamma((k^0)^2+\vec{k}^2)}}{(k^0)^2+\vec{k}^2+m^2}\frac{e^{-\gamma(-(p^0+ik^0)^2+(\vec{p}-\vec{k})^2)}}{-(p^0+ik^0)^2+(\vec{p}-\vec{k})^2+m^2}\,,
\end{eqnarray}
whose complex conjugate is given by
\begin{eqnarray}
\mathcal{M}^*_{\mathcal{I}}(p^2)&=&\lambda^2\int_{\mathbb{R}^4} \frac{{\rm d}^4k}{(2\pi)^4} \frac{e^{-\gamma((k^0)^2+\vec{k}^2)}}{(k^0)^2+\vec{k}^2+m^2}\frac{e^{-\gamma(-(p^0-ik^0)^2+(\vec{p}-\vec{k})^2)}}{-(p^0-ik^0)^2+(\vec{p}-\vec{k})^2+m^2}\nonumber\\[2mm]
&=&\lambda^2\int_{\mathbb{R}^4} \frac{{\rm d}^4k}{(2\pi)^4} \frac{e^{-\gamma((k^0)^2+\vec{k}^2)}}{(k^0)^2+\vec{k}^2+m^2}\frac{e^{-\gamma(-(p^0+ik^0)^2+(\vec{p}-\vec{k})^2)}}{-(p^0+ik^0)^2+(\vec{p}-\vec{k})^2+m^2}\nonumber\\[2mm]
&=&\mathcal{M}_{\mathcal{I}}(p^2)\,,
\end{eqnarray}
where in the last step we have made the change of variable $k^0\rightarrow -k^0.$

We now evaluate the contribution coming from $\mathcal{C}_r$ by using the residue theorem applied to the pole $Q_2=p^0-\omega_{\vec{p}-\vec{k}}$:
\begin{eqnarray}
\mathcal{M}_{\mathcal{C}_r}(p^2)= -\lambda^2\int_{\mathbb{R}^3} \frac{{\rm d}^3k}{(2\pi)^3}\frac{\Theta(p^0-\omega_{\vec{p}-\vec{k}})}{2\omega_{\vec{p}-\vec{k}}} \frac{\left[e^{-\gamma(k^2)}\right]_{k^0=Q_2}}{p^0+\omega_{\vec{k}}-\omega_{\vec{p}-\vec{k}}}\frac{\left[e^{-\gamma((p-k)^2)}\right]_{k^0=Q_2}}{p^0-\omega_{\vec{k}}-\omega_{\vec{p}-\vec{k}}+i\epsilon}\,;
\end{eqnarray}
where we remind the reader that the Heaviside theta function $\Theta\big(p^0-\omega_{\vec{p}-\vec{k}}\big)$ takes into account that the integral over $\mathcal{C}_r$ is non-zero only when ${\rm Re}[Q_2]\geq 0,$ and it is defined in Eq.~\eqref{theta-H}.

By using the identity~\eqref{identity-delta} we can write
\begin{eqnarray}
\mathcal{M}_{\mathcal{C}_r}(p^2)&=&-\lambda^2\int_{\mathbb{R}^3} \frac{{\rm d}^3k}{(2\pi)^3}\frac{\Theta(p^0-\omega_{\vec{p}-\vec{k}})}{2\omega_{\vec{p}-\vec{k}}}  \frac{\left[e^{-\gamma(k^2)}e^{-\gamma((p-k)^2)}\right]_{k^0=Q_2}}{p^0+\omega_{\vec{k}}-\omega_{\vec{p}-\vec{k}}}\,{\rm P.V.}\left(\frac{1}{p^0-\omega_{\vec{k}}-\omega_{\vec{p}-\vec{k}}}\right)\nonumber\\[2mm]
&&+i\pi\lambda^2\int_{\mathbb{R}^3} \frac{{\rm d}^3k}{(2\pi)^3}\frac{1}{2\omega_{\vec{k}}2\omega_{\vec{p}-\vec{k}}}\left[e^{-\gamma(k^2)}e^{-\gamma((p-k)^2)}\right]_{k^0=Q_2}\delta(p^0-\omega_{\vec{k}}-\omega_{\vec{p}-\vec{k}})\,.
\end{eqnarray}
Let us evaluate the imaginary part (second line) explicitly to check the consistency of the Euclidean prescription with unitarity also in nonlocal theories. 

By using $\int\frac{{\rm d}^3k}{(2\pi)^3}\frac{1}{2\omega_{\vec{k}}}=  \int \frac{{\rm d}^4k}{(2\pi)^4}2\pi\theta(k^0)\delta(k^2+m^2)$ and $\delta(x^2-y^2)=[\delta(x+y)+\delta(x-y)]/2|y|,$ we obtain 
\begin{eqnarray}
{\rm Im}[\mathcal{M}_{\mathcal{C}_r}(p^2)]&=& \pi^2\lambda^2\int\frac{{\rm d}^4k}{(2\pi)^4} \left[e^{-\gamma(k^2)}e^{-\gamma((p-k)^2)}\right]_{k^0=Q_2}
\nonumber\\[2mm]
&&\times \theta(k^0)\delta(k^2+m^2)\theta(p^0-k^0)\delta((p-k)^2+m^2)\nonumber\\[2mm]
&=&\pi^2\lambda^2\int\frac{{\rm d}^4k}{(2\pi)^4} \theta(k^0)\delta(k^2+m^2)\theta(p^0-k^0)\delta((p-k)^2+m^2)\,,
\end{eqnarray}
where we have used the normalization $\gamma(-m^2)=0.$ Thus, we can write 
\begin{eqnarray}
{\rm Im}[\mathcal{M}(p^2)]={\rm Im}[\mathcal{M}_{\mathcal{C}_r}(p^2)]&=&\pi\lambda^2\int_{\mathbb{R}^3} \frac{{\rm d}^3k}{(2\pi)^3}\frac{1}{2\omega_{\vec{k}}2\omega_{\vec{p}-\vec{k}}}\delta(p^0-\omega_{\vec{k}}-\omega_{\vec{p}-\vec{k}})\nonumber\\[2mm]
&=& \frac{\lambda^2}{16\pi}\frac{\sqrt{-p^2-4m^2}}{\sqrt{-p^2}}\theta(-p^2-4m^2) \,.
\label{imag-eucl-NL}
\end{eqnarray}
The last equations show that the standard Cutkosky rules are still satisfied also in nonlocal quantum field theories\footnote{For the bubble diagram~\eqref{1-loop-nonlocal} under investigation not only is unitarity respected but the expression of the imaginary part coincides with the local result (up to a normalization factor). However, for more complicated diagrams -- such as the triangle and box -- unitarity is still respected but the expressions of the imaginary parts are non-trivially modified by nonlocality.}; see Refs.~\cite{Briscese:2018oyx,Chin:2018puw,Koshelev:2021orf}, and also App.~\ref{sec-app}.

We have performed the computation only for the choice of internal momenta $k$ and $p-k;$ it should now be clear how to implement the Euclidean prescription for the choice $k+p/2$ and $k-p/2$ also in the nonlocal case. Actually, we will use the latter choice of internal momenta to perform a full analytic computation for a nonlocal model in Sec.~\ref{sec-explicit}.

Before concluding this Subsection, let us emphasize that the Euclidean prescription is very powerful because it allows one to identify and compute the imaginary part of an amplitude for a generic entire function $\gamma(k^2).$ In this respect, in the presence of generic nonlocal vertices it turns out to be more suitable than the Schwinger prescription as we now explain.

\subsection{Schwinger prescription}\label{sec-schwinger-NL}

All the assumptions required for the use of the Schwinger prescription are also satisfied by the amplitude~\eqref{1-loop-nonlocal} with nonlocal vertices. By using the Schwinger parametrization for the propagators in Eq.~\eqref{schwinger-param}, we can rewrite the integral~\eqref{1-loop-nonlocal} as
\begin{eqnarray} 
\mathcal{M}(p^2)=-i\lambda^2 \displaystyle \int_{0}^{\infty}{\rm d}t_1 \int_{0}^{\infty}{\rm d}t_2 \int \frac{{\rm d}^4k}{(2\pi)^4}e^{-t_1(k^2+m^2)}e^{-t_2((p-k)^2+m^2)}e^{-\gamma(k^2)}e^{-\gamma((p-k)^2)}\,,
\label{schwinger-NL}
\end{eqnarray}
which is convergent in the double limit $t_1,t_2\rightarrow \infty$ as long as the Schwinger parametrization is well-defined, namely if ${\rm Re}[k^2+m^2]>0$ and ${\rm Re}[(p-k)^2+m^2]>0.$ Also in this case, one can make the contour deformation $[0,\infty]\rightarrow [0,t_0]\cup[t_0,i\infty]$ for the integrals over $t_1,t_2$ as shown in Fig.~\ref{fig-schwinger}, and analytically continue internal and external energies in such a way that  ${\rm Im}[-(k^0)^2]<0$ and ${\rm Im}[-(p^0-k^0)^2]<0.$ This procedure will ensure the convergence of the integral for large values of $t_i,$ and will take care of possible pinching singularities.

By looking at Eq.~\eqref{schwinger-NL} we understand that further manipulations require the explicit form of the entire function $\gamma(k^2)$, even to extract information about the imaginary part. This shows why the Schwinger prescription is less suitable than the Euclidean to perform loop computations and prove unitarity in generic nonlocal theories.

It is still instructive to work with a specific nonlocal vertex, simple enough to evaluate the imaginary part explicitly and check the consistency with unitarity. Thus, in the remainder  of this Subsection we assume the entire function to be a polynomial of degree one of $k^2,$ i.e. 
\begin{eqnarray} 
\gamma(k^2)=(k^2+m^2)/M_s^2\quad {\rm and}\quad \gamma((p-k)^2)=((p-k)^2+m^2)/M_s^2\,,
\label{string-entire}
\end{eqnarray}
and thus the one-loop amplitude~\eqref{schwinger-NL} reads
\begin{eqnarray} 
\mathcal{M}(p^2)&=&  \nonumber\lambda^2 \displaystyle \int_{0}^{\infty}{\rm d}t_1 \int_{0}^{\infty}{\rm d}t_2 \int \frac{{\rm d}^4k}{(2\pi)^4}e^{-t_1(k^2+m^2)}e^{-t_2((p-k)^2+m^2)}e^{-(k^2+m^2)/M_s^2}e^{-((p-k)^2+m^2)/M_s^2}\\[2mm]
&=& \frac{\lambda^2}{16\pi^2}\int_0^\infty{\rm d}t_1\int_0^\infty{\rm d}t_2\,\frac{e^{-p^2\frac{(t_1+1/M_s^2)(t_2+1/M_s^2)}{t_1+t_2+2/M_s^2}}e^{-m^2(t_1+t_2+2/M_s^2)}}{(t_1+t_2+2/M_s^2)^2}\,,
\label{t1t2-NL-string}
\end{eqnarray}
where we have made the change of variable $k^0\rightarrow ik^0,$ and then used spherical coordinates. Note that, unlike the case of local vertices, we chose $\mathcal{C}=\mathcal{I}=[-i\infty,i\infty]$ as the initial integration contour for the $k^0$-integral because starting from $\mathbb{R}$ is not well-defined as discussed above; see also footnote~\ref{footnote-2}. 
As already mentioned in the Introduction, the type of entire functions in Eq.~\eqref{string-entire} are typical of string field theory~\cite{Witten:1985cc,Eliezer:1989cr,Arefeva:2001ps,Zwiebach:2011rg,Sen:2015uaa,Pius:2016jsl,DeLacroix:2018arq} and $p$-adic string~\cite{Freund:1987kt,Brekke:1988dg,Freund:1987ck,Frampton:1988kr,Dragovich:2020uwa}. 

We can notice a similarity between the expressions in~\eqref{t1t2-NL-string} and~\eqref{t1t2}, the only difference is that $t_1,t_2$ are both shifted by $1/M_s^2.$ This suggests the additional change of variable
\begin{eqnarray} 
s_1=t_1+1/M_s^2\quad {\rm and}\quad s_2=t_2+1/M_s^2\,,
\end{eqnarray}
which gives
\begin{eqnarray} 
\mathcal{M}(p^2)= \frac{\lambda^2}{16\pi^2}\int_{\frac{1}{M_s^2}}^\infty{\rm d}s_1\int_{\frac{1}{M_s^2}}^\infty{\rm d}s_2\,\frac{e^{-p^2\frac{s_1s_2}{s_1+s_2}}e^{-m^2(s_1+s_2)}}{(s_1+s_2)^2}\,.
\label{s1s2-NL-string}
\end{eqnarray}
Let us remind the reader that the ultraviolet divergence in the local case was associated with the limits $t_1,t_2\rightarrow 0.$ If we compare~\eqref{s1s2-NL-string} with~\eqref{t1t2} we notice that the only difference lies in the lower ends of the integration contour, and indeed the main role of nonlocality in this case is to improve the ultraviolet behavior of the amplitude since $s_1,s_2$ never take zero values. Instead, the behavior of the amplitude in the double limit $s_1,s_2\rightarrow \infty$ is the same as in the local theory, and a divergence appears for $-p^2>4m^2,$ i.e. above the threshold.

All the steps from Eq.~\eqref{uv-coord} to Eq.~\eqref{uv-int-4} performed in Sec.~\ref{sec-sch}  can be identically repeated for the integral~\eqref{s1s2-NL-string} by replacing $t_i$ with $s_i$, and in the end we would get an expression for the imaginary part of the amplitude that coincides with the one obtained through the Euclidean prescription in Eq.~\eqref{imag-eucl-NL}.

\subsection{(In)equivalence between the prescriptions}

Unlike the case of local vertices in Sec.~\ref{sec-local-vertices}, the Minkowski prescription is pathological in the presence of nonlocal vertices. Thus, of the three prescriptions we introduced only Euclidean and Schwinger can be used to compute amplitudes in a consistent way. This also implies that we need to prove only the  equivalence between Euclidean and Schwinger.

The same proof presented in Sec.~\ref{sec-EP-SP-equiv} applies to the case of nonlocal vertices too. Starting with the Euclidean prescription we can always find a contour deformation such as the one in Fig.~\ref{fig-equiv-4}, such that the Schwinger parametrization remains well-defined and the conditions~\eqref{condition-E-P-equiv} can be satisfied. The contour in Fig.~\ref{fig-equiv-4} is well-defined for the one-loop amplitude~\eqref{1-loop-nonlocal} because the entire functions $\gamma(k^2)$ are defined such that $e^{-\gamma(k^2)}$ converges to zero in the limit $k^0\rightarrow A+i\infty,$  $A$ being a finite real constant. 

Hence, Euclidean and Schwinger are equivalent also in the presence of nonlocal vertices. It is worth mentioning that such an equivalence was proven in the context of string theory first at one loop~\cite{Sen:2016gqt}, and later at all orders in perturbation theory~\cite{Sen:2016ubf}.

\subsection{An explicit example}\label{sec-explicit}

We now analyse a  nonlocal scalar model whose Lagrangian is inspired by the truncated open tachyon in string field theory~\cite{Witten:1985cc,Eliezer:1989cr,Arefeva:2001ps}:
\begin{equation}
\mathcal{L}=\frac{1}{2}\phi(\Box-m^2)\phi-\frac{\lambda}{3!}\left(e^{(\Box-m^2)/2M_s^2}\phi\right)^3\,,
\label{lagr-string}
\end{equation}
where the chosen entire function is $\gamma(\Box)=-(\Box-m^2)/M_s^2,$ and we work with a positive non-tachyonic real mass. We do not worry about whether the Lagrangian~\eqref{lagr-string} is interesting from a physical point of view but we just use it as a mathematical toy model to apply the contour prescriptions and the techniques discussed above\footnote{We could also work with a quartic nonlocal potential $\frac{\lambda}{4!}\big(e^{(\Box-m^2)/2M_s^2}\phi\big)^4,$ or even higher order $\frac{\lambda}{n!}\big(e^{(\Box-m^2)/2M_s^2}\phi\big)^n.$ In all these cases the bubble diagram in Eq.~\eqref{self-en} will still be the same, with the only difference that the number of external legs would be $2n-4$.}. Of course, we should expect that the amplitudes for the model~\eqref{lagr-string} respect unitarity, and indeed we will show it explicitly at one-loop order for the bubble diagram.

The  corresponding one-loop bubble amplitude reads
\begin{equation}
\mathcal{M}(p^2)=-i\lambda^2\int_{\mathcal{C}}\frac{{\rm d}k^0}{2\pi}\int \frac{{\rm d}k^3}{(2\pi)^3} \frac{e^{-(k^2+m^2)/M_s^2}}{k^2+m^2-i\epsilon}\frac{e^{-[(p-k)^2+m^2]/M_s^2}}{(p-k)^2+m^2-i\epsilon}\,.\label{self-en}
\end{equation}

Our aim is to perform a full analytic computation of the one-loop integral~\eqref{self-en} by using both Euclidean and Schwinger prescriptions, and as well as show explicitly that they give the same result for both real and imaginary parts of the amplitude. Although we start with a non-vanishing mass parameter, we are going to take the massless limit ($m=0$) to evaluate also the real part analytically for both prescriptions. Thus, in end we are going to work with the massless version of the Lagrangian~\eqref{lagr-string}. This explicit computation will be very instructive to clarify several aspects discussed in this paper so far.

\subsubsection{Euclidean prescription}

We start with the Euclidean prescription, and choose to work with the internal momenta $k+p/2$ and $k-p/2,$ thus we apply the steps in Sec.~\ref{sec-euclid-k+p/2} to the integral
\begin{equation}
\mathcal{M}(p^2)=-i\lambda^2\int_{\mathcal{C}}\frac{{\rm d}k^0}{2\pi}\int \frac{{\rm d}k^3}{(2\pi)^3} \frac{e^{-[(k+p/2)^2+m^2]/M_s^2}}{(k+p/2)^2+m^2-i\epsilon}\frac{e^{-[(k-p/2)^2+m^2]/M_s^2}}{(k-p/2)^2+m^2-i\epsilon}\,.\label{self-en-k+p/2}
\end{equation}
In this case the poles and the pinching singularity condition are given by Eqs.~\eqref{real poles-2} and~\eqref{pinching-real-2}, respectively; see Fig.~\ref{fig2a-2} for the location of the poles, and Figs.~\ref{fig4b-2},~\ref{fig4f-2},~\ref{fig-cauchy} for the choice of the integration contour $\mathcal{C}$ in the complex $k^0$ plane.

According to the Euclidean prescription we write the amplitude as done in Eq.~\eqref{M-splitted-2}:
\begin{equation}
\mathcal{M}(p^2)=\mathcal{M}_{\mathcal{I}}(p^2)+\mathcal{M}_{\mathcal{C}_{r,2}}(p^2)+\mathcal{M}_{\mathcal{C}_{r,3}}(p^2)\,,\label{2-terms-ampl}
\end{equation}
where $\mathcal{I}=[-i\infty,i\infty]$, and $\mathcal{C}_{r,2}$ and $\mathcal{C}_{r,3}$ are the contours around the poles $P_2$ and $P_3.$

For simplicity we work in the centre-of-mass frame ($\vec{p}=0$), i.e. we implement the contour prescription in Fig.~\ref{fig-cauchy} (see also the remark at the end of Sec.~\ref{sec-euclid-k+p/2}).

\paragraph{Evaluation of $\mathcal{M}_{\mathcal{I}}.$} By making the change of variable $k^0\rightarrow ik^0,$ and going to $4$-dimensional spherical coordinates ${\rm d}^4k=\rho^3\sin\alpha^2\sin\theta\,{\rm d}\rho{\rm d}\alpha{\rm d}\theta{\rm d}\varphi,$ with $0\leq k\leq \infty,$ $0\leq \alpha,\theta\leq \pi,$ $0\leq \varphi \leq 2\pi,$ and $k^0=\rho\cos\alpha,$ we obtain
\begin{eqnarray}
\mathcal{M}_{\mathcal{I}}(p^2)\!\!&=&\!\! \lambda^2\int \frac{{\rm d}k^4}{(2\pi)^4} \frac{e^{-[(k+p/2)^2+m^2]/M_s^2}}{(k+p/2)^2+m^2}\frac{e^{-[(k-p/2)^2+m^2]/M_s^2}}{(k-p/2)^2+m^2}\nonumber\\[2mm]
&=&\lambda^2\frac{e^{(p^0)^2/2M^2_s-2m^2/M_s^2}4\pi}{(2\pi)^4}\!\!\int_0^\infty{\rm d}\rho\,\rho^3\int_0^\pi \!\!\!{\rm d}\alpha \frac{\sin^2\alpha \,e^{-2\rho^2/M_s^2}}{(\rho^2-(p^0)^2/4+m^2)^2+(p^0)^2\rho^2\cos^2\alpha}\nonumber \\ [2mm]
\!\!&=&\!\! -\frac{\lambda^2\, e^{(p^0)^2/2M^2_s-2m^2/M_s^2}}{4\pi^2(p^0)^2}\!\!\int_0^\infty \!\!\!{\rm d}\rho \,\rho e^{-2\rho^2/M_s^2}\!\left[1\!-\!\sqrt{\frac{(\rho^2-(p^0)^2/4+m^2)^2+(p^0)^2\rho^2}{(\rho^2-(p^0)^2/4+m^2)^2}}\,\right]\,.\qquad\label{integ}
\end{eqnarray}
In the massless case we can evaluate the integral~\eqref{integ} analytically. Indeed, by setting $m=0$ and reinstating $(p^0)^2\rightarrow -p^2,$ if we  make the additional change of variables $z=\rho^2$ we get
\begin{eqnarray}
\mathcal{M}_{\mathcal{I}}(p^2)=\lambda^2\frac{e^{-p^2/2M^2_s}}{8\pi^2 p^2}\left[ \int_0^{-p^2/4-\varepsilon}{\rm d}z \frac{2ze^{-2z/M_s^2}}{z+p^2/4}+\int_{-p^2/4+\varepsilon}^\infty {\rm d}z \frac{(p^2/2)e^{-2z/M_s^2}}{z+p^2/4} \right]\,,\qquad
\end{eqnarray}
where the limit $\varepsilon\rightarrow 0$ is understood. We can explicitly evaluate the two integrals, and up to leading order in $\varepsilon$ (in the limit $\varepsilon\rightarrow 0$) we obtain 
\begin{eqnarray}
\mathcal{M}_{\mathcal{I}}(p^2)= \frac{\lambda^2}{16\pi^2}\left[\frac{2M^2_s}{p^2}\left(e^{-p^2/2M_s^2}-1\right)+{\rm Ei}\left( \frac{-p^2}{2M^2_s}\right)-2\,{\rm Ei}\left(\varepsilon\right) \right]\,,\qquad\label{imag-axis-part}
\end{eqnarray}
where we have absorbed constant factors in the infinitesimal parameter $\varepsilon,$ and introduced the exponential integral function 
\begin{eqnarray} 
{\rm Ei}(x)=-\int_{-x}^\infty{\rm d}t\, \frac{e^{-t}}{t}\,.
\end{eqnarray}
The divergent term ${\rm Ei}\left(\varepsilon\right)$ in Eq.~\eqref{imag-axis-part} will be canceled by an equal and opposite contribution coming from $\mathcal{M}_{\mathcal{C}_r}(p^2)$ as we now show.

\paragraph{Evaluation of $\mathcal{M}_{\mathcal{C}_{r,2}}+\mathcal{M}_{\mathcal{C}_{r,3}}.$} By computing the residues at the poles $P_2=p^0/2-\omega_{\vec{k}-\vec{p}/2}$ and $P_3=-p^0/2+\omega_{\vec{k}+\vec{p}/2},$ we obtain
\begin{eqnarray}
\mathcal{M}_{\mathcal{C}_{r,2}}(p^2)
&=&  -\lambda^2 {\rm P.V.} \left[\int \frac{{\rm d}^3k}{(2\pi)^3}\frac{\Theta(p^0/2-\omega_{\vec{k}-\vec{p}/2})\,e^{\big[(p^0-\omega_{\vec{k}-\vec{p}/2})^2-\omega_{\vec{k}+\vec{p}/2}^2\big]/M_s^2}}{2\omega_{\vec{k}-\vec{p}/2}(p^0-\omega_{\vec{k}-\vec{p}/2}+\omega_{\vec{k}+\vec{p}/2})(p^0-\omega_{\vec{k}-\vec{p}/2}-\omega_{\vec{k}+\vec{p}/2})} \right]\nonumber\\[2mm]
&&+ i\pi \lambda^2 \int \frac{{\rm d}^3k}{(2\pi)^3} \frac{\Theta(p^0/2-\omega_{\vec{k}-\vec{p}/2})}{2\omega_{\vec{k}-\vec{p}/2}2\omega_{\vec{k}+\vec{p}/2}}\delta(p^0-\omega_{\vec{k}-\vec{p}/2}-\omega_{\vec{k}+\vec{p}/2}) \,,
\label{pole-2}
\end{eqnarray}
and
\begin{eqnarray}
\mathcal{M}_{\mathcal{C}_{r,3}}(p^2)
&=&  -\lambda^2 {\rm P.V.} \left[\int \frac{{\rm d}^3k}{(2\pi)^3}\frac{\Theta(p^0/2-\omega_{\vec{k}+\vec{p}/2})\,e^{\big[(p^0-\omega_{\vec{k}+\vec{p}/2})^2-\omega_{\vec{k}-\vec{p}/2}^2\big]/M_s^2}}{2\omega_{\vec{k}+\vec{p}/2}(p^0+\omega_{\vec{k}-\vec{p}/2}-\omega_{\vec{k}+\vec{p}/2})(p^0-\omega_{\vec{k}-\vec{p}/2}-\omega_{\vec{k}+\vec{p}/2})}  \right]\nonumber\\[2mm]
&&+ i\pi \lambda^2 \int \frac{{\rm d}^3k}{(2\pi)^3} \frac{\Theta(p^0/2-\omega_{\vec{k}+\vec{p}/2})}{2\omega_{\vec{k}-\vec{p}/2}2\omega_{\vec{k}+\vec{p}/2}}\delta(p^0-\omega_{\vec{k}-\vec{p}/2}-\omega_{\vec{k}+\vec{p}/2}) \,,
\label{pole-3}
\end{eqnarray}
where the Heaviside theta function $\Theta(x)$ was defined in Eq.~\eqref{theta-H}.

To evaluate the Cauchy principal value integrals we go to the centre-of-mass frame for simplicity, i.e. we set $\vec{p}=0$ for the time being, and we rewrite the result in a Lorentz invariant form after. In the centre-of-mass frame the two contributions in the first lines of Eqs.~\eqref{pole-2} and~\eqref{pole-3} turn out to be equal. Thus, by going to spherical coordinates and making the change of variable $z=\omega_{\vec{k}} p^0/2,$ we can write
\begin{eqnarray}
\mathcal{M}^{\rm PV}_{\mathcal{C}_{r,2}}(p^2)=\mathcal{M}^{\rm PV}_{\mathcal{C}_{r,3}}(p^2)&=&  -\frac{\lambda^2e^{(p^0)^2/M_s^2}}{4}\int \frac{{\rm d}^3k}{(2\pi)^3}\frac{\Theta(p^0-2\omega_{\vec{k}}) e^{-2p^0\omega_{\vec{k}}/M_s^2}}{p^0\omega_{\vec{k}}(p^0/2-\omega_{\vec{k}})}\nonumber\\[2.5mm]
&=&  -\frac{\lambda^2e^{(p^0)^2/M_s^2}}{8\pi^2(p^0)^2}\int_{\frac{p^0m}{2}}^\infty {\rm d}z \sqrt{4z^2-(p^0)^2m^2}\frac{\Theta(p^0-4z/p^0) e^{-4z/M_s^2}}{(p^0)^2/4-z}\,.\,\,\,\quad\,\,\label{integ2}
\end{eqnarray}
In the massless case we can evaluate the integral~\eqref{integ2} analytically. Indeed, by setting $m=0$ and reinstating $(p^0)^2\rightarrow -p^2,$ up to leading order in $\varepsilon$ (in the limit $\varepsilon\rightarrow 0$) we obtain
\begin{eqnarray}
\mathcal{M}^{\rm PV}_{\mathcal{C}_{r,2}}(p^2)=\mathcal{M}^{\rm PV}_{\mathcal{C}_{r,3}}(p^2)&=&  -\frac{\lambda^2e^{-p^2/M_s^2}}{4\pi^2p^2}\int_0^{-p^2/4-\varepsilon} \frac{ze^{-4z/M_s^2}}{p^2/4+z}\nonumber\\[2mm]
&=&  \frac{\lambda^2}{32\pi^2}\left[-\frac{2M^2_s}{p^2}\left(e^{-p^2/M^2_s}-1 \right)  - 2\,{\rm Ei}\left(-\frac{p^2}{M^2_s}\right) +2\,{\rm Ei}(\varepsilon) \right]\,,\quad
\label{real-from-poles}
\end{eqnarray}
where again we absorbed constant factors in the infinitesimal parameter $\varepsilon.$ The last term in Eq.~\eqref{real-from-poles} is divergent and, as promised, it will be crucial for the cancellation of the other divergent term in Eq.~\eqref{imag-axis-part}. 

We now compute the two contributions in the second lines of Eqs.~\eqref{pole-2} and~\eqref{pole-3} that are responsible for a non-vanishing imaginary part. As discussed at the end of Sec.~\ref{sec-euclid-k+p/2}, in the centre-of-mass frame we have (under the integral sign)
\begin{equation}
\Theta(p^0/2-\omega_{\vec{k}})\delta(p^0-2\omega_{\vec{k}})\quad\rightarrow\quad \Theta(0)\delta(0)=\frac{1}{2}\,,
\end{equation}
namely the Heaviside theta function contributes with a factor of $1/2$. Therefore, the imaginary parts of Eqs.~\eqref{pole-2} and~\eqref{pole-3} are given by
\begin{equation}
{\rm Im}[\mathcal{M}_{\mathcal{C}_{r,2}}(p^2)]={\rm Im}[\mathcal{M}_{\mathcal{C}_{r,3}}(p^2)]= \frac{\pi \lambda^2}{2}\int\frac{{\rm d}^3k}{(2\pi)^3} \frac{1}{4\omega_{\vec{k}}^2}\delta(p^0-2\omega_{\vec{k}}) =\frac{\lambda^2}{32\pi}\frac{\sqrt{(p^0)^2-4m^2}}{p^0}\theta(p^0-2m) \,.
\end{equation}
The Lorentz invariant form can be obtained by replacing $(p^0)^2\rightarrow -p^2,$ and the massless limit reads
\begin{eqnarray}
{\rm Im}[\mathcal{M}_{\mathcal{C}_{r,2}}(p^2)]={\rm Im}[\mathcal{M}_{\mathcal{C}_{r,3}}(p^2)]= \frac{\lambda^2}{32\pi}\theta(-p^2) \,.
\end{eqnarray}

\paragraph{Full amplitude.} We now have all the ingredients to write down the full expressions for both real and imaginary parts of the amplitude~\eqref{self-en} in the massless case:
\begin{eqnarray}
\mathcal{M}(p^2)={\rm Re}[\mathcal{M}(p^2)]+i\,{\rm Im}[\mathcal{M}(p^2)] \,,
\end{eqnarray}
where
\begin{eqnarray}
{\rm Re}[\mathcal{M}(p^2)]&=& \mathcal{M}_\mathcal{I}(p^2)+\mathcal{M}^{\rm PV}_{\mathcal{C}_{r,2}}(p^2)+\mathcal{M}^{\rm PV}_{\mathcal{C}_{r,3}}(p^2) \nonumber\\[2mm]
&=&  \frac{\lambda}{16\pi^2}\left[\frac{2M^2_s}{p^2}\left(e^{-p^2/2M^2_s}-e^{-p^2/M_s^2} \right)+ {\rm Ei}\left(-\frac{p^2}{2M^2_s}\right) -2\,{\rm Ei}\left(-\frac{p^2}{M^2_s}\right) \right]\,,\,\,\,
\label{real-part-amplitude}
\end{eqnarray}
and
\begin{eqnarray}
{\rm Im}[\mathcal{M}(p^2)]={\rm Im}[\mathcal{M}_{\mathcal{C}_{r,2}}(p^2)]+{\rm Im}[\mathcal{M}_{\mathcal{C}_{r,3}}(p^2)]= \frac{\lambda^2}{16\pi}\theta(-p^2) \,.
\label{imag-part-ampl}
\end{eqnarray}

It is worthwhile to mention that the same one-loop computation was performed in Ref.~\cite{Koshelev:2021orf} for a nonlocal model whose Lagrangian is the same as~\eqref{lagr-string} with the only difference that a quartic potential was considered instead of a cubic (the bubble diagram~\eqref{self-en} is the same for both models). Our results~\eqref{real-part-amplitude} and~\eqref{imag-part-ampl} for the final expression of the amplitude agree with the ones in~\cite{Koshelev:2021orf}. The authors in~\cite{Koshelev:2021orf} also correctly noticed some of the problems related to the Minkowski prescription but then, to evaluate the bubble diagram they assumed that a contour such as the one in Fig.~\ref{fig2a-wick} and the Wick rotation in Fig.~\ref{fig2b-wick} can be ``formally'' used in the presence of nonlocal vertices. Although this formal procedure happens to give the correct result for the bubble diagram, it can be misleading and hides essential details. In this respect, we believe that our analysis -- especially the explanation in Sec.~\ref{sec-euclid-k+p/2} -- is more  rigorous from a mathematical point of view and fully clarifies why the expressions~\eqref{real-part-amplitude} and~\eqref{imag-part-ampl} for real and imaginary parts of the amplitude are correct. In particular, our analysis clearly explains why the imaginary contribution to the amplitude is non-zero when the poles lie on the imaginary axis in the centre-of-mass frame.

\subsubsection{Schwinger prescription}

We now perform the computation of the same one-loop amplitude~\eqref{self-en} in the massless case by using the Schwinger prescription introduced in Secs.~\ref{sec-sch} and~\ref{sec-schwinger-NL}. In particular, as explained in the remark at the end of Sec.~\ref{sec-sch} we implement the prescription in a more practical way: we perform the full integral working with imaginary external energies, i.e. $p^0=ip^4$ with $p^4\in \mathbb{R}$ such that $p^2>0,$ and analytically continue to real physical energies at the end.

We focus directly on the massless limit of the one-loop amplitude~\eqref{self-en},
\begin{equation}
\mathcal{M}(p^2)=\lambda^2\int \frac{{\rm d}k^4}{(2\pi)^4} \frac{e^{-k^2/M_s^2}e^{-(p-k)^2/M_s^2}}{k^2(p-k)^2}\,, \label{self-en-euclid}
\end{equation}
where the integration measure is ${\rm d}k^1{\rm d}k^2{\rm d}k^3{\rm d}k^4,$ with $k^4=-ik^0.$ Some of the initial steps to evaluate the above integral were already shown in Sec.~\ref{sec-schwinger-NL}, but for the sake of clarity we now repeat them and show all the details of the calculation.

By using the Schwinger parametrization for the two Euclidean propagators,
\begin{equation} 
\begin{array}{rl}
\displaystyle \frac{1}{k^2}= \displaystyle \int_{0}^{\infty}dt_1e^{-t_1k^2}\,,\qquad \frac{1}{(p-k)^2}= \displaystyle \int_{0}^{\infty}dt_2e^{-t_2(p-k)^2}\,,
\end{array}\label{schw}
\end{equation}
the integral~\eqref{self-en-euclid} can be recast as
\begin{eqnarray} 
\mathcal{M}(p^2)&=&  \nonumber\lambda^2 \displaystyle \int_{0}^{\infty}{\rm d}t_1 \int_{0}^{\infty}{\rm d}t_2 \int \frac{{\rm d}^4k}{(2\pi)^4}e^{-(t_1+1/M_s^2)k^2}e^{-(t_2+1/M_s^2)(p-k)^2}\\
&=& \lambda^2 \displaystyle \int_{0}^{\infty}{\rm d}t_1 \int_{0}^{\infty}{\rm d}t_2\, e^{-p^2(t_2+1/M_s^2)}\int \frac{{\rm d}^4k}{(2\pi)^4}e^{-k^2(t_1+t_2+2/M_s^2)}e^{2k\cdot p(t_2+1/M_s^2)}\,.\label{self-ener-t1t2}
\end{eqnarray}
By going to four-dimensional spherical coordinates ${\rm d}^4k=k^3\sin\alpha^2\cos\theta\,{\rm d}k{\rm d}\alpha{\rm d}\theta{\rm d}\varphi,$ with $0\leq k\leq \infty,$ $0\leq \alpha,\theta\leq \pi,$ $0\leq \varphi \leq 2\pi,$ and $p\cdot k=|p||k|\cos\alpha,$ we obtain
\begin{equation}
\int \frac{{\rm d}^4k}{(2\pi)^4}e^{-k^2(t_1+t_2+2/M_s^2)}e^{2k\cdot p(t_2+1/M_s^2)}=\frac{M_s^4}{16\pi^2}\frac{e^{\frac{p^2}{M_s^2}\frac{(M_s^2t_2+1)^2}{M_s^2(t_1+t_2)+2}}}{\left[M_s^2(t_1+t_2)+2\right]^2}\,.
\end{equation}
Performing the additional change of integration variables
\begin{equation}
s_1=t_1+\frac{1}{M_s^2}\,,\qquad s_2=t_2+\frac{1}{M_s^2}\,,
\end{equation}
the integral~\eqref{self-ener-t1t2} becomes
\begin{eqnarray} 
\mathcal{M}(p^2)= \frac{\lambda^2}{16\pi^2} \int_{\frac{1}{M_s^2}}^{\infty}{\rm d}s_1 \int_{\frac{1}{M_s^2}}^{\infty}{\rm d}s_2\,\frac{e^{-p^2\frac{s_1s_2}{s_1+s_2}}}{(s_1+s_2)^2}\,.
\label{self-ener-s1s2}
\end{eqnarray}
One can notice that the last integral is convergent as $p^2\geq 0$ in Euclidean signature. By performing the double integration we get
\begin{eqnarray} 
\mathcal{M}(p^2)&=& \frac{\lambda^2}{16\pi^2}\left[\frac{2M_s^2}{p^2}\left(e^{-p^2/2M_s^2}-e^{-p^2/M_s^2}\right)\right.\nonumber\\[2.5mm]
&&\left.+{\rm Ei}\left(0,-\frac{p^2}{2M_s^2}\right)-{\rm Ei}\left(0,-\frac{p^2}{M_s^2}\right)+\Gamma\left(0,\frac{p^2}{M_s^2}\right)\right]\,,
\label{self-en-eucl-final}
\end{eqnarray}
where $\Gamma(a,z)$ is the incomplete Gamma function defined as
\begin{eqnarray} 
\Gamma(a,z)=\int_z^\infty{\rm d}t\, t^{a-1}e^{-t}\,;
\end{eqnarray}
in our case $a=0$.

The compact expression in Eq.~\eqref{self-en-eucl-final} is still defined for space-like external momenta, i.e. $p^2>0.$ However, we are interested in physical energies and time-like momenta $p^2<0.$ 
For negative arguments ($p^2<0$) the incomplete Gamma function $\Gamma(0,x)$ satisfies the following relation:
\begin{eqnarray} 
\lim\limits_{\epsilon\rightarrow 0^+}\Gamma(0,x\pm i\epsilon)=-{\rm Ei}\left(-x\right)\mp i\pi\,,\quad {\rm for}\quad x<0\,. \label{analytic-gamma}
\end{eqnarray}
In our case we need to use Eq.~\eqref{analytic-gamma} with $-i\epsilon,$ so that the one-loop amplitude in Minkowski signature acquires an imaginary component:
\begin{eqnarray} 
\mathcal{M}(p^2)={\rm Re}[\mathcal{M}(p^2)]+i\,{\rm Im}[\mathcal{M}(p^2)]\,,
\end{eqnarray}
where 
\begin{eqnarray} 
{\rm Re}[\mathcal{M}(p^2)]=\frac{\lambda^2}{16\pi^2}\left[\frac{2M_s^2}{p^2}\left(e^{-p^2/2M_s^2}-e^{-p^2/M_s^2}\right)+{\rm Ei}\left(-\frac{p^2}{2M_s^2}\right)-2\,{\rm Ei}\left(-\frac{p^2}{M_s^2}\right)\right]\,,
\end{eqnarray}
which coincides with ~\eqref{real-part-amplitude}, and
\begin{eqnarray} 
{\rm Im}[\mathcal{M}(p^2)]=\frac{\lambda^2}{16\pi}\theta(-p^2)\,.
\end{eqnarray}
which matches the expression~\eqref{imag-part-ampl}. 

Hence, we have verified analytically that Schwinger and Euclidean prescriptions give the same result for the computation of the amplitude~\eqref{self-en} (in the massless case). Let us emphasize again that the results are consistent with the Cutkosky rules and unitarity; see also App.~\ref{sec-app}.

\subsubsection{Local limit}

As a further consistency check of the Euclidean and Schwinger prescriptions applied to the amplitude~\eqref{self-en} (in the massless case) we can take the local limit ($|p^2|/M_s^2\ll 1$) of~\eqref{self-en-eucl-final}, and verify that the known local result in Eq.~\eqref{full-amplitude-massless} is recovered.

By expanding in Taylor series the amplitude~\eqref{self-en-eucl-final} for small $|p^2|/M_s^2\ll 1$ we obtain
\begin{eqnarray} 
\mathcal{M}(p^2)=\frac{\lambda^2}{16\pi^2}\left[ 1-\frac{\gamma_{\rm E}}{2}-\log 2-\log\left(\frac{p^2-i\epsilon}{M_s^2}\right) \right]\,,
\label{local-limit}
\end{eqnarray}
which coincides with~\eqref{full-amplitude-massless} once we identify $M_s^2=2\Lambda^2e^{1+\gamma_{\rm E}/2},$  $\gamma_{\rm E}$ being the Euler-Mascheroni constant. Therefore, we verified that our result is consistent with the local limit, as expected.

It is worth  mentioning that our computation of the bubble diagram~\eqref{self-en} can also be seen as a regularization method to compute the bubble diagram in the local case, where $M_s$ is now interpreted as a regulator instead of a fundamental energy scale. Indeed, the result in Eq.~\eqref{local-limit} is the same that we would get by using dimensional regularization or Pauli-Villars.

\section{Discussion \& conclusions}\label{sec-discussion}

In Secs.~\ref{sec-local-vertices} and~\ref{sec-nonlocal-vertex} we discussed in great detail three ``different'' prescriptions for the evaluation of the bubble diagram in the presence of local~\eqref{1-loop-local} and nonlocal vertices~\eqref{1-loop-nonlocal}. We noticed that, while all the prescriptions are valid and equivalent in  the presence of local vertices, the Minkowski prescription fails when nonlocal vertices are introduced. Moreover, the Euclidean prescription turns out to be more suitable than Schwinger to compute the imaginary part of an amplitude in a generic nonlocal model. We also performed an explicit and fully analytic calculation in a string-inspired nonlocal model, and confirmed that Euclidean and Schwinger prescriptions give the same result.

\paragraph{Other diagrams and higher loops.}  One natural question to ask is how and whether Euclidean and Schwinger can be generalized to more complicated amplitudes and to higher loops. Both prescriptions have been proven to be valid at all orders in perturbation theory in the context of string theory~\cite{Sen:2016ubf}. For more general nonlocal theories the Euclidean prescription has been used for the verification of Cutkosky rules at all order in perturbation theory~\cite{Briscese:2018oyx,Chin:2018puw}, and explicit computations of imaginary parts have been performed for other types of one-loop (box and triangle)~\cite{Pius:2018crk,Briscese:2018oyx} and two-loop diagrams (sunset)~\cite{Briscese:2018oyx}. For example, in the case of the triangle diagram it was shown that the anomalous threshold does not contribute to the imaginary part of the amplitude, i.e. it does not affect the optical theorem, similarly to the local case. To extend our analysis, one should study the triangle and box diagrams in nonlocal theories by means of the Schwinger prescription and show that the result coincides with the one obtained through the Euclidean prescription. A proof of the equivalence between Euclidean and Schwinger at higher order in perturbation theory and for generic nonlocal theories is still lacking. More explicit computations of both real and imaginary parts for more complicated diagrams and at higher loops in specific models (as we did in Sec.~\ref{sec-explicit}) are also still missing. Therefore, future investigations are surely needed along these directions.

\paragraph{Minkowski vs Euclidean.} One interesting feature that needs further discussion is the failure of the Minkowski prescription in the presence of nonlocal vertices. We learned that the integration contour in the complex $k^0$ plane cannot be $\mathbb{R}=[-\infty,\infty]$ because of unwanted divergences. Indeed, we emphasized that the initial contour must be chosen to coincide with the imaginary axis $\mathcal{C}=\mathcal{I}=[-i\infty,i\infty],$ and that the amplitude must initially depend on purely imaginary energies in order for the poles to have a certain location relative to the contour. In other words, the amplitude must be defined in Euclidean signature, and analytically continued to Minkowski after. The inverse procedure, from Minkowski to Euclidean, is \textit{not} well-defined.

This is not the only known example of quantum field theories that are pathological if initially defined in Minkowski. For instance, in theories with higher-derivative kinetic terms such as sixth-order Lee-Wick models~\cite{Lee:1969fy,Cutkosky:1969fq} -- whose propagators contain additional pairs of complex conjugate poles -- the amplitudes need to be initially defined in Euclidean signature otherwise unitarity would be violated and standard renormalizability properties would be spoiled~\cite{Aglietti:2016pwz,Anselmi:2017yux,Anselmi:2017lia,Anselmi:2018kgz}. The same happens in theories with fakeon propagators~\cite{Anselmi:2018kgz}. In all these cases, the analytic structure of the integrands (i.e. propagators and vertices) is somehow modified: (i) in the presence of nonlocal vertices essential singularities at infinity are introduced; (ii) in Lee-Wick theories pairs of complex conjugate poles appear and their locations discriminate between computations performed in the Minkowski or Euclidean signature, preventing the use of the usual Wick rotation~\cite{Anselmi:2017yux,Anselmi:2017lia}; (iii) in fakeon models the Feynman shift in the propagator is replaced by an alternative prescription that introduces fictitious pairs of complex conjugate poles such that only a one-way non-analytic Wick rotation from Euclidean to Minkowski can be well-defined~\cite{Anselmi:2018kgz}. 

An important message we would like the reader to appreciate is the following: the fact that  amplitudes can be defined in both Minkowski and Euclidean, and that they can be analytically continued from one signature to the other either way, is a very special feature of quantum field theories with two-derivative kinetic operators and local vertices. However, this happens to be just a coincidence. Indeed, when going beyond standard theories and working in more complicated setups, e.g. with higher powers of inverse momentum in the propagator and/or nonlocal vertices, it becomes clear that amplitudes must be initially defined in Euclidean signature, and that a consistent set of (alternative) rules must be prescribed in order to define the analytic continuation to Minkowski. This also means that the definition of an amplitude and the prescription to deform the integration contour can be model-dependent.

There are, in fact, at least two further reasons to justify the formulation of a generic  quantum field theory (local or nonlocal) and the definition of its amplitudes in the Euclidean signature. In addition to the fact that in more general theories unitarity can be respected only if amplitudes are initially defined in Euclidean space, we also point out that:
\begin{itemize}
	
	\item Typically the functional path integral is convergent only if the theory is initially defined in Euclidean signature;
	
	\item Strictly speaking the power counting analysis and the discussion/proof of renormalizability of a theory is well-defined only in the Euclidean signature.
	
\end{itemize}
Therefore, the failure of the Minkowski prescription in alternative theories, such as the ones investigated in Sec.~\ref{sec-nonlocal-vertex}, is not really a problem because the corresponding quantum field theory can be consistently formulated in Euclidean space. Physical observables depending on real momenta can be obtained via analytic continuation according to a certain prescription.

Let us also clarify that our discussion does not imply that string theory must be initially defined in Euclidean space. In fact, in the context of string perturbation theory the failure of the Minkowski prescription should not be seen as a problem of the general framework, but only as an issue of the string field theory formulation in which the exponential vertices appear. In fact, in other formulations amplitudes might be well-defined in Minkowksi already from the beginning, e.g., this can be the case in the worldsheet formulation.

\paragraph{Outlook.} In this paper, we only worked with propagators possessing the standard real poles. However, more general nonlocal theories admitting extra pairs of complex conjugate poles have been proposed~\cite{Buoninfante:2018lnh,Buoninfante:2020ctr}. It will be very interesting to investigate whether the prescriptions analysed in this work can be generalized in the presence of extra complex conjugate poles, and whether a unitary $S$-matrix can be formulated for these more complicated theories too~\cite{Buoni-Yama}. 

Furthermore, we only focused on scalar theories but it would be interesting to apply Euclidean and Schwinger prescriptions to amplitudes involving more complicated tensorial structures.
In the recent years several approaches to formulate a unitary and renormalizable quantum field theory of the gravitational interaction have been proposed. In particular, interesting ideas to solve the ghost problem~\cite{Stelle:1976gc} in higher-derivative gravity  were put forward in both local~\cite{Anselmi:2017ygm,Anselmi:2017lia,Modesto:2015ozb,Modesto:2016ofr,Donoghue:2019fcb,Salvio:2018crh,Mannheim:2021oat} and nonlocal theories~\cite{Tomboulis:1997gg,Biswas:2005qr,Modesto:2011kw,Biswas:2011ar}, and also at the non-perturbative level~\cite{Becker:2017tcx,Platania:2020knd}. In the context of nonlocal gravity explicit computations of amplitudes, e.g. at one loop, are still lacking in the literature. For future works it will be very interesting  to perform these computations and extract from them information on physical observable such as decay rates and cross sections for processes involving (nonlocal) gravitons.

Hence, many interesting aspects concerning scattering amplitudes and unitarity in nonlocal quantum field theories deserve further investigation, and indeed more work is expected to come in the future.


\subsection*{Acknowledgments}
The author is grateful to Sravan Kumar and Ashoke Sen  for comments.
Nordita is supported in part by NordForsk.


\appendix

\section{Unitarity \& optical theorem}\label{sec-app}

We briefly review the notion of unitarity of the $S$-matrix, and verify explicitly some of the statements made in the main text about optical theorem and Cutkosky rules in the context of nonlocal field theories.  This Appendix is also important to fix notations and conventions used for propagators, vertices, and amplitudes in the main text.

Given a Fock space $W,$  the $S$-matrix operator can evolve an initial state $\left| a\right\rangle\in W$  into a final state $\left| b\right\rangle= S\left| a\right\rangle \in W.$ Then, probability conservation  $\left\langle b|b\right\rangle = \left\langle a|a\right\rangle$ implies the unitarity condition on the $S$-matrix~\cite{tHooft:1973wag}:
\begin{equation}
S^\dagger S= \mathds{1}\,.\label{unitarity-S}
\end{equation}
By writing $S=\mathds{1}+iT,$ where $T$ is the so-called transfer matrix, we can recast the unitarity condition as 
\begin{equation}
i\left(T^\dagger-T\right)= T^\dagger T\,,\label{opt-the}
\end{equation}
which is the operatorial form of the \textit{optical theorem}. The  identity~\eqref{opt-the} is very useful to prove unitarity in perturbative quantum field theories~\cite{Anselmi:2016fid}, and to compute cross sections and decay rates~\cite{Itzykson:1980rh}.

Let us now introduce the matrix $\mathcal{M},$ whose elements are defined through the following relation
\begin{equation}
\left\langle b|T|a \right\rangle= (2\pi)^4\delta^{(4)}(P_b-P_a)\left\langle b|\mathcal{M}|a \right\rangle\,,
\label{feynm-diagr}
\end{equation}
where $P_b$ and $P_a$ are the total outgoing and ingoing momenta, respectively. Moreover, we use the completeness relation
\begin{equation}
\mathds{1}=\sum\limits_{\left\lbrace n \right\rbrace}\prod\limits_{l=1}^n\int \frac{{\rm d}^3k_l}{(2\pi)^3}\frac{1}{2\omega_l} \left|\left\lbrace k_l\right\rbrace  \right\rangle \left\langle \left\lbrace k_l\right\rbrace \right|\,,
\label{complet-integ}
\end{equation}
where the summation is over all possible sets $\left\lbrace n \right\rbrace $ of intermediate states $\left|\left\lbrace k_l\right\rbrace\right\rangle $ containing $l$  momenta, and $\omega_l=\sqrt{\vec{k}_l^2+m_l^2}$ are the frequencies (energies) for each single momentum $k_l.$ 

By using Eqs.~\eqref{feynm-diagr} and~\eqref{complet-integ}, we can recast Eq.~\eqref{opt-the} as follows
\begin{equation}
 i\left[\left\langle b|\mathcal{M}^{\dagger}|a \right\rangle - \left\langle b|\mathcal{M}|a \right\rangle \right]\!=\!\sum\limits_{\left\lbrace n \right\rbrace}\prod\limits_{l=1}^n\int\! \frac{{\rm d}^3k_l}{(2\pi)^3}\frac{1}{2\omega_l} 
(2\pi)^4 \delta^{(4)}\left(P_a-\sum_{l=1}^nk_l\right) \left\langle b|\mathcal{M}^{\dagger}|\left\lbrace k_l\right\rbrace  \right\rangle \left\langle \left\lbrace k_l\right\rbrace|\mathcal{M}|a \right\rangle\,,
\label{optical theorem-feynman}
\end{equation}
which must hold order by order in perturbation theory. 

The simplest application of the formula~\eqref{optical theorem-feynman} is the case in which the l.h.s. is equal to $2\,{\rm Im} \left\lbrace \left\langle b|\mathcal{M}|a \right\rangle  \right\rbrace .$ Some examples are given by the $2\rightarrow 2$ one-loop diagram constructed with quartic vertices, and the $1\rightarrow 1$ one-loop diagram constructed with cubic vertices; see Fig.~\ref{fig1} for an illustration of such diagrams.

The diagrammatic form of the optical theorem for the amplitudes associated with the Feynman diagrams in Fig.~\ref{fig1} reads
\begin{eqnarray}
&&2\hspace{0.01in}\text{Im}\left[ (-i)\raisebox{-1mm}{\scalebox{2}{$\rangle%
		\hspace{-0.06in}\bigcirc\hspace{-0.06in}\langle$}}\,\right]=\hspace{0.01in}%
\raisebox{-1mm}{\scalebox{2}{$\rangle\hspace{-0.06in}\bigcirc\hspace{-0.1955in}%
		{\tiny\begin{array}{c} \begin{sideways} -\,-\,-\,-\, \end{sideways}\end{array}}\hspace{-0.039in}\langle$}} =\int \mathrm{d}\Pi _{b}\hspace{0.01in}%
\left\vert \raisebox{-1mm}{\scalebox{2}{$\rangle\hspace{-0.035in}\langle$}}^b_b%
\right\vert ^{2}\,, \label{cutd2}\\[2mm]
&&2\hspace{0.01in}\text{Im}\left[ (-i)\raisebox{-1mm}{\scalebox{2}{$-%
		\hspace{-0.06in}\bigcirc\hspace{-0.06in}-$}}\right]=\hspace{0.01in}%
\raisebox{-1mm}{\scalebox{2}{$-\hspace{-0.06in}\bigcirc\hspace{-0.1955in}%
		{\tiny\begin{array}{c} \begin{sideways} -\,-\,-\,-\, \end{sideways}\end{array}}\hspace{-0.035in}-$}} =\int \mathrm{d}\Pi _{b}\hspace{0.01in}%
\left\vert \raisebox{-1mm}{\scalebox{2}{$-\hspace{-0.035in}\langle$}}^b_b%
\right\vert ^{2}\,, 
\label{cutd3}
\end{eqnarray}
where $\int {\rm d}\Pi_b$ is a short notation for the phase-space integral in Eq.~\eqref{optical theorem-feynman}, and the sum is taken over the final states labeled by $b.$ The intermediate steps of the above two equations take into account of the so-called cutting rules (or Cutkosky rules)~\cite{Cutkosky:1960sp}, where the matrix elements $\left\langle \left\lbrace k_l\right\rbrace|\mathcal{M}|a \right\rangle$ and $\left\langle b|\mathcal{M}^{\dagger}|\left\lbrace k_l\right\rbrace  \right\rangle=(\left\langle \left\lbrace k_l\right\rbrace|\mathcal{M}|b \right\rangle)^*$ contribute to the left and to the right of the cut dashed line, respectively. For instance, the square on the r.h.s. of Eq.~\eqref{cutd3} means $\small{\hspace{0.01in}\Big| \raisebox{-1mm}{\scalebox{2}{$-\hspace{-0.035in}\langle$}}\Big| ^{2}= \left( \raisebox{-1mm}{\scalebox{2}{$-\hspace{-0.035in}\langle$}}\right) \left( \raisebox{-1mm}{\scalebox{2}{$-\hspace{-0.035in}\langle$}}\right)^*},$ with $\left( \raisebox{-1mm}{\scalebox{2}{$-\hspace{-0.035in}\langle$}}\right)^*=\raisebox{-1mm}{\scalebox{2}{$%
		\rangle\hspace{-0.035in}-$}},$  and similarly for~\eqref{cutd2}. 

In this work we were interested in investigating contour prescriptions to evaluate one-loop integrals such as~\eqref{1-loop-nonlocal} where interaction vertices are nonlocal (i.e. non-polynomial in momenta). Let us now check the validity of the two identities~(\ref{cutd2}) and~(\ref{cutd3}) for a generic class of nonlocal field theories, and thus prove one-loop unitarity explicitly.
For instance, we can work with a nonlocal model such as the one in Eq.~\eqref{nonlocal-lag-1} with cubic potential, thus the bubble diagram is the one in Fig.~\ref{fig1b} and in Eq.~\eqref{cutd3}; the same result will be valid for Fig.~\ref{fig1a} and Eq.~\eqref{cutd2}.  See also Ref.~\cite{Pius:2016jsl} for the first satisfactory discussion, to our knowledge,  of unitarity with nonlocal vertices in the context of string field theory. 

The Feynman rules for the Lagrangian~\eqref{nonlocal-lag-1} with cubic potential are:
\begin{eqnarray}
{\rm Propagator:}&&\qquad \Pi(k^2)=\frac{-i}{k^2+m^2-i\epsilon}\,;\nonumber\\[2mm]
{\rm Vertex:}&&\qquad V(k_1,k_2,k_3)=-i\lambda\,e^{-\frac{1}{2}\big[\gamma(k_1^2)+\gamma(k_2^2)+\gamma(k_3^2)\big]}\,.
\end{eqnarray}
Given an external momentum $p$ and a loop momentum $k,$ the bubble amplitude reads
\begin{eqnarray}
	\mathcal{M}(p^2)\equiv \left\langle p |\mathcal{M}|p\right\rangle 
	=(-i)(-i\lambda)^2\int_\mathcal{C} \frac{{\rm d}k^0}{2\pi}\int_{\mathbb{R}^3} \frac{{\rm d}^3k}{(2\pi)^3} \frac{(-i)e^{-\gamma(k^2)}}{k^2+m^2-i\epsilon}\frac{(-i)e^{-\gamma((p-k)^2)}}{(p-k)^2+m^2-i\epsilon}\,,
	\label{1-loop-nonlocal-app}
\end{eqnarray}
Obviously, if $\gamma=0$ we recover the local case in Eq.~\eqref{1-loop-local}.
To prove that the amplitude~\eqref{1-loop-nonlocal-app} satisfies the optical theorem~\eqref{optical theorem-feynman}, and thus unitarity, we should verify the identity~\eqref{cutd3}, i.e. the validity of the Cutkosky rules for the diagram in Fig.~\ref{fig1b}.

In Sec.~\ref{sec-nonlocal-vertex} we computed the imaginary part of $\mathcal{M}(p^2),$ see Eq.~\eqref{imag-eucl-NL}. Thus, we already know the result for the l.h.s. of the optical theorem~\eqref{optical theorem-feynman}:
\begin{eqnarray}
	{\rm l.h.s.}&=&i\left[\mathcal{M}^*(p)-\mathcal{M}(p)\right]\nonumber\\[2mm]
	&=& 2{\rm Im}[\mathcal{M}(p^2)]\nonumber\\[2mm]
	&=&2\pi\lambda^2 \int_{\mathbb{R}^3} \frac{{\rm d}^3k}{(2\pi)^3}\frac{1}{2\omega_{\vec{k}}2\omega_{\vec{p}-\vec{k}}}\delta(p^0-\omega_{\vec{k}}-\omega_{\vec{p}-\vec{k}}) \,,
	\label{lhs-one-loop}
\end{eqnarray}
where we have used the relation $\left\langle p |\mathcal{M}^\dagger|p\right\rangle=(\left\langle p |\mathcal{M}|p\right\rangle)^*=\mathcal{M}^*(p^2).$

Let us now evaluate the r.h.s. of~\eqref{optical theorem-feynman}. For the bubble diagram, we have one set of intermediate states characterized by $n=2,$ $l=1,2,$ internal momenta $k_1,$ $k_2,$ frequencies $\omega_1=\sqrt{\vec{k}_1^2+m^2},$ $\omega_2=\sqrt{\vec{k}_2^2+m^2},$ and $P_a=p:$
\begin{equation}
	{\rm r.h.s.}=  \int_{\mathbb{R}^3} \frac{{\rm d}^3k_1}{(2\pi)^3}\int_{\mathbb{R}^3} \frac{{\rm d}^3k_2}{(2\pi)^3}\frac{1}{2\omega_1 2\omega_2} (2\pi)^4\delta^{(4)}(p-k_1-k_2) \left\langle p|\mathcal{M}^{\dagger}|k_1,k_2 \right\rangle \left\langle k_1,k_2|\mathcal{M}|p \right\rangle\,.\label{rhs-two-loop}
\end{equation}
By using the relations
\begin{equation}
	\left\langle k_1,k_2|\mathcal{M}|p \right\rangle=(-i)\raisebox{-1mm}{\scalebox{2}{$-\hspace{-0.035in}\langle$}}=(-i)V(p,k_1,k_2)=-\lambda \,e^{-\frac{1}{2}\big[\gamma(k_1^2)+\gamma(k_2^2)\big]}\,
\end{equation}
and
\begin{equation}
	\left\langle p |\mathcal{M}^\dagger|k_1,k_2\right\rangle=\left(\left\langle k_1,k_2 |\mathcal{M}|p\right\rangle\right)^*= (i)\left( \raisebox{-1mm}{\scalebox{2}{$-\hspace{-0.035in}\langle$}}\right)^*=(i)V(p,k_1,k_2)^*=-\lambda\, e^{-\frac{1}{2}\big[\gamma(k_1^2)+\gamma(k_2^2)\big]}\,,
\end{equation}
where we imposed the normalization $\gamma(-m^2)=0,$ and going from a three-dimensional to a four-dimensional integral $\int \frac{{\rm d}^3k_i}{(2\pi)^3} \frac{1}{2\omega_i}= \int \frac{{\rm d}^4k_i}{(2\pi)^4}2\pi \theta(k^0_i) \delta(k_i^2+m^2),$ we can recast~\eqref{rhs-two-loop} as 
\begin{eqnarray}
	{\rm r.h.s.}&=& (2\pi)^2 \lambda^2\int \frac{{\rm d}^4k}{(2\pi)^4}e^{-\gamma(k^2)}e^{-\gamma((p-k)^2)}\theta(k^0)\delta(k^2+m^2)\theta(p^0-k^0)\delta((p-k)^2+m^2) \nonumber \\[2mm]
	&=& (2\pi)^2 \lambda^2\int \frac{{\rm d}^4k}{(2\pi)^4}\theta(k^0)\delta(k^2+m^2)\theta(p^0-k^0)\delta((p-k)^2+m^2)
	\,,\label{loop-two-cutk-form}
\end{eqnarray}
where we replaced $k_2=p-k_1,$ redefined $k\equiv k_1,$ and imposed again $\gamma(-m^2)=0.$

Then, by using the property $\delta(x^2-y^2)=[\delta(x+y)+\delta(x-y)]/2|y|$ Eq.~\eqref{loop-two-cutk-form} becomes
\begin{eqnarray}
	{\rm r.h.s.}&=& (2\pi)^2\lambda^2 \int \frac{{\rm d}^4k}{(2\pi)^4} \theta(k^0)\theta(p^0-k^0) \frac{1}{2\omega_{\vec{k}}2\omega_{\vec{p}-\vec{k}}}\left[\delta(k^0-\omega_{\vec{k}})+\delta(k^0+\omega_{\vec{k}})\right]\nonumber\\[2mm]
	&&\qquad\times \left[\delta(p^0-k^0-\omega_{\vec{p}-\vec{k}})+\delta(p^0-k^0+\omega_{\vec{p}-\vec{k}})\right]   \nonumber\\[2mm]
	&=&2\pi\lambda^2 \int_{\mathbb{R}^3} \frac{{\rm d}^3k}{(2\pi)^3}\frac{1}{2\omega_{\vec{k}}2\omega_{\vec{p}-\vec{k}}}\delta(p^0-\omega_{\vec{k}}-\omega_{\vec{p}-\vec{k}}) \,,
	\label{rhs-two-loop-2}
\end{eqnarray}
which coincides with Eq.~\eqref{lhs-one-loop}. Thus, we have shown that the optical theorem is satisfied for the bubble diagram in the nonlocal theory~\eqref{nonlocal-lag-1},  i.e. we verified the identities~\eqref{cutd2} and~\eqref{cutd3}.

Furthermore, the Cutkosky rules~\cite{Cutkosky:1960sp} are manifest, especially from Eq.~\eqref{loop-two-cutk-form}. This means that when computing the imaginary part, or more precisely the anti-hermitian part of the amplitude (i.e., l.h.s. of the optical theorem), each internal propagator can be replaced as 
\begin{equation}
\frac{-i}{k^2+m^2-i\epsilon}\quad\rightarrow\quad  2\pi\theta(k^0)\delta(k^2+m^2)\,. \label{cut-rule}
\end{equation}
%

\bibliographystyle{utphys}
\bibliography{References}


\end{document}